\RequirePackage{fix-cm}
\documentclass[smallextended]{svjour3}       % onecolumn (second format)
\smartqed  % flush right qed marks, e.g. at end of proof
\usepackage{hyperref}
\usepackage{graphicx}
\usepackage{amssymb}
\usepackage{natbib}
\usepackage{xcolor}
\usepackage{pdflscape}
\def\hst        {{HST}\/}
\def\chandra    {{Chandra}\/}
\def\xmm        {{XMM-Newton}\/}
\def\rosat      {{ROSAT}\/}
\def\galex      {{GALEX}\/}
\def\her        {{Herschel}\/}

\newcommand{\hi}{H\,{\sc i}}
\newcommand{\hii}{H\,{\sc ii}}

\newcommand{\oii}{[O\,{\sc ii}]}

\newcommand{\ovi}{[O\,{\sc vi}]}
\newcommand{\nii}{[N\,{\sc ii}]}
\newcommand{\cii}{[C\,{\sc ii}]}
\newcommand{\civ}{[C\,{\sc iv}]}

\hypersetup{
     colorlinks=true,
     linkcolor=blue,
     filecolor=blue,
     citecolor = blue,      
     urlcolor=cyan,
     }
\begin{document}

\title{Ram Pressure Stripping in High-Density Environments}

\author{Alessandro Boselli         \and
        Matteo Fossati             \and
        Ming Sun
}


\institute{A. Boselli \at
               Aix Marseille Univ, CNRS, CNES, LAM, Marseille, F-13013 France \\
              \email{alessandro.boselli@lam.fr}           %  \\
           \and
           M. Fossati \at
               Universit\'a di Milano-Bicocca, Piazza della Scienza 3, I-20100 Milano, Italy \\
               INAF-Osservatorio Astronomico di Brera, via Brera 28, I-20121 Milano, Italy \\
            \email{matteo.fossati@unimib.it}
            \and
            M. Sun \at
            Department of Physics \& Astronomy, University of Alabama in Huntsville, 301 Sparkman Drive, Huntsville, AL 35899, USA \\
            \email{ms0071@uah.edu}
}

\date{Received: date / Accepted: date}

\maketitle

\begin{abstract}
Galaxies living in rich environments are suffering different perturbations able to 
drastically affect their evolution. Among these, ram pressure stripping, i.e. the 
pressure exerted by the hot and dense intracluster medium (ICM) on galaxies moving at high velocity 
within the cluster gravitational potential well, is a key process able to remove their 
interstellar medium (ISM) and quench their activity of star formation. This review is aimed at
describing this physical mechanism in different environments, from rich clusters of galaxies 
to loose and compact groups. We summarise the effects of this perturbing process on the baryonic
components of galaxies, from the different gas phases (cold atomic and molecular, ionised, hot)
to magnetic fields and cosmic rays, and describe their induced effects on the different stellar 
populations, with a particular attention to its role in the quenching episode generally observed
in high density environments. We also discuss on the possible fate of the stripped material 
once removed from the perturbed galaxies and mixed with the ICM, and we try
to estimate its contribution to the pollution of the surrounding environment. Finally, combining 
the results of local and high redshift observations with the prediction of tuned models and
simulations, we try to quantify the importance of this process on the evolution of galaxies
of different mass, from dwarfs to giants, in various environments and at different epochs.

\end{abstract}

\tableofcontents

\newpage

\section{Introduction}
\label{sec:intro}

The environment plays a major role in shaping galaxy evolution. Since the seminal work of \citet{Dressler80}, it is now well established that galaxies in high density regions have undergone a different evolution than their counterparts in the field. Rich environments are indeed dominated by early-type galaxies (morphology-density relation, e.g., \citealt{Dressler80, Postman84, Postman05, Whitmore93,Dressler97}) and quiescent systems \citep[e.g.,][]{Lewis02, Gomez03, Peng10}.
Systematic differences between cluster and field galaxies are also observed in the late-type population. Indeed, in high density environments spirals have, on average, a lower atomic \citep[e.g.,][]{Haynes84, Cayatte90, Solanes01, Gavazzi05, Catinella13} and molecular \citep[e.g.,][]{Fumagalli09, Boselli14a} gas content than similar objects in the field. As a consequence, their star formation activity is also reduced, especially when the molecular gas reservoir needed to feed star formation is affected \citep[e.g.,][]{Kennicutt83, Gavazzi98, Boselli14b, Boselli16}.

Various perturbing mechanisms have been proposed to explain the different evolution of galaxies in high density regions \citep[e.g.][]{Boselli06, Boselli14c}. These can be divided into two main families, 
namely the gravitational perturbations with other group or cluster members \citep[e.g.][]{Merritt83}, with the gravitational potential well of the high density region itself \citep[e.g.][]{Byrd90}, or their combined effect in multiple fly-by encounters, generally called galaxy harassment \citep[e.g.][]{Moore96,Moore98}, or the hydrodynamic interaction of galaxies moving at high velocity ($\sigma$ $\simeq$ 500-1000 km s$^{-1}$) within the hot ($T_{\rm ICM}$ $\simeq$ 10$^7$-10$^8$ K) and dense ($\rho_{\rm ICM}$ $\simeq$ 10$^{-3}$ cm$^{-3}$) intracluster medium (ICM) permeating high density regions \citep[e.g.][]{Sarazin86}. The hydrodynamic interaction of galaxies
with the hot surrounding ICM can make the cold interstellar medium (ISM) evaporate (thermal evaporation, e.g. \citealt{Cowie77}), or simply remove it because of the external pressure exerted on the galaxy ISM by the surrounding medium (ram pressure, e.g., \citealt{Gunn72}). The hydrodynamic friction at the interface between the cold ISM and the hot ICM in galaxies moving at high velocities can create different physical instabilities in the gas 
%(viscosity, e.g., \citealt{Nulsen82}, 
(Rayleigh-Taylor, e.g., \citealt{Roediger08a}, and Kelvin-Helmholtz instabilities, e.g. \citealt{Livio80, Nulsen82, Mori00}) which contribute to the removal of gas from the galaxy disc.
Finally, it has also been proposed that once galaxies become satellites of a larger halo, the gas infall is suppressed, thus reducing the supply of fresh material to the galaxy disc on longer timescales and thus reducing the rate of formation of new stars (starvation or strangulation, e.g.,  \citealt{Larson80, Bekki02, van-den-Bosch08, Peng15}). \textit{The main difference between these two families of mechanisms is that while gravitational perturbations act indifferently on all the different galaxy components (dark matter, stars, gas, dust), hydrodynamic interactions affect only the diffuse interstellar medium \citep[e.g.][]{Boselli06}.} Although second order effects such as induced star formation events in the compressed gas at the interface between the hot ICM and the cold ISM (see Sec. \ref{subsec:indsfr}) or the perturbation of the wind produced during the mass loss of individual evolved stars (see Sec. \ref{subsec:stars}) are possible, this remains the main difference between hydrodynamic and gravitational perturbations, and will be used throughout this work to identify galaxies undergoing an RPS process.

%\begin{figure}
%\centering
%\includegraphics[width=1.0\textwidth, angle=0]{N1265_google.pdf}
%\vspace{-0.5cm}
%\caption{The 20 cm (1.4 GHz) radio continuum image of the radio galaxy NGC 1265 in the %Perseus cluster, a typical example of head tail radio galaxy in a nearby cluster formed by %the ram pressure exerted on the radio jets by the surrounding ICM \citep[e.g.][]{Ryle68, %Miley75}. Despite its large radial velocity with respect to the cluster ($\sim$ 2300 km %s$^{-1}$), ram pressure is not able to displace the old stellar distribution, which remains %perfectly symmetric.
%}
%\label{NGC1265}       
%\end{figure}

First proposed in the seventies to explain the peculiar morphology of head-tail radio galaxies \citep[e.g.][]{Gunn72}, %(see Fig. \ref{NGC1265})
RPS is often considered as the dominant perturbing mechanism in nearby rich clusters \citep[e.g.][]{Vollmer01, Boselli06, Gavazzi13a, Boselli14b} where the high velocity of satellite galaxies moving within the dense ICM make the ideal conditions for a stripping process to be effective. Tuned models and simulations indicate that this mechanism can remove the cold gas reservoir of galaxies of different mass depending on their local environment, from more massive galaxies in clusters down to dwarf systems orbiting within the halo of a more massive galaxy whenever combined with tidal forces \citep[e.g.][]{Mayer06}. It is, however, still unclear whether ram pressure, or more in general the hydrodynamic interaction of the cold galaxy ISM with the hot surrounding ICM, is the most effective perturbing mechanism in lower mass structures such as loose \citep[e.g.][]{Catinella13, Brown17} and compact groups \citep[e.g.][]{Verdes-Montenegro01}, or in high redshift clusters still under formation through the assembly of smaller groups, where gravitational perturbations were probably present (pre-processing, e.g. \citealt{Dressler04, Fujita04}). The identification of the dominant perturbing mechanism in galaxies of different mass (from giant to dwarfs) and belonging to different environments (from massive clusters to loose groups) at different epochs is also made difficult by the fact that the different perturbing mechanisms described above, and now commonly included in hydrodynamic cosmological simulations or semi-analytic models, are all active at the same time with additive effects \citep[e.g.][]{Boselli06, Weinmann10, Bahe15, Marasco16, Trayford16, van-de-Voort17, Henriques15, Henriques17, Cortese21}. Furthermore, the fate of the stripped material once removed from the galaxy disc remains unclear. Observations and simulations indicate that this gas can either fall back on the disc once the galaxy reaches the outskirts of the cluster along its orbit \citep[e.g.][]{Vollmer01}, or produce long cometary tails where, under some still unclear conditions, star formation can or not take place \citep[e.g.][]{Boissier12, Jachym14, Fossati16, Boselli16, Poggianti19}.

The recent availability of increasingly sensitive ground- and space-based imaging and spectroscopic instruments allowed the completion of large and deep multifrequency surveys both of local clusters: Virgo (e.g. ACSVCS \citealt{Cote06}, VIVA \citealt{Chung09}, HeViCS \citealt{Davies10}, GUViCS \citealt{Boselli11a}, NGVS \citealt{Ferrarese12}, VESTIGE \citealt{Boselli18}), Fornax (e.g. FCOS \citealt{Mieske04}, ACSFCS \citealt{Jordan07}, HeFoCS \citealt{Davies13}, 
NGFS \citealt{Munoz15}, FDS \citealt{Iodice16}, the MeerKAT Fornax Survey \citealt{Serra16}, Fornax3D \citealt{Sarzi18}, AlFoCS, \citealt{Zabel19}), Coma \citep{Carter08, Yagi10, Smith10, Chiboucas11, Koda15, van-Dokkum15a, Yagi16, Gavazzi18a, Zaritsky19, Chen20, Lal20}, A1367 \citep{Yagi17, Scott18}, and of higher redshift clusters \citep[e.g.][]{Muzzin12, van-der-Burg13, van-der-burg18, Balogh14, Balogh17, Nantais16, Rudnick17, Galametz18}. These large observational efforts aimed at building statistical samples have been ideally complemented with targeted observations at extraordinary angular resolution of representative objects \citep[e.g.][]{Sun10, Fumagalli14, Poggianti17, Fossati18} where the role of the environment on their evolution can be studied in great detail. The increasing level of detail and complexity of these observational datasets required a similar advancement of theoretical and numerical models needed for their interpretation. This lead to a new class of hydrodynamic simulations \citep[e.g.][]{Vogelsberger14, Schaye15} and semi-analytic models of galaxy formation \citep[e.g.][]{Weinmann10, DeLucia12, Guo13a, Henriques15, Henriques17, Hirschmann16} which are now able to reproduce with exquisite resolution the dark matter and baryonic components (stars, dust, cold, ionised, and hot gas) of galaxies in different density regions \citep[e.g.][]{Luo16, Xie18, Stevens19, Yun19, Troncoso-Iribarren20}, offering the community with a unique set of tools to significantly advance in the study of the role of the environment on galaxy evolution.
The synergy of the multifrequency data and of the simulations allow us to quantify the importance of the different perturbing mechanisms on large statistical samples and at the same time to study the detailed physical process down to the scale of giant molecular clouds and star forming \hii\ regions. For these several reasons, significant improvements have been done in the recent years in the study of the RPS process. In particular, a growing interest on this mechanism comes from the discovery of spectacular tails of gas associated to some nearby cluster galaxies seen in cold atomic \citep[e.g.][]{Chung07}, ionised \citep[e.g.][]{Gavazzi01, Yagi10, Boselli16}, and hot \citep[e.g.][]{Sun06, Sun07} gas, which identified with no ambiguity several objects now undergoing a stripping event. These observational results have been mirrored by high resolution hydrodynamic simulations now able to reproduce the different gas phases and the star formation process in the stripped material down to the scale of individual \hii\ regions \citep[e.g.][]{Roediger05, Tonnesen09, Tonnesen12, Roediger15}.

It is thus time to summarise these important results reached by the astrophysical community in a review entirely dedicated to the RPS process. The purpose of this work is giving a general view of this major perturbing mechanism, summarising the properties of the physical process by means of simple analytical prescriptions, models, and more complex simulations and explain how multifrequency data can be used to identify galaxies undergoing an RPS event. The same data can be used to derive important physical parameters necessary to describe the process and to constrain tuned models and simulations. Within this review we also try to explain which are the major effects of the perturbation on the evolution of galaxies and study the fate of the stripped gas. And finally, we try to combine all this information to derive the relationship between ram pressure and galaxy mass, halo mass, and redshift to understand and quantify which is the statistical importance and contribution of this mechanism on the evolution of galaxies. We try to do this exercise by summarising and combining the results of the growing number of papers dedicated to this interesting process, and combine their results in a coherent and self consistent picture on galaxy evolution in dense environments. 
%Despite we try to do this exercise as objectively as possible, this work might reflect our personal view and might thus be biased vs. the results which we obtained during more than twenty years of work on this topic. We tried to give credit to most of the main references available in the literature on  this subject, but given their more and more growing number, we have certainly missed some of them, and we apologise for that.
%{\color{red}{MS: Do we really need to say the last two sentences? In the end, it won't change anything --- people still think we missed their papers. And we are supposed to be unbiased as much as we can, which goes without explanation.}}

The paper is structured as follows: in Sec. 2 we describe the physical process and its relations to the properties of galaxies (mass and morphological type) and of the high density regions (mass, velocity dispersion, gas density and temperature). In Sec. 3 we summarise the multifrequency observational evidence indicating that RPS is now at place or it has contributed to modify the evolution of galaxies in dense regions. In Sec. 4 we discuss the impact of RPS on galaxy evolution, while in Sec. 5 the impact of the process on the surrounding environment. Finally, in Sec. 6 we try to derive the statistical importance of this mechanism in shaping the evolution of galaxies of different mass, in different density regions, and at different epochs. The final discussion and the conclusions are given in Sec. 7 and 8, where we compare the effects of RPS to those induced by other main perturbing mechanisms in high density regions.

%Kravtsov \& Borgani 2012 for clusters and groups formation

\section{The physical process}
\label{sec:physicalproc}

High density regions are generally characterised by a hot ($T \sim 10^7~-~10^8$ K) and dense ($n_{\rm ICM}$ $\sim$ 10$^{-4}$ - 10$^{-2}$ cm$^{-3}$) ICM trapped within their gravitational potential well \citep[e.g.][]{Sarazin86}. Given its temperature, the ICM emits X-ray emission with a distribution roughly corresponding to the density square projected along the line of sight, 
%emits for bremsstralung in X-rays with a smooth, diffuse distribution picked in the highest density region and 
extending up to the virial shock.
%(see Sect. \ref{subsec:prophidens}).
A gas cloud (e.g., a galaxy with the ISM remained) moving with a velocity $V$ relative to the ICM suffers a drag force. The drag force (by the ions in the ICM) exerts the ram pressure on the cloud that can be written as (when the relative velocity is normal to the cloud surface)\footnote{Ram pressure of the ICM comes from ions in the ICM (mainly protons), while the contribution from electrons is tiny (similar to the ICM viscosity). On the other hand, the thermal pressure of the ICM has roughly the equal contribution from free electrons and ions.}:

\begin{equation}
P = \rho_{\rm ICM} V^2
\end{equation}

\noindent
which is able to remove the gaseous component of its ISM whenever it overcomes the gravitational forces keeping the gas anchored to the stellar disc of the galaxy.
According to \citet{Kuzmin56} the gravitational potential $\Phi$ of an infinitely thin stellar disc of radius $R$ at a distance $z$ perpendicular to the disc is:
\begin{equation}
\Phi(R,z) = -\frac{GM_{\rm star}}{\sqrt{R^2+ (a+z)^2}}
\end{equation}
\noindent
where $a \geq 0$ is the radial scale-length. The condition for stripping the gas at a given radius $R$ are satisfied whenever \citep{Koppen18}:
\begin{equation}
P \geq \Sigma_{\rm gas} \Big|\frac{\partial \Phi(R,z)}{\partial z}\Big|_{\rm max} = G  \Sigma_{\rm gas} M_{\rm star} \Big| \frac{(a+z)}{(R^2+(a+z)^2)^{3/2}} \Big|_{\rm max} 
\end{equation}
\noindent
which occurs at a height $z_{\rm max}$ from the disc given by $\partial^2 \Phi/\partial z^2 = 0$. \citet{Roediger05} analysed the variation of $z_{\rm max}$ as a function of radius showing that for increasing $R$ the maximum gravitational resistance occurs at larger distances from the disc. Assuming an exponential profile for the disc density, and defining the maximum restoring force above the disc (at given R) as a threshold gives the traditional
\citet{Gunn72} criterion for instantaneous stripping \citep[see also][]{Fujita99,Yamagami11}: 
\begin{equation}
\rho_{\rm ICM} V_{\perp}^2 > 2 \pi G \Sigma_{\rm star} \Sigma_{\rm gas} = \frac{v_{\rm rot}^2 \Sigma_{\rm gas}}{R_{\rm gal}}
\end{equation}

\noindent
where $\Sigma_{\rm star}=v_{\rm rot}^2/2 \pi G R_{\rm gal}$ (from Virial equilibrium), and $\Sigma_{\rm gas}$ are the stellar and total gas surface densities, $v_{\rm rot}$ the rotational velocity
of the galaxy, and $R_{\rm gal}$ is the disc scale radius. The RPS thus depends on both the properties of the high density region ($\rho_{\rm ICM}$),
on the motion of the galaxy within it ($V$, where $V_{\perp}$ indicate the component perpendicular to the galaxy disc plane), and on the physical properties of the galaxies themselves ($\Sigma_{\rm star}$, $\Sigma_{\rm gas}$, $v_{\rm rot}$, and $R_{\rm gal}$).
It is important to underline that ram pressure is an hydrodynamic interaction between two different gas phases, the (mainly cold) ISM of the perturbed galaxy with
the (mainly hot) ICM. For this reason stars, which have a very small cross section to the gas flow and much higher internal pressure than the external ram pressure, are unperturbed during the interaction. Second order effects, due to the heating of the stellar disc after the stripping of the gaseous disc, can be present (see Sec. \ref{subsec:kinematics}). They can induce a mild change of the orbits of the stars which produces ultra diffuse discs \citep{Safarzadeh17}. These effects on the structure of the stellar disc are expected to be more important in dwarf systems, where the gravitational potential well of the galaxy is the shallower. 

RPS is a general process in astronomy whenever there is a relative bulk motion between a moving object and its surroundings, e.g., 
mergers of galaxy clusters and groups (elongation and trails on the X-ray morphology), bent radio jets and the stripped ISM at the scale of galaxies, pulsar wind nebulae, stellar wind (including the heliosphere) at the scale of stars and compact objects. RPS can also work on individual stars, especially on those with strong winds.
%(the analogy with galaxy would be galaxies with strong galactic wind in stripping). 
One of the best examples at the scale of stars is the Mira star with a 4 pc tail \citep{Martin07} that is believed to be mainly composed of H$_{2}$ gas\footnote{The ram pressure Mira experiences is only $\sim 1.3 \times 10^{-11}$ dyn (assuming \hi\ gas), which is only typical for the RP experienced by cluster galaxies at the cluster outskirts. Thus, an AGB star like Mira soaring in the ICM with a velocity of $\sim$ 2000 km s$^{-1}$ can have a tail of $\sim$ 0.06 kpc.}.
In this review, however, we focus on the RPS process occurring on galaxy scales.

%Probably can discuss the effect of RP on individual stars (with winds or not, stars with strong winds are like galaxies with strong outflows) somewhere --- compare the typical ICM RPS with the ISM RPS of Mira star etc. 
%For Mira, $n_{\rm ISM}$ = 0.05 cm$^{-2}$ and the velocity of the star relative to the surrounding ISM is 125 km/s. Thus, the ram pressure Mira experiences is only $\sim 1.3 \times 10^{-11}$ dyn (assuming \hi\ gas), which is only typical for the RP experienced by cluster galaxies at the cluster outskirts.
%Mira has a 4 pc tail that is believed to be mainly composed of H$_{2}$ gas, with the {\em Galex} emission from the fluorescent H$_{2}$ lines after collisional excitation at the interface of the cool AGB wind and the surrounding hot gas.
%{\color{red}{do we have sth. similar in RPS tails? only by Galex FUV}}
%Thus, such AGB stars soaring in the ICM with a velocity of $\sim$ 2000 km/s can have tails of $\sim$ 0.06 kpc.
%https://ui.adsabs.harvard.edu/abs/2019ApJ...887...41L/abstract
%https://ui.adsabs.harvard.edu/abs/2007Natur.448..780M/abstract
%This example suggests that individual stars can leave sizable trails while soaring in the ICM.

\subsection{Properties of high-density regions}
\label{subsec:prophidens}

The external pressure acting on the disc of spiral galaxies inhabiting clusters depends on two main parameters, the ICM density $\rho_{\rm ICM}$ that is tied to the ICM gas fraction of the cluster and the velocity of the galaxy with respect to the ICM $V$ that is tied to the cluster mass.
%are tightly connected with the overall properties of the massive dark matter halo.}
We thus summarize some cluster properties and relations useful for this work, while readers are referred to dedicated cluster review papers \citep[e.g.][]{Sarazin86,Voit05,Kravtsov12} for more detail.
Two fundamental properties of clusters are mass and size, which are often connected. While new characteristic sizes of clusters like the mean matter density radius and splashback radius have been proposed \citep[e.g.][]{Kravtsov12}, we still use the virial radius in this work. Particularly, the virial radius adopted in this work is $r_{200}$ ($r_{\Delta}$ as the radius enclosing an overdensity of $\Delta$ times the critical density of the Universe at that redshift). $r_{500}$ and $r_{101}$ are also widely used in cluster studies. Typically $r_{500} / r_{200} \sim$ 2/3 and $r_{500} / r_{101} \sim$ 1/2 for clusters.\footnote{Different definitions of the virial radius were used in different works, e.g., $r_{200}$, $r_{180}$ and $r_{101}$ (see e.g., \citealt{Bryan98} for detail).} The relation between any two over-density radii can be derived by assuming a total mass model. For the NFW model, $r_{2500}/r_{500} = 0.405$, $r_{500}/r_{200} = 0.638$ and $r_{200}/r_{101} = 0.730$ for $c_{200} = 3$, where $c_{200} = r_{200} / r_{\rm s}$, with $r_{\rm s}$ as the characteristic radius in the NFW profile.
$r_{2500}/r_{500} = 0.479$, $r_{500}/r_{200} = 0.677$ and $r_{200}/r_{101} = 0.755$ for $c_{200} = 8$.

By definition:

\begin{equation}
M_{\Delta} = \Delta \: \frac{4}{3} \, \pi \, r_{\Delta}^{3} \: \rho_{\rm crit}(z)
\end{equation}

\noindent
where $\rho_{\rm crit}(z)$ is the critical density of the Universe at redshift $z$.
For example, $M_{200}$ = 1.14$\times10^{14}$ M$_{\odot}$ ($r_{200}$/Mpc)$^{3} E(z)^{2}$, where:
\begin{equation}
E(z)^{2} = \Omega_{\rm M} (1+z)^{3} + \Omega_{\Lambda}
\end{equation}

\begin{table}
\caption{Cluster parameters used in this work}

\label{TabNFW}
{
\[
\begin{tabular}{lccccc}
\hline
\noalign{\smallskip}
\hline
Cluster		& Luminosity distance  & $r_{200}$	& $M_{200}$		& Ref	\\
 		& (Mpc)       & (Mpc)		& (10$^{14}$ M$_{\odot}$)		&	\\
\hline
\noalign{\smallskip}
Virgo		& 16.5      & 0.974		& 1.06	& 1, 2  \\
Norma       & 69.6      & 1.80      & 6.75  & 3, 4  \\
Coma		& 100       & 1.97		& 8.90	& 5, 6	\\
A1367       & 92.2      & 1.41      & 3.26  & 7, 4  \\
Group       & -         & 0.70      & 0.39  & 8     \\

%Virgo	     & 16.5$^a$  &799  & 8.6	& 0.974		& 1.06	& 1	\\
%Norma       & 69.6      &925  &       & 1.80      &  6.75     &       \\
%Coma		 & 101       &873  & 4.0  & 1.97		& 8.90	& 2, 3	\\
%A1367       & 95.8      &726  &       & 1.41  & 3.26  &       \\
%Fornax (Group)  & 20.0  &374  & 8.0  & 0.70  & 0.39      &  4 \\
%A1689		 &   888     &                         & 7.3	& 2.1 & 13 & 5,6	\\
%Update Sept 21, Figures updated to this values, The scale radius of the NFW profile, $r_s$ = $R_{200}$ / $c_{200}$.
\hline
\end{tabular}
\]
%\scriptsize
References: 1) \citet{Gavazzi99} and \citet{Mei07} 2) Adapted from \citet{Simionescu17} using a distance of 16.5 Mpc; 3) \citet{Woudt08}; 4) The mass values of the Norma cluster and A1367 are estimated with the $M-T$ relation from \cite{Sun09} with $T$ of 5.6 keV and 3.6 keV respectively. 5) \citet{Colless96}; 6) \citet{Planck13} and \citet{Churazov21}; 7) \citet{Cortese04}; 8) \citet{Sun09}, galaxy group with a mass similar to Fornax.
Notes: all values are measured using H$_0$ = 70 km s$^{-1}$ Mpc$^{-1}$, $\Omega_{\rm M}$ = 0.3, $\Omega_{\rm \Lambda}$ = 0.7.}
\end{table}

% now discussion on mass function and mass profile
In this work, we use several systems with different mass as examples (see Table~\ref{TabNFW}).
%1) Coma: $R_{200}$ = 1.97 Mpc and $M_{200} = 8.90\times10^{14}$ M$_{\odot}$ \cite{Planck13,Churazov21};
%2) Virgo: $R_{200}$ = 974 kpc and $M_{200} = 1.06\times10^{14}$ M$_{\odot}$ \cite{Simionescu17}.
The dominant mass component in clusters is dark matter ($\sim$ 85\% of the total mass).
The dark matter distribution in clusters is often approximated with a NFW radial profile \citep{Navarro97} of the form:
\begin{equation}
\rho(r) = \rho_{\rm crit}\frac{\delta_{\rm c}}{r/r_{\rm s}(1+r/r_{\rm s})^2}
\label{NFW0}
\end{equation}
\noindent
with
\begin{equation}
\delta_{\rm c} = \frac{200}{3}\frac{c^3}{\ln(1+c)-c/(1+c)}
\label{NFW1}    
\end{equation}
\noindent
where $c$ = $r_{\rm 200}/r_{\rm s}$ is the concentration parameter (or $c_{200}$ here as $r_{200}$ is adopted), $r_{\rm s}$ the scale radius of the density distribution and $\rho_{\rm crit}$ is again the critical density of the Universe. For this analytic density distribution,
the total mass within a radius $r$ is given by the relation:
\begin{equation}
M(<r) = M_0 \left[\ln\left(1+\frac{r}{r_s}\right) - \frac{r/r_{\rm s}}{1+r/r_{\rm s}}\right]  ~~~ \rm{for} ~~~\it{r\leq r_{\rm s}c}
\label{NFW2}
\end{equation}
with
\begin{equation}
M_0 = 4\pi\frac{3H_0^2}{8\pi G}E(z)^2\frac{200c^3r_{\rm s}^3}{3\left[\ln(1+c)-c/(1+c)\right]}
\label{NFW3}
\end{equation}

Other forms of mass distribution for dark matter halos have also been suggested (see e.g., \citealt{Merritt06} for details).

\begin{figure}
\centering
\includegraphics[width=1.0\textwidth]{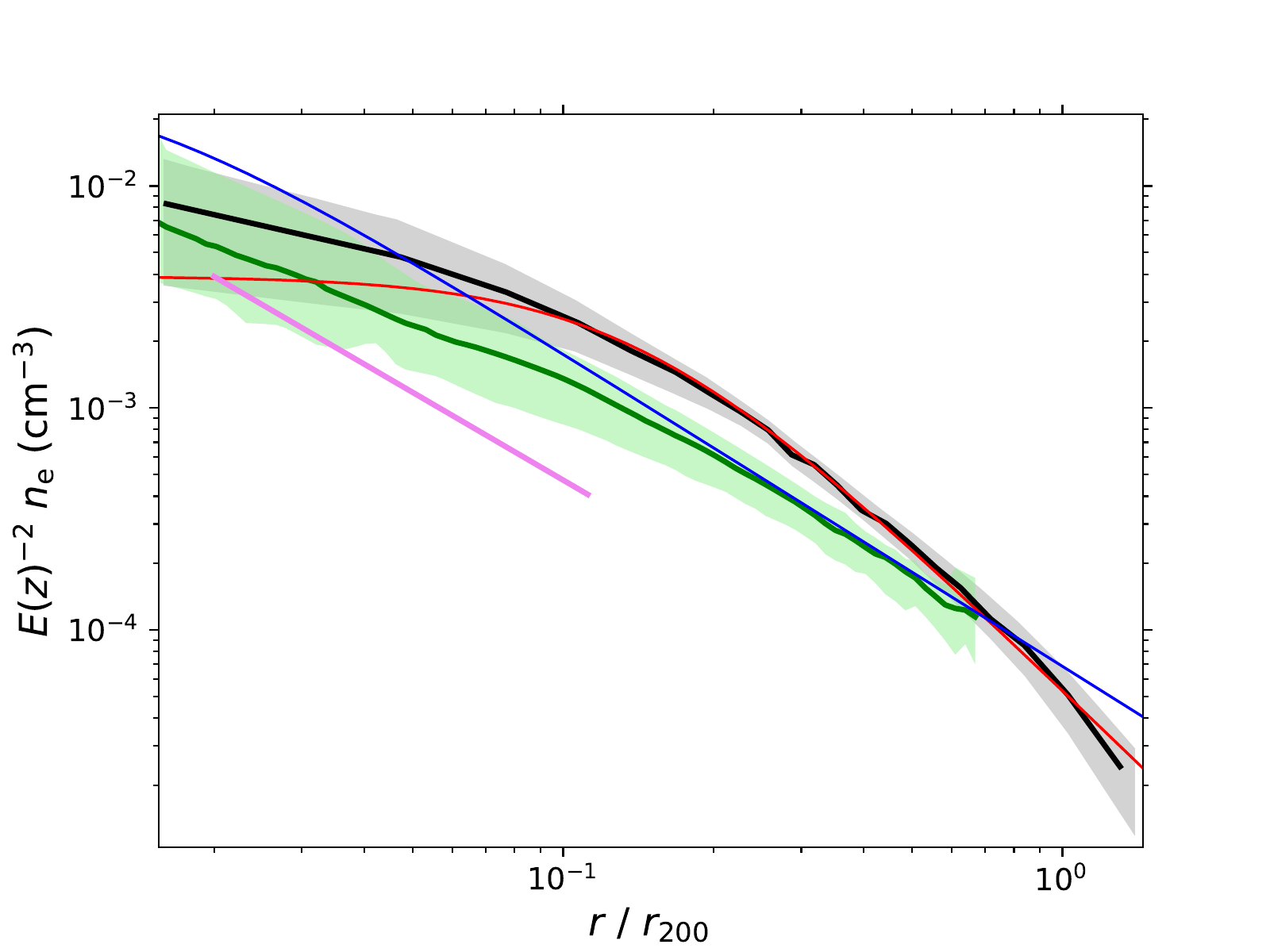}
\vspace{-0.5cm}
\caption{The typical electron density radial profile of galaxy clusters, groups and massive spiral galaxies. The black profile and the grey shaded area show the median density and 1$\sigma$ scatter for a sample of 320 galaxy clusters from \cite{Morandi15}.
The red curve shows the electron density radial profile of the Coma cluster from the eRosita data \citep{Churazov21}, as an example of clusters with a flat central core lacking a large, dense cool core}. The blue curve shows the electron density radial profile of the Virgo cluster from the \rosat{} data \citep{Schindler99}, as an example of clusters hosting a dense, peaked cool core.
The green profile and the lightgreen shaded area show the median density and 1$\sigma$ scatter for a sample of 43 galaxy groups from \cite{Sun09}.
The magenta profile shows the the electron density radial profile of a massive spiral galaxy NGC~6753 from \cite{Bogdan17}.
The hot gas content in the inner regions generally decreases with the decreasing mass, likely demonstrating  the increasing impact of baryon physics with decreasing mass.
\label{ne_all}       
\end{figure}

Most baryons in clusters are in the hot ICM emitting X-rays. The classical model on the ICM density profile is the $\beta$-model (see e.g., \citealt{Sarazin86} for details), with the ICM density proportional to $r^{-3 \beta}$ asymptotically beyond the central core. However, a single $\beta$-model (or even a double $\beta$-model) has been shown to be too simple to describe the ICM density profile, with $\beta$ typically increasing with radius at the cluster outskirts \citep[e.g.,][]{Croston08,Morandi15}.
A typical ICM density profiles is shown in Fig.~\ref{ne_all}, from stacking the \chandra{} data of 320 clusters. This median density profile can be approximated with a form first introduced by \cite{Patej15}: 

\begin{equation}
E(z)^{-2} n_{\rm e} (x) = 0.00577 (\frac{x}{0.201})^{-0.150} (\frac{x}{0.265})^{-0.0638} [1+0.759(\frac{x}{0.201})^{0.949}]^{-2.936}
\end{equation}

\noindent
where $x = \frac{r}{r_{200}}$. This fit is good at $x = 0.02 - 1.3$ and $\beta$ increases to 1 at the cluster outskirts ($n_{\rm e} \propto r^{-3 \beta}$), same as the asymptotic behaviour of the NFW profile.
\footnote{Note that many recent cluster works adopted the ICM density profile first proposed by \cite{Vikhlinin06} (their equation 3). However, unlike the model proposed by \cite{Patej15}, this model is not physically motivated. Nevertheless, the best fit is also given here: $E(z)^{-2} n_{\rm e} (x) = 0.00472 \Big(0.0818(\frac{x}{0.0735})^{-0.769} [1+(\frac{x}{0.0735})^{2}]^{-0.897} [1+(\frac{x}{0.280})^{3}]^{-0.570}\Big)^{1/2}$, good for $x = 0.02 - 1.1$. Note that the central component of the \cite{Vikhlinin06} model is not included as the density profile does not include the very central core.
}
There is significant scatter on the ICM density profiles (as well as azimuthal variations not shown here), especially around the cluster centre (e.g., cool core as in the Virgo cluster vs. non-cool core as in the Coma cluster). With the relations presented above we can write:
\begin{equation}
 \rho_{\rm ICM} = 1.92 ~ \mu ~ n_{\rm e} ~ m_{\rm p} = 1.15 \mbox{ } n_{\rm e} ~ m_{\rm p}   
 \label{eq:rhoicm}
\end{equation}
where $\mu=0.60$ is the mean molecular weight for a metallicity of 0.3 - 1.0 $\rm Z_\odot$ with the solar abundance table from \cite{Asplund09}.

The self-similar relations for properties of galaxy clusters, originally proposed by \cite{Kaiser86} \citep[also see reviews in e.g.,][]{Bohringer12,Kravtsov12}, treat galaxy groups and clusters of different masses as identical objects after scaled by their mass. The useful self-similar relations for this work are summarized here (at a fixed over-density $\Delta$). \\
Total mass and the ICM mass: 
\begin{equation}
M_{\Delta} \mbox{  }(\mbox{or } M_{\rm ICM, \Delta}) \propto E(z)^{-1}
\label{eq:Mdelta}
\end{equation}
However, this evolution solely comes from the definition of $M_{\Delta}$ while observations suggest stronger evolution than $E(z)^{-1}$ from the cluster growth \citep[e.g.,][]{Kravtsov12}. The ICM properties typically depend on both mass and redshift.\footnote{We did not use the equation~\ref{eq:Mdelta} in the following equations as we want to keep the mass dependency and the equation~\ref{eq:Mdelta} is not correct for real clusters.}

Radius: 
\begin{equation}
r_{\Delta} \propto M_{\Delta}^{1/3} E(z)^{-2/3} %\propto E(z)^{-1} 
\end{equation}

%X-ray bolometric luminosity: 
%\begin{equation}
%L_{{\rm X}, \Delta} \propto M_{\Delta}^{4/3} %E(z)^{7/3} %\propto E(z)
%\end{equation}
%
ICM density: 
\begin{equation}
%n_{\rm e} \propto M_{\Delta}^{0} E(z)^2
n_{\rm e} \propto E(z)^2
\end{equation}

ICM temperature: 
\begin{equation}
T \propto M_{\Delta}^{2/3} E(z)^{2/3}
%\propto E(z)^{0}
\end{equation}

ICM pressure: 
\begin{equation}
P \propto n_{\rm e} T \propto M_{\Delta}^{2/3} E(z)^{8/3} 
\label{eq:P_ICM}
%\propto E(z)^{2}
\end{equation}

Note that the ICM density is independent of mass with the self-similar assuption.
Assuming dark matter particles, ICM and cluster galaxies are all virialized in the same potential,

Galaxy velocity dispersion:
\begin{equation}
\sigma_{\rm V} \propto T^{1/2} \propto M_{\Delta}^{1/3} E(z)^{1/3}
%\propto E(z)^{0}
\end{equation}

The galaxy velocity dispersion is a robust measure of the cluster mass, with the expected self-similar relation \citep[e.g.,][]{Armitage18}. On the other hand, the observed $\sigma_{\rm V} - T$ correlation is typically steeper than the self-similar relation with an average slope of 0.6 - 0.7 \citep[e.g.,][]{Lovisari21}. Using $\sigma_{\rm V}$ as the typical velocity of cluster galaxies, ram pressure should follow the same relation as the ICM pressure.
The above relation would also suggest the crossing time in the cluster, 2$R_{\Delta} / \sigma_{\rm V} \propto E(z)^{-1}$, and is independent of the mass. However, the $E(z)^{-1}$ factor simply comes from the definition of the cluster size ($r_{\Delta}$), also as the free-fall time of the cluster has the same $E(z)^{-1}$ dependency.

The above self-similar relations serve as the simplest model for the ICM mass scaling and evolution, while the actual ICM scaling and evolution can be different from scale-dependent physical processes (e.g., baryonic processes like radiative cooling, AGN feedback and galactic wind) and the fact that some assumptions for the self-similar model are not accurate \citep[e.g.,][]{Kravtsov12}. Observations have shown the importance of baryon physics, most significant in low-mass systems, e.g., elevated hot gas entropy in low-mass systems, a significant mass dependency on the hot gas fraction from massive galaxies, to galaxy groups and clusters and an observed $L_{\rm X} - M$ relation steeper than the self-similar relation \citep[e.g.,][]{Sun12,Lovisari21,Eckert21}. We also plot the median density profile of local galaxy groups from \cite{Sun09} in Fig.~\ref{ne_all}, as well as the density profile of a massive spiral. One can clearly see the mass dependency of the density profile, unlike the simple self-similar relation.
% no need to show the simple powerlaw fit of the group density profile
The mass dependency of the ICM density likely varies with radius \citep[e.g.,][]{Sun12}.
If we assume $n_{\rm e} \propto M^{\sim 0.3}$ from \cite{Sun12} at $r \sim r_{2500}$ and $\sigma_{\rm V} \propto M^{1/3}$ (still follow the self-similar relation), $P_{\rm RP} \propto M^{\sim 1.0}$, steeper than the expected relation from the self-similar relations.
The observed $T - M$ relations typically have a slope of $\sim$ 0.6, somewhat smaller than 2/3 \citep[e.g.,][]{Sun12,Lovisari21}. Thus, observations reveal departure from the self-similar relations of equations 15 - 18, as the ICM properties are not only determined by gravitational processes. 

The redshift evolution of the ICM and cluster properties demonstrates the roles of RPS with redshift. 
% $n_{\rm e} \propto E(z)^2$ (as the Universe was smaller in the past)
Recent sample studies suggested self-similar evolution for the ICM density at $r >$ 0.2 $r_{500}$, but weak to no evolution within 0.2 $r_{500}$ \citep{McDonald17,Ghirardini21}, which only occupies 0.24\% of the cluster volume within $r_{200}$. Simulations also suggest self-similar evolution to $z \sim 2.5$ \citep[e.g.,][]{Mostoghiu19}.
%Better and large cluster samples at high $z$ (mass selected) would be required
The evolution is significant, $E(z)^{2}$=3.1 and 8.8 at $z=1$ an $z=2$ respectively (for $\Omega_{\rm M} = 0.3$ and $\Omega_{\rm \Lambda} = 0.7$).
This suggests a stronger role of RPS at higher $z$, at fixed cluster mass, although the abundance of massive clusters decreases with $z$.

Besides the relations of global properties of the ICM, the efficiency of the micro transport processes in the ICM is also important in the studies of RPS and the associated tails.
Giving the low ICM density, the mean free path of particles in the ICM is large \citep{Sarazin86}: 
\begin{equation}
\lambda = 23\mbox{ } (\frac{kT_{\rm ICM}}{10^{8}\mbox{ } \rm{K}})^2 \mbox{ }(\frac{n_{\rm e}}{10^{-3} \mbox{ }\rm{cm^{-3}}})^{-1} ~\rm{kpc}
\end{equation}
This is comparable to the sizes of galaxies. However, magnetic field in the ICM significantly affects the micro transport processes in the ICM \citep[e.g.,][]{Donnert18}. While studies with the Chandra and XMM data on the ICM suggest suppression of heat conductivity and viscosity in the ICM \citep[e.g.,][]{Markevitch07,Simionescu19}, the most definite constraints require high energy resolution X-ray spectroscopy at arcsec scales, coupled with the high-angular resolution X-ray morphological studies. On the other hand, potentially kinematic and morphological studies of RPS tails can also put constraints on the efficiency of the micro transport processes in the ICM.

\subsection{Galaxies physical parameters}
\label{subsec:galpars}

The gravitational forces keeping the ISM anchored to the stellar disc of the galaxies can be derived from the observed radial variation of gas and stellar surface density
or of the rotation curve. 
The radial variation of these three variables in unperturbed late-type systems is well known, and can be derived from observational data or analytically predicted using simple relations. 
The stellar and total gas components can be well represented by an exponentially declining disc, in particular in the outer regions, those most easily perturbed by
an RPS event \citep{Freeman70, Bigiel12, Wang14}. For these distributions the gas and stellar density profiles \citep{Domainko06, Cortese07} are:

\begin{equation}
\Sigma_{gas,star}(r) = \frac{M_{gas,star}}{2\pi R_{0 gas,star}^2} ~ \exp \left(-\frac{R}{R_{0 gas,star}}\right)
\end{equation}

\noindent
where $R_{0 gas,star}$ is the scale-length of the exponential profile of the gas and stellar distribution (for alternative scaling relations, see \citealt{Wang16, Stevens19}). Similarly, the radial variation of the rotational velocity of galaxies of different mass or luminosity can be parametrised using the Polyex function \citep{Catinella06} or the universal rotation curve function \citep{Persic91}.
From the above relations \citet{Domainko06} also derived the radius outside which gas can be stripped $r_{\rm strip}$:

\begin{equation} \label{eq:rstrip}
R_{\rm strip} \geq \left(1 + \frac{R_{\rm 0,star}}{R_{\rm 0,gas}}\right)^{-1} \ln \left(\frac{G M_{\rm star} M_{\rm gas}}{2\pi \rho_{\rm ICM} V_{\rm perp}^2 R_{\rm 0,star}^2 R_{\rm 0,gas}^2}\right)
\end{equation}

\noindent
By analysing a sample of 33 nearby galaxies with available \hi\ and CO resolved maps from the THINGS and HERACLES surveys, \citet{Bigiel12} have shown that 
the total gas surface density (total gas = 1.36(HI+H$_2$) to account for Helium) radial profile has a fairly universal form given by the relation:

\begin{equation}
\Sigma_{\rm gas}(R) = 2.1 \Sigma_{\rm trans} ~ \exp \left(-\frac{1.65 r}{R_{25}}\right)
\end{equation}

\noindent 
where $\Sigma _{\rm trans}$ is the surface density of the total gas when $\Sigma_{\rm HI}$ = $\Sigma_{\rm H_2}$ and $r_{25}$ is 
the isophotal radius measured in the $B$-band at 25 mag arcsec$^{-2}$. Given the shape of the radial profile, they have also derived 
that the total gas mass is given by the relation \citep{Bigiel12}:

\begin{equation}
M_{\rm gas} = 2 \pi \times 2.1 \times \Sigma_{\rm trans} \times R_{25}^2 \times X
\end{equation}

\noindent 
with $X$ changing with radius ($X$ = 0.31 at 2$\times$ $R_{25}$). They have also shown that $\Sigma _{\rm trans}$ is fairly constant in galaxies
of different morphological type:

\begin{equation}
\Sigma_{\rm trans} \simeq 14 ~~\rm{M_{\odot} pc^{-2}}
\end{equation}
 
\begin{figure}
\includegraphics[width=0.99\textwidth]{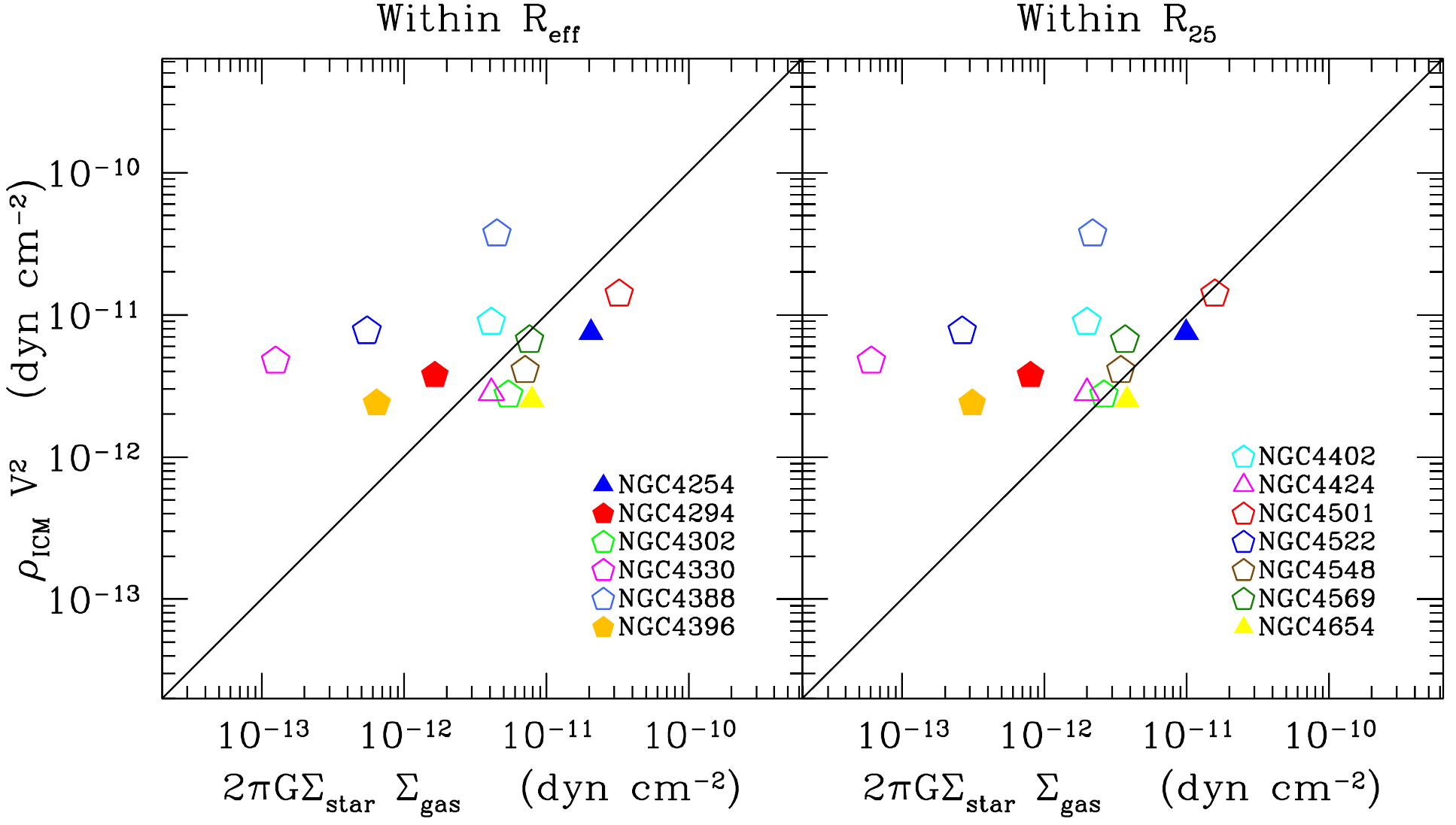}
\caption{Comparison between the external pressure exerted on different Virgo cluster galaxies suffering an RPS event (listed in Table \ref{tab:RPSgalaxies}) and moving at a velocity $V$ within the ICM ($\rho_{\rm ICM}V^2$, where $\rho_{\rm ICM}$ is the density of the ICM from Equation \ref{eq:rhoicm} here measured at the projected distance of the galaxy from the cluster centre) and the gravitational forces keeping 
the gas anchored to the stellar disc 2$\pi G \Sigma_{\rm star} \Sigma_{\rm gas}$, in units of dyn cm$^{-2}$. $V$ is the 3D velocity of galaxies within the cluster, and it is here defined as $V$ = $\sqrt(3)$ $\times$ $|v_{\rm Galaxy}-v_{\rm Virgo}|$, where $v_{\rm Galaxy}$ and $v_{\rm Virgo}$ are the observed heliocentric velocities of the galaxies and of the Virgo cluster (955 km s$^{-1}$), whenever $|v_{\rm Galaxy}-v_{\rm Virgo}|$ $>$ $\sigma_{\rm Virgo}$ = 799 km s$^{-1}$, otherwise $V$ = $\sqrt(3)$ $\times$ $\sigma_{\rm Virgo}$.} In the left panel the gravitational forces are measured within the effective radius, in the right panel within the 25 mag arcsec$^{-2}$ radius.
Empty symbols are for HI-deficient galaxies, filled ones for HI-normal
objects. Pentagons indicate galaxies undergoing an RPS event, triangles objects suffering RPS combined with a gravitational perturbation.
\label{ramnew}       % Give a unique label
\end{figure}

\noindent
Using standard scaling relations such as those reported in \citet{Boselli14a} we can also estimate the total gas mass for a galaxy of given stellar mass $M_{\rm star}$:

\begin{equation}
M_{\rm gas} = 10^{\left(0.26 \log M_{\rm star} + 7.03\right)}
\end{equation}

\noindent
with both entities expressed in solar units, when $M_{\rm gas}$ has been derived using the luminosity dependent $X_{\rm CO}$ conversion factor given in \citet{Boselli02a}.
Using the relations above and considering that in an exponentially declining stellar disc with a typical central surface brightness of $\mu_0(B)$ = 21.65 mag arcsec$^{-2}$
\citep{Freeman70} $r_{25}$ = 3.1 $R_{\rm 0,star}$, we can derive the typical scalelength of the stellar disc expressed as a function of the total gas mass:

\begin{equation}
R_{\rm 0,star}  [\rm{pc}] = \left(\frac{\it{M_{\rm gas}}}{550}\right)^{0.5}
\end{equation}

\noindent
with $M_{\rm gas}$ in solar units. These leads to:

\begin{equation}
\Sigma_{\rm gas} [\rm{M_{\odot}~{\rm pc^{-2}}}]= 29.4 ~ \exp \left(-\frac{R}{R_{\rm 0,gas}}\right)
\end{equation}

\noindent
with $R_{\rm 0,gas}$ = 1.89 $R_{\rm 0,star}$\footnote{We recall that in such a configuration $R_{\rm 0,star}$ = 0.32 $R_{25}$, the effective radius (radius including half of the total galaxy luminosity),
$R_{eff}$ = 0.59 $R_{25}$, and $R_{\rm 0,gas}$ = 0.61 $R_{25}$.}, consistent with the assumption made by \citet{Cortese07} based on the observed relation between isophotal optical and \hi\ radii derived by \citet{Cayatte94}  ($R_{25}/r_\mathrm{HI}\simeq$ 1.8 measured at a column density of 10$^{20}$ cm$^{-2}$, corresponding to 0.8 M$_{\odot}$ pc$^{-2}$).
As an example, Fig. \ref{ramnew} shows the relation between the gravitational forces derived using resolved stellar mass surface densities (from \citet{Cortese12}) and \hi\ column densities (from the VIVA survey, \citet{Chung09}) and the ram pressure measured at the cluster projected distance for the sample of Virgo cluster late-type galaxies undergoing an RPS event
listed in Table \ref{tab:RPSgalaxies}. The left panel shows the relation derived within the effective radius, while the right panel within the 25 mag arcsec$^{-2}$ isophotal radius.
Figure \ref{ramnew} clearly shows that the external pressure is sufficient to remove the diffuse ISM in the outer disc in all galaxies, but also within the effective radius in most of the objects (NGC 4294, 4330, 4388, 4396, 4402, 4522).

All these scaling relations allow us to express $R_{\rm strip}$, $M_{\rm strip}$ (mass of gas stripped during the interaction):

\begin{equation}
M_{\rm strip} = M_{\rm gas}\left[\left(\frac{R_{\rm strip}}{R_{\rm 0,gas}} + 1\right) ~ \exp \left(-\frac{R_{\rm strip}}{R_{\rm 0,gas}}\right)\right]
\end{equation}

\noindent
and $M_{\rm retain}$ (mass of gas retained on the galaxy disc):

\begin{equation}
M_{\rm retain} = M_{\rm gas}\left[1 - \left(\frac{R_{\rm strip}}{R_{\rm 0,gas}} + 1\right) ~ \exp \left(-\frac{R_{\rm strip}}{R_{\rm 0,gas}}\right)\right]
\end{equation}

\noindent
as a function of a single galaxy parameter $M_{\rm star}$. Despite being a simplification of a more complex reality (galaxy discs are not infinitely thin and stable and we neglected the role of bulges), these relations are successful in reproducing the stripping radius of objects which are clearly undergoing RPS \citep[see e.g.][]{Arrigoni-Battaia12, Fossati12}.

\subsection{Region of action}
\label{subsec:regaction}

Taking advantage of the galaxy scaling relations and properties shown in Sections \ref{subsec:galpars} and \ref{subsec:prophidens} and of the typical density profiles of the ICM in clusters we can now study the impact of RPS on galaxies as a function of their position and orbit within the cluster, of their stellar mass, and of the mass of the host halo. 

\begin{figure}
\centering
\includegraphics[width=0.75\textwidth]{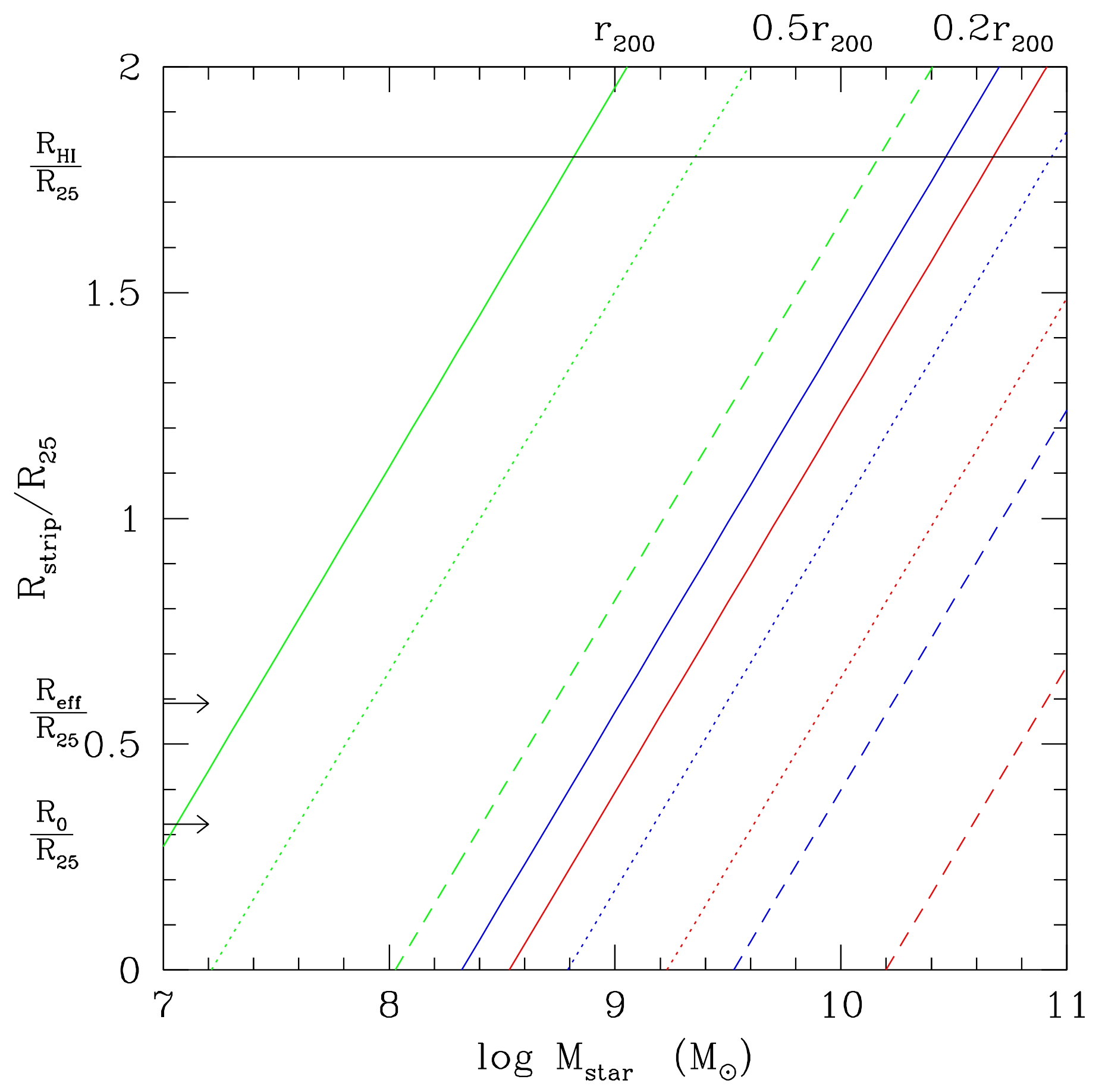}
\caption{Expected variation of the stripping radius $R_{\rm strip}$ normalised to the $B$-band isophotal radius at 25 mag arcsec$^{-2}$ as a function of the stellar mass $M_{\rm star}$ for galaxies
in the Coma cluster (red), in the Virgo cluster (blue), and in a representative group of total mass $M_\mathrm{tot}\simeq 4 \times 10^{13}~\mathrm{M}_{\odot}$ (green)
derived using the scaling relations given above. The external pressure is estimated assuming as typical velocity of galaxies within the different clusters $V$ = $\sqrt{3}$ $\times$ $\sigma_{LT}$, with $\sigma_{LT}$ the mean velocity dispersion of late-type systems.}
The expected variations are measured at $0.2 \times r_{200}$ (dashed), $0.5 \times r_{200}$ (dotted), and $r_{200}$ (solid). All gas outside the predicted lines
can be potentially removed during an RPS event. The arrows on the Y-axis indicate the effective radius-to-isophotal radius $R_\mathrm{eff}$/$R_{25}$ and the exponential scalelength-to-isophotal radius $R_{\rm 0,star}$/$R_{25}$ for an exponentially declining stellar profile of central surface brightness $\mu_0(B)$ = 21.65 mag arcsec$^{-2}$. The black horizontal solid line indicates
the typical HI-to-isophotal radius $R_{HI}$/$R_{25}$ = 1.8 measured at a column density of $\Sigma_{\rm HI} = 10^{20}~{\rm cm}^{-3}~(0.8~{\rm M}_{\odot}~{\rm pc}^{-2})$ by \citet{Cayatte90}
\label{modellorho}       
\end{figure}

Figure \ref{modellorho} shows the expected variation of the stripping radius as a function of the stellar mass for galaxies in three different environments, the Coma cluster, the Virgo cluster, and a representative group of total mass $M_{\rm group}$ $\simeq$ 4 $\times$ 10$^{13}$ M$_{\odot}$ comparable to the Fornax cluster (whose parameters are given in Table \ref{TabNFW}) derived using the scaling relations given above. The expected variations are measured at 0.2 $\times$ $r_{200}$, 0.5 $\times$ $r_{200}$, and $r_{200}$, where the density of the ICM, $\rho_{\rm ICM}$, is derived from the X-rays radial profile given in Fig. \ref{ne_all} assuming equ.~\ref{eq:rhoicm} 
%$\rho_{\rm ICM} = 1.15$ $n_{\rm e}$ $m_{\rm p}$ (or $\mu$ = 0.60)
and a velocity within the cluster $V$ = $\sqrt{3}$ $\times$ $\sigma_{LT}$, where $\sigma_{LT}$ is the line of sight velocity dispersion of the high density region for the late-type galaxy population.

Figure \ref{modellorho} clearly shows that, in a massive cluster such as Coma, all star forming galaxies of stellar mass $M_{\rm star}$ $\leq$ 10$^{8.3}$ M$_{\odot}$ 
can be completely stripped of their total gas content up to the virial radius of the cluster $r_{200}$, those of stellar mass $M_{\rm star}$ $\leq$ 10$^{10.2}$ M$_{\odot}$ 
only if they cross the cluster in the inner regions ($r$ $\leq$ 0.2$\times$$r_{200}$).
In the same cluster stripping is also able to remove all the gas located outside the optical disc ($\simeq$ 50 \% of the total gas content) 
in galaxies of stellar mass $M_{\rm star}\leq 10^{9.5}$ M$_{\odot}$ up to the cluster virial radius.
As expected, the gas stripping efficiency decreases in lower-density environments but it is still dominant in clusters of the mass of Virgo ($M_{\rm cluster}\simeq 10^{14}$ M$_{\odot}$), and can be important
in groups in low mass galaxies crossing the ICM in the inner regions. In this configuration ($r \leq 0.2 \times r_{200}$), all galaxies of stellar mass $M_{\rm star} \leq 10^{8.0}$ M$_{\odot}$
can be completely stripped of their gas, while those with $M_{\rm star} \leq 10^{9.1}$ M$_{\odot}$ stripped of their gas located outside the optical radius. For a more accurate analytical description of the process, we refer the reader to \citet{Hester06}. 

\begin{figure*}
    \centering
    \includegraphics[width=1.0\textwidth]{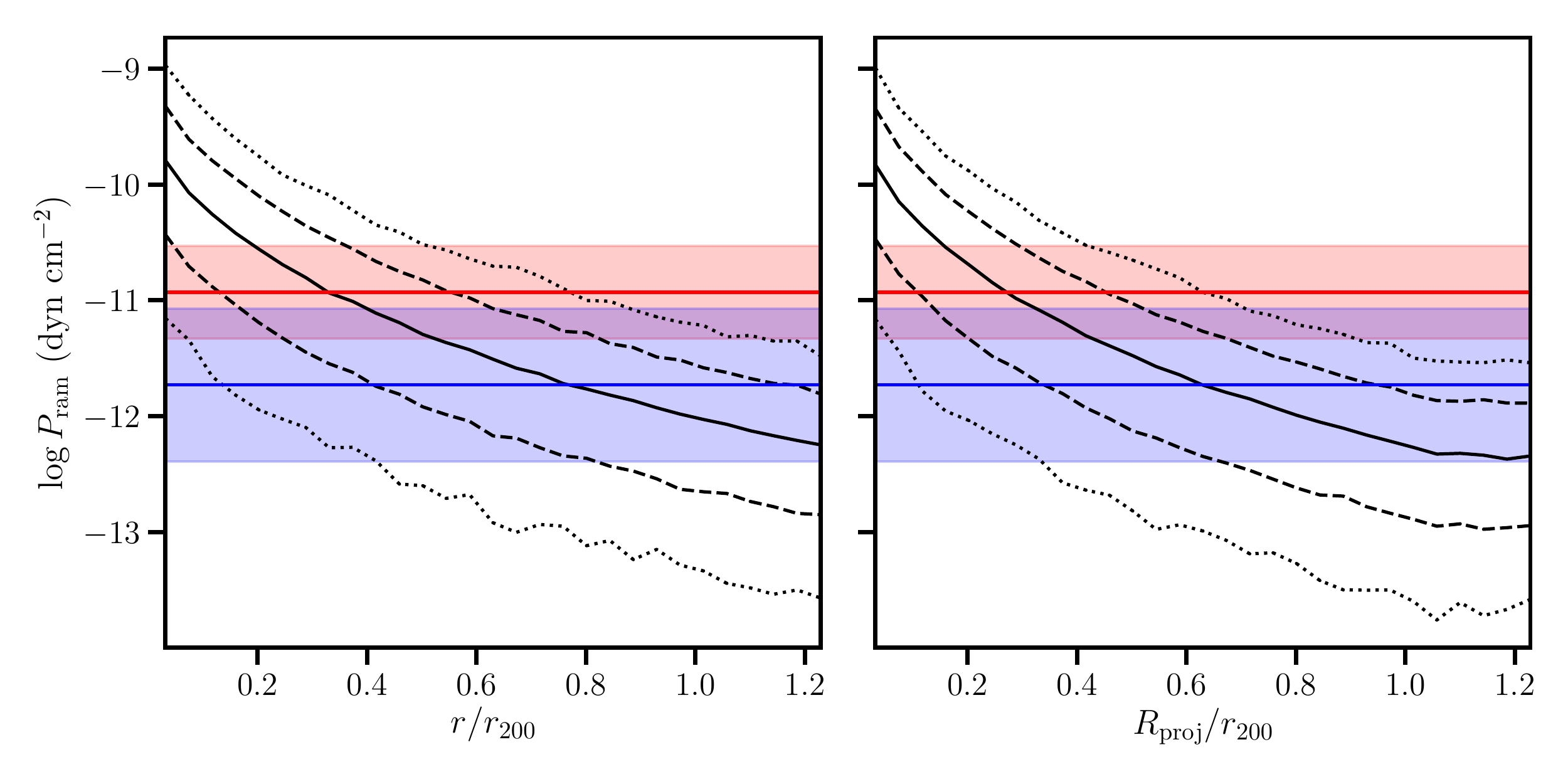}
    \caption{The ram pressure force profile as a function of the 3D distance of each galaxy from the cluster centre normalised to $r_{200}$ (Left panel) and as a function of the projected distance on the plane of the sky normalized to the same quantity (Right panel). The RPS force is obtained from the 3D peculiar velocity of each galaxy in the cluster multiplied by the ICM density at its position as derived from the equation given in the caption of Fig. \ref{ne_all}. The solid black line marks the median ram pressure force profile, while the dashed and dotted lines mark the $10^{\rm th}$, $90^{\rm th}$ and $1^{\rm st}$, $99^{\rm th}$ percentiles, respectively. The red (blue) solid lines and the associated shaded areas show the average and 1$\sigma$ dispersion of the gravitational restoring force (at the stellar effective radius) for galaxies with $M_{\rm star} > 10^{10} \rm{M_\odot}$ ($10^9 < M_{\rm star} < 10^{10} \rm{M_\odot}$), where $<\Sigma_{\rm star}>$ is derived using galaxies extracted from the \her{} Reference Survey as described in \citet{Boselli14} and $<\Sigma_{\rm gas}>$ from the VIVA sample \citep{Chung09} both limited to those objects with a normal \hi\ content. The gas surface density is derived assuming that $M_{\rm H_2}$ = $M_{\rm HI}$ within the \hi\ effective radius.}
    \label{fig:rps_profile_sam}
\end{figure*}

In Fig. \ref{modellorho} the truncation radius has been determined assuming as typical velocity of cluster galaxies the 3D projection of the line of sight velocity dispersion of late-type systems. This is a good approximation which takes into account the fact that late-type galaxies have, on average, a velocity dispersion which is $\sigma_{LT}$ $\simeq$ $\sqrt{2}$ $\times$ $\sigma_{ET}$ \citep{Colless96, Biviano09} typical of an infalling population. Infalling systems have radial orbits which make RPS much more efficient than circular orbits typical of the virialised early-type component since they brings the gas rich systems in the inner cluster regions with higher velocities. To take into account this effect, we can derive a more realistic
distribution of ram pressure forces acting on a mix of star-forming galaxies using the orbital velocity parameters and positions obtained for $z=0$ clusters in the simulated framework of semi-analytic models of galaxy formation. We use the model presented in \citet{Henriques15} and based on the Millennium simulation \citep{Springel05}. We select as clusters all haloes more massive than $M_{\rm cluster} > 10^{14} \rm{M_\odot}$ and the galaxy sample is made by satellites of these haloes with $M_{\rm star} > 10^9 \rm{M_\odot}$. The left panel of Figure \ref{fig:rps_profile_sam} shows the ram pressure force profile as a function of the three dimensional distance of each galaxy from the cluster centre ($r$) normalized to $r_{200}$. The median ram pressure force can increase by two orders of magnitude going from the virial radius to the inner core of the clusters, mostly due to the increase in local ICM density. When these profiles are compared to the typical gravitational restoring force at the effective radius for more ($M_{\rm star} > 10^{10} \rm{M_\odot}$, red lines) and less massive ($10^9 < M_{\rm star} < 10^{10} \rm{M_\odot}$, blue lines) galaxies, we find that on average only 1\% of the massive galaxies are subject to a force barely able to strip them at the virial radius, while this fraction increases above 50\% within 0.5$\times$ $r_{200}$. Conversely, lower mass galaxies are efficiently stripped when they cross the virial radius and the stripping force is greater than the restoring force for more than 90\% of them within half of the virial radius. We stress that these values are to be considered only as average quantities as they do not take into account local variations in the ICM density, the stellar and dark matter profile of galaxy discs, the presence of bulges, and the clumpiness of the cold gas components. The right panel of Figure \ref{fig:rps_profile_sam} shows the same ram pressure force but as a function of the projected distance of galaxies from the cluster centre on the plane of the sky ($R_{\rm proj}$). This figure can thus directly be compared to the observations. With projected positions we notice a mild increase in the scatter of the ram pressure force distribution and a slightly steeper profile but the main trend remains unchanged. 

%How far from the central regions, where both the density of the ICM and the velocity of infalling systems reach their maximum, can ram pressure strip the ISM and thus efficiently perturb the evolution of galaxies in high density environments? DONE

%How does this radial variation of the efficiency of the RPS process compare with the one expected for other mechanisms? 

\section{Observational evidence}
\label{sec:obsevid}

\subsection{Identification of the dominant perturbing mechanism}
\label{subsec:mainmech}

%Rich environments are characterised by a high density of galaxies and by the presence of a hot and dense ICM \citep[e.g.][]{Sarazin86}. As extensively described in \citet{Boselli06}, galaxies belonging to these environments can be affected by different mechanisms which can be broadly divided in two main families: those related to the gravitational interaction with nearby companions, with the potential well of the cluster itself, and their joint effect \citep[tidal interactions, tidal steering, harassment -][]{Merritt83,Byrd90,Moore98}, and those related to the hydrodynamic interaction of the cold ISM of galaxies with the hot ICM (RPS, viscous stripping, thermal evaporation, starvation - \citealt{Gunn72}; \citealt{Cowie77}; \citealt{Larson80}; \citealt{Nulsen82}). Each mechanism has different effects on the physical and kinematical properties of galaxies as suggested by the accurate study of representative objects and by more and more sophisticated tuned models and simulations.

\begin{figure}
\centering
\includegraphics[width=0.99\textwidth]{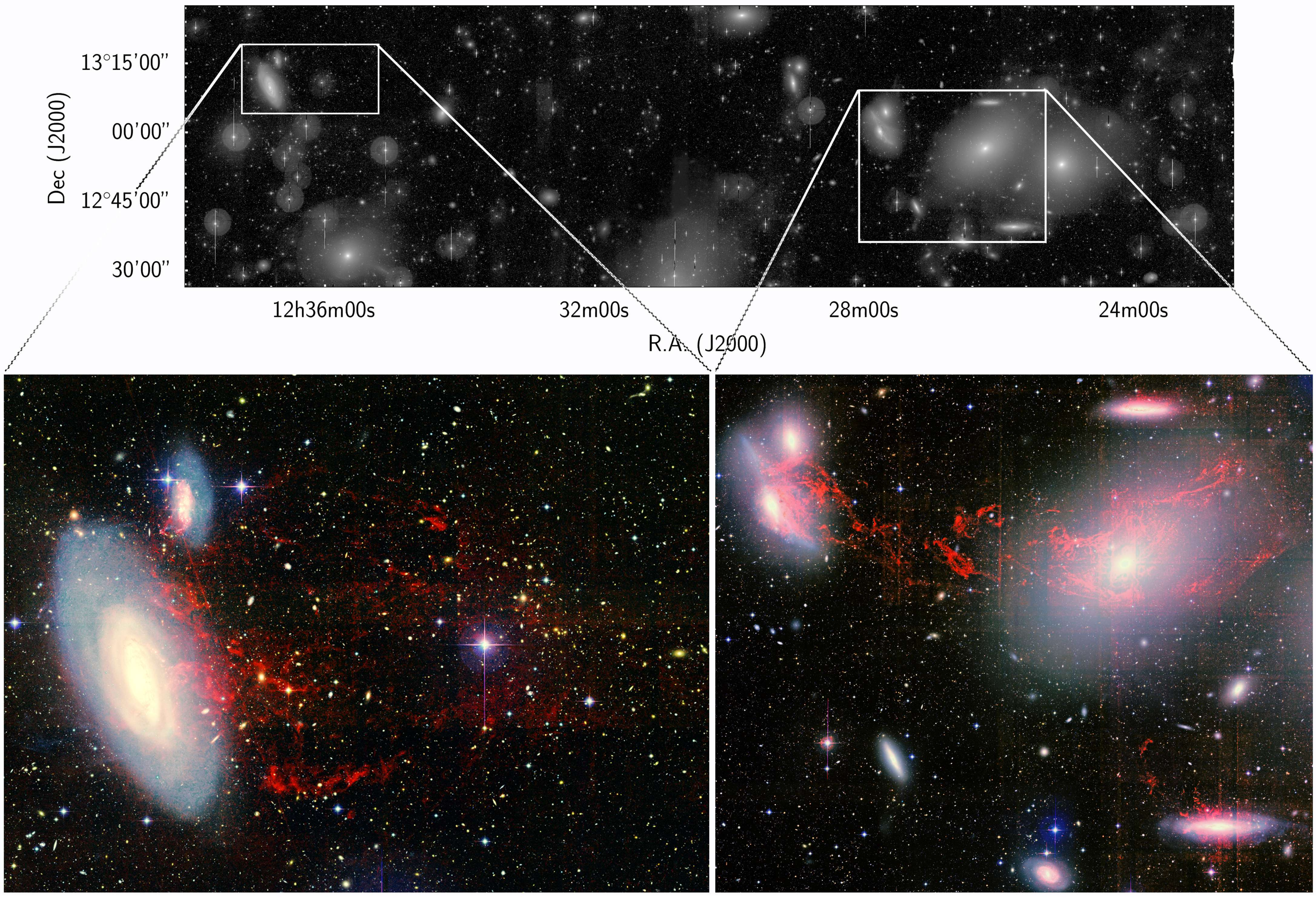}\\
\includegraphics[width=0.99\textwidth]{NGC4569_rgb_cut.pdf}
\caption{Upper panel: The pseudo-colour image of part of the Markarian chain in the Virgo cluster (from right to left NGC 4402 (top), M86 (bottom), NGC 4435 (top), and NGC 4438 (bottom)) 
obtained combing the CFHT MegaCam NGVS \citep{Ferrarese12} optical 
images with the VESTIGE \citep{Boselli18} H$\alpha$+\nii{} narrow-band image. North is up, east left. The size of the image is of $\simeq$ 40$^\prime$ $\times$ 20$^\prime$, corresponding to 190 kpc $\times$ 95 kpc (projected scale)
at the distance of the Virgo cluster. The diffuse red filamentary structures escaping from the galaxies 
is ionised gas (adapted from \citet{Boselli18}). The galaxy NGC 4438 in the upper left corner of the image shows extended filaments of ionised gas but also asymmetric 
extended features in the stellar continuum. These features are tidal tails formed after the gravitational interaction of the galaxy with M86 and NGC 4435 \citep[e.g.][]{Combes88, Boselli05, Kenney08}.
Although the gas can be partly removed by ram pressure \citep{Vollmer09}, the dominant perturbing mechanism is gravitational since able to act on the stellar component.
Lower panel: The pseudo-colour image of NGC 4569 and IC 3583 in the Virgo cluster. The size of the image is of $\simeq$ 50$^\prime$ $\times$ 25$^\prime$, corresponding to 250 kpc $\times$ 125 kpc (projected scale)
at the distance of the Virgo cluster. The diffuse red emission escaping from the galaxy
in the western direction is ionised gas extending up to $\simeq$ 150 kpc (projected distance from the stellar disc). The cometary shape of the gaseous component and the lack of any stellar
counterpart indicate that the galaxy NGC 4569 is undergoing an RPS event (adapted from \citet{Boselli16}).}
\label{fig:N4569}	    
\end{figure}

A critical aspect in environmental studies is the identification of the dominant perturbing mechanism on galaxies of different mass, in different high-density regions, and at different epochs.
The main difference between gravitational perturbations and hydrodynamic interactions between the cold ISM of galaxies and the hot ICM of high-density regions is that while the former acts indistinctly 
on all galaxy components (gas, stars, dark matter...), the latter affects only the diffuse ISM (gas in all its phases, dust). Gravitational interactions are thus able to produce low surface brightness tidal tails
in the stellar component, truncate stellar discs and induce gas infall in the nuclear regions because of their effect on the gravitational potential well of the perturbed galaxy, for instance through the formation of
dynamical instabilities such as bars or spiral arms \citep[see Fig. \ref{fig:N4569}, e.g.][]{Valluri93, Miwa98,Ellison11, Skibba12, Sellwood14}. The stellar component is 
insensitive to hydrodynamic interactions which act only on the diffuse ISM. Asymmetries in the old smooth and diffuse stellar component such as tidal tails, shells, etc
are undoubted signs of an ongoing or a recent gravitational perturbation, while marked asymmetries only present in the different components of the ISM are clearly witnessing an hydrodynamic interaction (Fig.
\ref{fig:N4569}). Second order (minor) effects on the gravitational potential well of the 
perturbed galaxies can be produced once most of the ISM is displaced from the stellar disc, in particular in dwarf systems characterised by a shallow potential well and a large fraction of gas 
\citep[e.g.][]{Boselli14c}. At the same time, 
new stars can be formed in the stripped gas, producing asymmetric, clumpy structures in the youngest stellar component.

It is harder to discriminate among the different hydrodynamic mechanisms since all of them act indiscriminately on the different diffuse phases of the ISM of galaxies\footnote{As specified in Sec. \ref{subsec:molecular}, there are indications that clumpy structures such as giant molecular clouds decouple from the ram pressure wind}. RPS, laminar viscous and turbulent stripping via Kelvin-Helmholtz
instabilities have very similar observational effects on the gaseous component of the ISM, removing it from the outer parts of the galactic disc where the gravitational potential well is not sufficient to overcome the external pressure 
exerted by the ICM on the galaxy moving at high velocity. The ISM is removed either by the external pressure or by the friction between the two media (ICM-ISM) once they get in contact during the crossing of the galaxy within
the cluster. All these hydrodynamic mechanisms are related to the motion the galaxies within the ICM, and can thus be identified with dynamical signatures such as cometary tails in the gaseous component.
Thermal evaporation occurs once the cold ISM ($T$ $\leq$ 10$^4$ K) gets in contact with the hot ICM ($T$ $\sim$ 10$^7$-10$^8$ K)
\citep{Cowie77}. The transfer of energy via electron heat conduction is able to make the cold gas
evaporate and change to phases of higher temperatures.
%becoming first ionised than hot gas (TRUE ?).
This mechanism is thus independent from the dynamic of the galaxy within the cluster, although its motion within the ICM grants a continuum supply of external energy. The identification of this mechanism
is challenging since it would imply the direct observation of a changing of phase of the ISM along the disc and in the surrounding regions of the perturbed galaxies.
Starvation or strangulation occurs once a galaxy enters within the halo of the high-density structure. As satellite of a larger halo, gas infall supplying fresh material to support star formation is stopped \citep[e.g.][]{Larson80}.
The perturbed galaxies thus consume the available cold gas reservoir gradually and uniformly reducing on long timescales ($\tau \simeq$ 3-4 Gyr; \citealt{Boselli14}) their star formation activity on the disc. Their stellar and gas
radial profiles are very symmetric but systematically reduced with respect to those of similar unperturbed objects in the field, where the gas infall continue to sustain star formation at all galactocentric distances up to the
present epoch \citep[e.g.][]{Boselli06a}. Starvation is thus not able to produce the typical truncated gaseous discs
observed in RPS galaxies (see Sec. \ref{subsec:truncdiscs}). To summarise, the identification between gravitational and hydrodynamic perturbing mechanisms in high-density regions through the observation of individual objects
is possible provided that high-quality, high-sensitivity multifrequency data are available. Hydrodynamic phenomena (RPS, Kelvin-Helmholtz and Rayleigh-Taylor instabilities, and viscous stripping) have very similar large scale observational signatures, and can only be disentangled whenever the perturbed regions at the interface between the cold ISM and the hot ICM are fully resolved with sufficient sensitivity \citep{Roediger08, Roediger13} as in Milky Way streams \citep{Barger20} or in a few large nearby galaxies\footnote{Kelvin-Helmholtz instabilities are due to the velocity difference across the interface between two fluids \citep[e.g.,][]{Livio80, Nulsen82, Mori00}. They manifest as small scales waves and vortexes as those observed in the atmosphere of Jupiter. Rayleigh-Taylor instabilities are due to the pressure of a light fluid pushing a heavy fluid, forming small scale spikes and bubbles at their interface \citep[e.g.,][]{Roediger08}}. We thus consider these different mechanism as different flavors of an hydrodynamic stripping process through this work, unless otherwise stated. They can be easily distinguished from starvation and thermal evaporation as mentioned above.

We recall, however, that all these mechanisms, gravitational and hydrodynamic, can act simultaneously on the gas-rich galaxies entering the cluster \citep[e.g.][]{Cortese21}. Their effect is additive, and can thus
increase the efficiency in removing the gas and quenching the activity of star formation of the perturbed systems. What is interesting in this context is understanding which, among these mechanisms, is 
dominant in different environments, on different galaxy populations, and at different epochs, in order to pose strong observational constrains to models of galaxy evolution and cosmological simulations.

\subsection{The presence of RPS tails}
\label{subsec:tails}

RPS has been first proposed to explain the nature of head-tail radio galaxies in nearby clusters %\citep[e.g.][]{Miley72, Rudnick76, Jones79, Miley80 }
\citep[e.g.][]{Miley72, Miley80}.
In these peculiar objects, generally associated to the massive dominant early-type galaxies located deeply within the gravitational potential well of massive clusters, the radio jet is bent because of the pressure exerted by the ICM on the relativistic electrons escaping from the active nucleus during the motion of the parent galaxy within the cluster. Since then, several morphological features peculiar of an ongoing RPS event have been discovered at different wavelengths in a large variety of galaxies populating nearby clusters. Among these, the first evidence of an hydrodynamic 
interaction between the hot ICM and the cold ISM of late-type galaxies in rich environments has been gathered by \citet{Gavazzi78} in the radio continuum and claimed by \citet{Shostak82} in HI. In RPS objects, the most unique and peculiar feature of the interaction are one-sided cometary tails of stripped ISM material combined with a normal and non-displaced stellar component, as now routinely observed in several nearby cluster galaxies. Since the first discoveries in radio observations, cometary tails without an associated evolved stellar component have been detected in the different gas phases (atomic, molecular, ionised, hot) and in the dust. It is worth noting that galaxies with a tail but first identified by young and blue extraplanar asymmetric features outside their discs are sometimes called ``jellyfish'' galaxies, after this name has been first introduced by \citet{Chung09}, \citet{Bekki09} and extensively used by \citet{Smith10}, \citet{Ebeling14}, and \citet{Poggianti17} (extraplanar features are sometimes also called ``fireballs'', \citealt{Yoshida08}). As described in more details in Sec. \ref{subsec:stars}, this nomenclature can be misleading to indicate galaxies undergoing an RPS event. Indeed, asymmetric features observed in broad-band optical images are tracing the distribution of stars and it cannot be ruled out they are produced by gravitational perturbations rather than RPS. We thus caution in using these ambiguous definitions to identify galaxies clearly undergoing an RPS event, and rather encourage the reader to call these objects ram pressure stripped galaxies. When these terms are used hereafter we only report the identification given in the original papers, without solving the ambiguity raised above and therefore not necessarily referring to galaxies purely undergoing RPS.

\begin{figure}
\centering
\includegraphics[width=0.95\textwidth, angle=0]{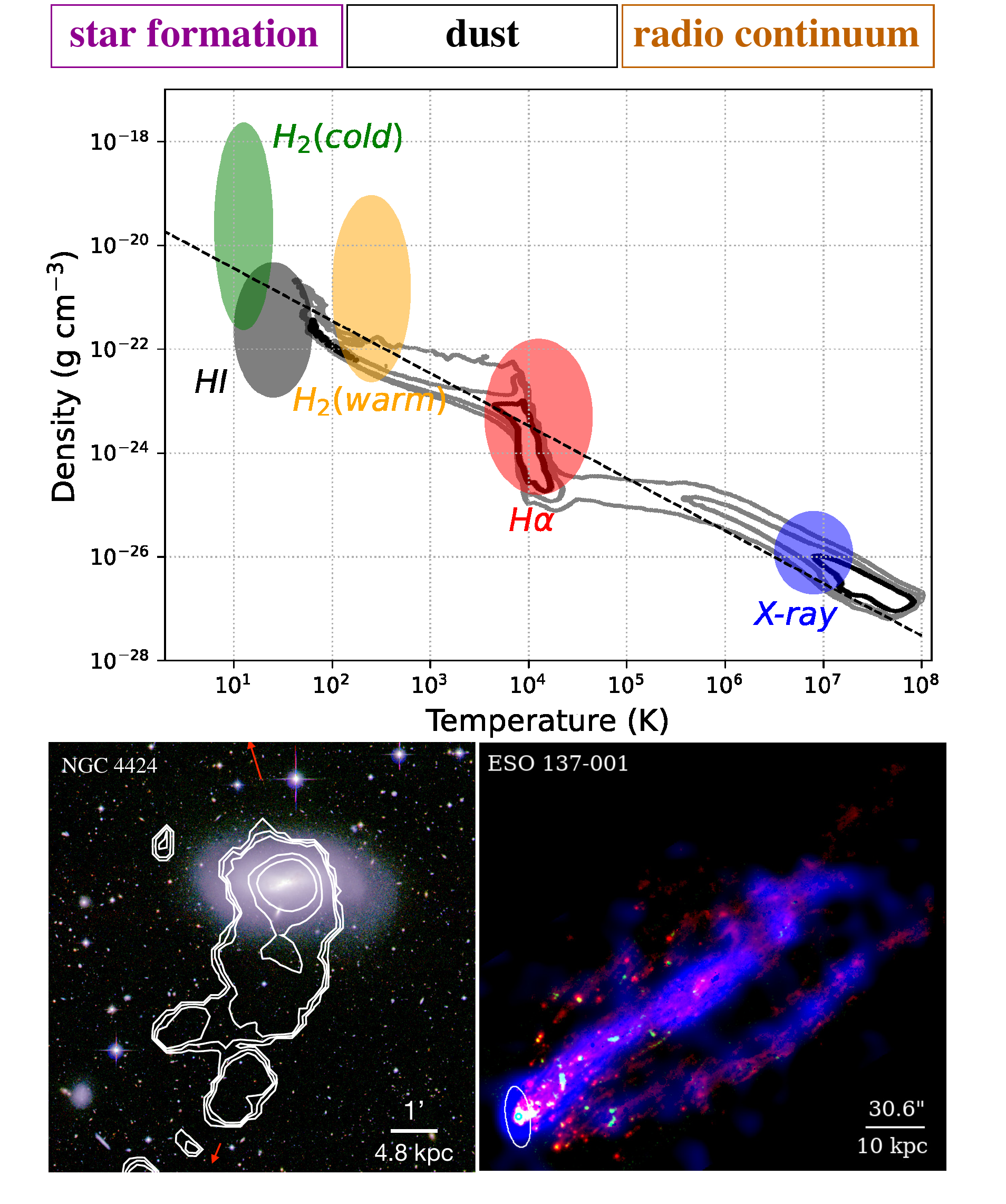}
\vspace{-0.4cm}
\caption{Phase diagram for the gas in the stripped tails. Current data have revealed the existence of cold molecular gas (traced by CO), warm molecular gas (traced by H$_{2}$ lines and other lines with the NIR/MIR spectroscopy), cold atomic gas (\hi{}), warm and inoised gas (H$\alpha$ and other optical emission lines) and hot gas traced by the X-ray emission.
The grey contours are the gas phase diagram from simulations by Stephanie Tonnesen. The black dashed line shows the isobaric condition. While cold molecular and atomic gas are not included, the distribution peaks around $T \sim 10^{7.5}$ K and 10$^{4}$ K are clear, as well as the continuous distribution with temperature and density.
Other potential tracers, already detected in X-ray cool cores and/or galactic winds, include coronal lines like \civ{} and \ovi{} to trace $T = 10^{5} - 10^{6}$ K gas, ro-vibrational H$_{2}$ lines at the NIR to trace hot molecular gas at $T \sim 10^{3}$ K and FIR lines like \cii{} to trace cold gas. We include examples for most gas phases: ESO~137-001 for CO from {ALMA} (green), H$\alpha$ from MUSE (red) and X-ray from \chandra{} (blue) (from \citealt{Sun22}), and NGC~4424 for \hi\ (white contours, from \citealt{Chung09}) on the optical image. Warm molecular gas was also detected in ESO~137-001 \citep{Sivanandam10}, as well as \hi\ from {MeerKAT} (Ramatsoku, private communication). 
We also list other components detected in stripped tails, young stars recently formed, dust and radio continuum indicative of the existence of relativistic electrons and likely CRs. }
\label{fig:gasphasediag}       
\end{figure}

\subsubsection{Radio continuum}
\label{subsec:radcont}

Tails in radio continuum are due to the synchrotron emission of the relativistic electrons of the cosmic rays spinning in week magnetic field. In early-type galaxies these relativistic electrons are accelerated by the radio jet. In late-types they are accelerated in the supernovae remnants present within
the star forming disc of the perturbed objects. They are swept away from the disc with the ISM during the interaction. Presence of radio continuum tails in late-type galaxies have been observed in
the centimetre spectral domain in the nearby clusters A1367 \citep{Gavazzi78, Gavazzi85, Gavazzi87, Gavazzi95}, Coma \citep{Miller09, Chen20, Lal20}, Virgo \citep{Vollmer04a, Vollmer07a, Vollmer10, Vollmer13, Crowl05, Chyzy07, Kantharia08}, Perseus \citep{Roberts22a} and in other nearby clusters \citep{Poggianti19a}
They are located within the inner regions of the cluster, up to $R/r_{200}$ $\simeq$ 1.4 in Coma \citep{Chen20,Roberts21} and $R/r_{200}$ $\simeq$ 0.6 in A1367 \citep{Gavazzi95}. There is also evidence of 
a radio continuum tail in a galaxy suffering an RPS episode within a nearby group \citep{Rasmussen06}. Recently similar tails have been also observed in the metre spectral domain (144 MHz LOFAR data taken during the LoTSS survey) in a significant fraction of late type galaxies in several nearby clusters (including Coma and A1367) and in groups of mass $10^{12.5} < M_{\rm group} < 10^{14} {\rm M_{\odot}}$ up to one virial radius \citep{Botteon20, Roberts21, Roberts21a}. It has been also observed that these objects 
have a radio-to-FIR ratio higher than similar objects in the field \citep{Gavazzi86, Gavazzi91, Niklas95, Andersen95, Rengarajan97, Gavazzi99a, Miller01, Reddy04, Boselli06, Murphy09, Vollmer10, Vollmer13,Chen20}. This evidence, statistically shared also by the other late-type galaxies
populating nearby clusters, has been first interpreted as due to the compression of the magnetic field frozen within the ISM during its hydrodynamic interaction with the ICM \citep{Gavazzi91, Volk94, Boselli06}. The synchrotron radio emissivity $\epsilon_{syn}(\nu)$ at a frequency $\nu$ is proportional to the magnetic field $B$ as \citep{Rybicki79}:

\begin{equation}
\epsilon_{\rm syn}(\nu) \propto n_{\rm e} B^{\alpha_{\rm syn}+1} \nu^{-\alpha_{\rm syn}}  ~~~~~ \rm{[ erg ~s^{-1} cm^{-3} sr^{-1} Hz^{-1}] }
\end{equation}

\noindent
where $n_{\rm e}$ is the density of relativistic electrons accelerated in the supernovae remnants and $\alpha_{\rm syn}$ the nonthermal synchrotron spectral index, 
which in late-type galaxies has a typical value of $\alpha_{\rm syn}$ $\simeq$ 0.8  at $\sim$ 1.4 GHz \citep{Gioia82, Tabatabaei17, Chyzy18}. Since both the density of relativistic electrons and the far-IR
emission are tightly related to the star formation activity of galaxies, as suggested by the very low dispersion observed in the FIR-radio correlation of unperturbed systems \citep{deJong85, Helou85}, 
the observed increase in the radio-to-FIR ratio of cluster galaxies ($\simeq$ 4-10) can be explained by an increase of the magnetic field of a factor of $\sim$ 2-3. More recently,
tuned modifications of this mechanism have been proposed to explain the increased radio continuum emissivity of cluster galaxies
(enhancement of the magnetic field strength caused by the ISM shear motion \citep{Murphy09}, compression of the ISM/magnetic field by the external ICM hydrodynamical and/or thermal pressure \citep{Murphy09},
and turbulence and ICM shocks caused by ram pressure \citep{Volk94, Kang11}). A local compression of the magnetic field in the leading edge of ram pressure stripped galaxies has been observed thanks to polarised
data in a few galaxies in the Virgo cluster \citep{Vollmer10, Vollmer13}. 

Comparing resolved far-IR Spitzer and radio continuum VLA images of Virgo cluster galaxies, \cite{Murphy09}
discovered the presence of radio deficient regions in the outer discs or RPS galaxies. These radio-deficient regions seem
to be situated just outside regions with an high radio polarisation and a flat spectral index, at the leading edge of the 
galaxy disc interacting with the surrounding ICM. This signature has been explained as due to a sweeping of the cosmic ray electrons and of the associated magnetic field on the leading side of the disc where gas has been removed during the interaction, a compression of the magnetic field at the edges of the ISM-ICM interface, with a re-acceleration of the cosmic rays able to enhance the global radio emissivity of the perturbed galaxies \cite{Murphy09}.

The properties of these radio continuum  tails can be used to derive interesting parameters necessary to constrain models and simulations. The synchrotron emission timescale is given 
by the relation \citep{Feretti08}:

\begin{equation}
\tau_{\rm syn} = 4.5 \times 10^7 \left(\frac{B}{10 {\rm \mu G}}\right)^{-3/2} \nu_{\rm GHz}^{-1/2}    ~~~~~~\rm{[yr]}
\end{equation}

\noindent
The deprojected physical length of the radio continuum tail measured from the observations can be used to derive a typical timescale for the motion of the galaxy within the cluster and thus
infer an estimate of the magnetic field intensity within the tail \citep[e.g.][]{Vollmer21, Chen20}. This number can be used as input to models and simulations now able to 
quantify the impact of magnetic fields on the stripped material \citep{Tonnesen14, Ruszkowski14, Shin14}. This estimate of the magnetic field in the radio continuum tail, however, can be 
done only in the hypothesis that the stripped electrons are neither re-accelerated nor produced within the tail. This might indeed be the case. Regions of recent star formation
in the stripped gas, although not ubiquitous (see Sec. \ref{subsec:sfrtail}), have been observed in several galaxies with tails in the ionised gas component such as ESO137-001 in the Norma cluster \citep{Sun07, Sun10, Fossati16},
in most of the jellyfish galaxies of the GASP survey  \citep{Poggianti17} 
%(see however Sec. \ref{subsec:stars}), and in the tail of UGC 6697 \citep{Consolandi17}, where radio continuum tails have been previously detected \citep{Gavazzi78, Gavazzi85, Gavazzi87, Gavazzi95}. If the tails are detected also in H$\alpha$, 
the contribution of star forming regions can be roughly quantified by comparing the star formation rate derived from the radio continuum emission to that determined using the ionised gas emission
by means of standard calibrators such as those proposed by \citet{Kennicutt98} (see \citealt{Chen20} for an example). At the same time, the electrons can 
be re-accelerated within the tail by turbulence and shocks produced during the RPS phenomenon \citep{Kang11, Pinzke13}.
If the radio spectral index and the volume of the emitting tail are known, the strength of the magnetic field within the tail of stripped gas can be inferred using the classical equipartition or minimum-energy estimates as suggested by \citet{Beck05}. A recent example of magnetic field estimated in a single RPS tail can be found in \citet{Ignesti22}.

It is also worth mentioning that the spectral slope of the radio continuum spectrum steepens with the age of the cosmic ray electrons. As indicated by \citet{Vollmer21},
for a purely aging population of cosmic ray electrons the evolution of flux density is given by:

\begin{equation}
f(t) = f_0 \times e^{-t/\tau_{\rm syn}}
\end{equation}

\noindent
If we define the spectral index $\alpha$ as:

\begin{equation}
\alpha = \frac{\ln(f_1/f_2)}{\ln(\nu_1/\nu_2)}
\end{equation} 

\noindent
where $f_1$, $f_2$, $\nu_1$, and $\nu_2$ are the flux densities (or surface brightnesses) and the frequencies in the two bands,
the timescale needed to steepen the spectra index from an initial value $\alpha_{\rm in}$ to a final value $\alpha_{\rm end}$ is given by the relation:

\begin{equation}
\tau_{\rm SI} = \frac{(\alpha_{\rm in} - \alpha_{\rm end}) \ln(\nu_1/\nu_2)}{1/\tau_{\rm syn,1} - 1/\tau_{\rm syn,2}}
\end{equation}

\noindent
A steepening of the spectrum should thus be observed with increasing distance from the galaxy (e.g. NGC 4522, \citealt{Vollmer04}).

\subsubsection{Atomic gas}
\label{subsec:atomic}

The first observation of a nearby galaxy with a \hi\ cometary tail is probably NGC 1961 in a nearby group \citep{Shostak82}.
The asymmetric \hi\ gas distribution of this galaxy was explained as due to an ongoing RPS event, but this interpretation has been later questioned given the lack of hot X-ray emitting gas in the associate group environment \citep{Combes09, Anderson16}. Since then, targeted or blind surveys done using interferometric observations 
revealed the presence of several galaxies with long tails of cold atomic hydrogen without any associated optical counterpart up to  
$R/r_{200}$ $\simeq$ 1 in the Virgo cluster \citep{Kenney04, Chung07, Sorgho17, Minchin19}, $R/r_{200}$ $\simeq$ 0.5
in the Coma cluster \citep{Gavazzi89, Bravo-Alfaro01, Healy21}, $R/r_{200}$ $\simeq$ 0.6 in A1367 \citep{Gavazzi89, Dickey91, Scott10}, where also a couple of objects suffering the joint effect of ram pressure and harassment have been 
detected well outside the virial radius ($R/r_{200}$ $\simeq$ 1.6-2.2; \citet{Scott12}), and in a few other nearby clusters and groups \citep[e.g.][]{Dickey97, Serra13, Lee18, Ramatsoku19, Deb20, Ramatsoku20, Reynolds20, Loni21, Wang21}.
%A beautiful example is NGC 4424 in the Virgo cluster shown in Fig. \ref{fig:N4424}. 
The number of objects with tails of \hi\ gas, however, is still relatively limited.
In Virgo, for instance, where the proximity of the cluster (16.5 Mpc) ideally combines the sensitivity and the angular resolution needed to detect these low column density features, they are detected in only seven galaxies out of the 53 mapped during the VIVA survey \citep{Chung07, Chung09}. This relatively low fraction ($\sim$ 15 \%)
can be due to different factors: i) a still limited sensitivity of radio observations of low column density extended structures, ii) RPS being not dominant in the Virgo cluster, and iii) a change in the gas phase
once stripped from the galaxy, making the \hi\ cometary tails as short lived structures. Point i) is probably the case for clusters located at distances of $\geq$ 100 Mpc, where
the required sensitivity is only reached in detriment of the angular resolution. Hopefully, the most recent generation of instruments such as ASKAP and MeerKAT, or the future ambitious projects such as SKA, will further extend our capability to detect these low column density features in clusters other than Virgo and Fornax. 
Very promising results, although based on stacking of large samples, are coming from the \hi\ survey of the Coma cluster done with the WSRT \citep{Healy21}, while direct detections of several late-type systems with extended \hi\ tails at column densities of $\Sigma_{\rm HI}$ $\sim$ 10$^{18}$ cm$^{-2}$ have been obtained in Fornax with the sensitive MeerKAT telescope (Serra, private communication). 
Concerning point ii), there is indeed evidence that a few galaxies with extended tails of \hi\ gas have also suffered gravitational perturbations (e.g. NGC 4254, \citealt{Haynes07, Duc08, Vollmer05, Boselli18a}; NGC 4424, \citealt{Sorgho17, Boselli18b}; NGC 4654,  \citealt{Vollmer03, Lizee21}). This is also the case for the newly detected galaxies with tails in \hi, which often have perturbed morphologies suggesting a combined effect of ram pressure and harassment as discussed in more details in Section \ref{subsec:stars}. iii) Similarly, we cannot rule out the heating of the cold gas, since a growing number of tails are being detected in multiple gas phases as described in Section \ref{sec:impactenv}.
This indicates that \hi\ tails are a short lived phenomenon witnessing an ongoing process (while the presence of truncated \hi\ discs are rather a proof of a past RPS event, see Sec. \ref{subsec:truncdiscs}).

%\begin{figure}
%\centering
%\includegraphics[width=1.0\textwidth]{rgb_HIcont_arrows.pdf}
%\caption{The pseudo-colour image of NGC 4424 in the Virgo cluster 
%obtained using the CFHT MegaCam NGVS \citep{Ferrarese12} optical 
%images with overplotted \hi\ contours derived from the VLA data of the VIVA survey \citep{Chung09}. The two red arrows indicate %the position of the core of the Virgo cluster A substructure dominated by M87 (upper arrow) and of cluster B with M49 (lower %arrow).
%{\color{red}{merge it to Fig. 7?}}
%}
%\label{fig:N4424}	    
%\end{figure}

\subsubsection{Molecular gas}
\label{subsec:molecular}

The hydrodynamic pressure exerted by the hot ICM acts regardless on all the different components of the ISM of galaxies moving at high velocity within the gravitational potential well 
of the high density structure. Models and simulations consistently suggests that the stripping efficiency is less severe on the molecular gas than on the diffuse \hi\ since H$_2$ 
is mainly located within dense giant molecular clouds which have a relatively small cross-section (10 pc) with respect to the diffuse atomic gas ($\simeq$ 100-500 pc), 
they are much denser ($n(\rm H_2)$ $\simeq$ 50-500 cm$^{-3}$, \citet[e.g.][]{Solomon87, Engargiola03, Bolatto08}, vs. $n(\rm HI)$ $\simeq$ 0.1-1 cm$^{-3}$), 
and are located deeper within the stellar disc where gravitational forces are strong \citep[e.g.][]{Yamagami11}. However, the ISM is turbulent and clumpy, and overedensities are constantly created and destroyed. It is thus conceivable that the diffuse molecular gas can be stripped during the interaction, and thus that the dense molecular gas within giant molecular cloud also decreases because not replenished by the diffuse component \citep{Tonnesen09}. Simulations indicate that most of the molecular gas is removed from the outer disc \citep{Lee20}.

Direct observational evidence of molecular gas stripping via ram pressure is still very limited not only because of the reasons mentioned above, but also because 
of the lack of high sensitivity panoramic detectors with a wide field of view in
the mm and sub-mm spectral domain, where molecular gas is generally looked for through the detection of the main $^{12}$CO(1-0) (2.6 mm, 115 GHz) and $^{12}$CO(2-1) (1.3 mm, 230 GHz) emission lines.
CO observations of late-type cluster galaxies characterised by extended tails at other frequencies revealed molecular gas to be present in the tails. The first of these evidences have been found in ESO137-001 in the Norma cluster \citep{Jachym14}, in NGC 4388 in Virgo by \citep{Verdugo15}, and in 
D100 in the Coma cluster \citet{Jachym17}. While in ESO137-001 and in D100 very large amounts of molecular gas ($\simeq$ 10$^9$ M$_{\odot}$) have been detected up to $\sim$ 40-60 kpc from the galaxy disc, in NGC 4388 a small amount of gas ($\simeq$ 10$^6$ M$_{\odot}$) has been detected in well defined regions associated with star forming blobs. More recently, clearer evidences of molecular gas stripping come from the detection of extended tails in CO observed in several late-type galaxies in the Fornax cluster \citep{Zabel19}, a high spatial resolution map of ESO137-001 with ALMA \citep{Jachym19}, and an extraplanar plume of molecular gas $\sim$ 1 kpc wide and $\sim$ 2 kpc above the stellar disc of NGC 4402 in Virgo associated to dusty filaments seen in absorption in the \hst{} images, kinematically dissociated from the disc \citep{Cramer20}. A further example in the Virgo cluster is NGC 4438, where molecular gas has been removed from the inner regions \citep{Combes88}, although this galaxy is mainly suffering a gravitational perturbation \citep{Vollmer05a, Vollmer09}.
%There are, however, several cluster galaxies characterised by extended tails at other frequencies where molecular gas has been searched for and detected in various CO lines. Clear examples areESO137-001 in the Norma cluster \citep{Jachym14, Jachym19} and D100 in the Coma cluster \citep{Jachym17}, where very large amounts of molecular gas ($\simeq$ 10$^9$ M$_{\odot}$) have been detected up to $\sim$ 40-60 kpc from the galaxy disc, or NGC 4388 in Virgo, where a small amount of gas ($\simeq$ 10$^6$ M$_{\odot}$) has been detected in well defined regions associated with star forming blobs \citep{Verdugo15}. 
A few other cases of CO detections are galaxies from the GASP  sample located in clusters further away \citep{Moretti18, Moretti20}. While in the case of ESO137-001 and D100 the molecular gas is dominant in mass in the stripped tails the Virgo examples show a smaller molecular fraction. Moreover many ionised gas tails in Virgo galaxies have not been detected in CO. It is therefore still unclear what drives the fraction of molecular gas in the stripped ISM and if these variations are due to molecular cloud formation in-situ, to the stripping conditions near the galaxy disc, or to the ICM pressure as suggested by \cite{Tonnesen12}.
Milder evidence such as asymmetric molecular line profiles or CO gas distributions in the outer regions of known perturbed galaxies
or other kinematical disturbances has been also searched for in several galaxies in the Virgo \citep{Kenney90, Vollmer99, Vollmer01a, Vollmer08, Vollmer12, Cortes06, Jachym13, Lee17} and A1367 \citep{Boselli95,Scott13, Scott15} clusters.
Among these, it is worth noticing the galaxy NGC 4522 in Virgo, where $^{13}$CO(1-0) has been detected in a few \hii\ regions located outside the stellar disc, suggesting that heavy elements produced in evolved
stars at the origin of the $^{13}$C of the ISM can be stripped from the stellar disc during a ram pressure episode \citep{Lee18}.

The tails of a few galaxies in RPS 
have been observed with \textit{Spitzer} in imaging and spectroscopic mode. 
These observations revealed the
presence of large amounts ($M_{\rm H_2}$ = 10$^6$-10$^8$ M$_{\odot}$ and column densities $\Sigma_{ \rm H_2}$ = 10$^{19}$-10$^{20}$ cm$^{-3}$) of warm 
($T$ = 115-160 K) and smaller amounts ($\simeq$ 1\% ~ of the warm component) of hot ($T$ = 400-600 K) molecular gas \citep{Sivanandam10, Sivanandam14}.
The large amount of warm molecular gas observed in these tails with respect to what generally observed in star forming regions has been interpreted as due to 
the presence of gas shock-heated during the hydrodynamic interaction with the ICM, consistently with a picture where the cold stripped gas is heated and changes phase once stripped from the galaxy.
Similar \textit{Spitzer} observations of the leading side of galaxies suffering an RPS event revealed an excess of the warm H$_2$/PAH ratio 
with respect to what observed in the disc of unperturbed star forming
systems. This excess of warm H$_2$ emission has been attributed to the contribution of shock-excited molecular hydrogen triggered during the interaction \citep{Wong14}.

\subsubsection{Ionised gas}
\label{subsec:ionised}

The first evidence of diffuse tails of ionised gas in ram pressure stripped galaxies has been gathered after the observations of two peculiar objects in the cluster A1367 with 
extended tails in radio continuum \citep{Gavazzi01}. These observations have been carried out through very deep exposures on 2m class telescopes using narrow-band (NB, $\sim$ 100 \AA\ wide) 
interferential filters centered on the emission of the redshifted H$\alpha$ Balmer line. The advent of large panoramic detectors coupled with high-transmissivity NB filters at larger 4m and 8m class telescopes opened a new era in the observation of nearby clusters. Since the discovery of \citep{Gavazzi01}, the detection of diffuse tails of 
ionised gas without any associated stellar component witnessing ongoing RPS events significantly increased. These tails are now commonly detected in most of the nearby clusters such as Virgo \citep{Yoshida02, Boselli16, Boselli18b, Fossati18}, Coma \citep{Yagi07,Yagi10,Fossati12,Yoshida12,Gavazzi18a}, A1367 \citep{Gavazzi01,Yagi17}, and Norma \citep{Sun07,Sun10} and have been detected using this technique in clusters up to $z \sim 0.4$ \citep{Yagi15}. One of the most spectacular cases is NGC 4569 in the Virgo cluster, shown in Fig. \ref{fig:N4569}. Ionised gas tails have been also observed in broad-band selected perturbed galaxies in $0.04 \lesssim z \lesssim 0.07$ clusters by \citet{Poggianti17} as part of the GASP  survey. They have been also predicted by wind-tunnel hydrodynamic simulations, although their physical properties (density, size, clumpiness etc) significantly change according to the adopted simulation and the details of the radiative cooling recipes implemented \citep{Kronberger08,Kapferer09,Tonnesen10}.
These ionised gas tails extend up to 100 kpc in projected distance from their parent galaxies, and have H$\alpha$ surface brightnesses of the order of $\Sigma_{\rm H\alpha}$ $\simeq$ 1-5 $\times$ 10$^{-18}$ erg s$^{-1}$ cm$^{-2}$ arcsec$^{-2}$.
The blind surveys of nearby clusters carried out so far, unfortunately only partially homogeneous and complete, consistently suggest that the frequency of galaxies showing extended diffuse structures in ionised gas is very high, reaching $\simeq$ 50 \%~ in the late-type systems \citep{Boselli14c, Gavazzi18a}, making at present narrow-band imaging on of the most efficient observational techniques to identify galaxies undergoing a perturbation. The ongoing deep blind surveys of nearby clusters such as VESTIGE \citep[A Virgo Environmental Survey Tracing Ionised Gas Emission,][]{Boselli18},
thanks to their excellent sensitivity and image quality, have been designed to accurately quantify on strong statistical basis the fraction of galaxies undergoing the different perturbing mechanisms in a 
typical cluster environment. The detailed study of the physical and kinematical properties of the stripped gas, which are necessary to understand the nature of the undergoing perturbing mechanism, 
might however require targeted spectroscopic observations, mostly obtained with single- or multi-objects slits \citep{Yoshida04,Yoshida12,Yagi13,Boselli16a,Fossati18}. Most recently, however, new IFU spectrographs with large fields of view coupled with 4-8m class telescopes such as MUSE at the VLT have provided us with resolved spectroscopic observations. These datasets are by far the most information rich, allowing a detailed study of the metal content and ionisation conditions of the stripped gas \citep{Merluzzi13,Fumagalli14,Fossati16,Fossati19,Consolandi17,Poggianti17,Poggianti19,Gullieuszik17,Boselli18b,Vulcani18,Bellhouse19,Liu21}. These instruments are also extremely efficient in detecting tails of ionised gas 
at higher redshift, where the field of view of the instrument ($\sim$ 1 arcmin$^2$) is sufficient to cover a large fraction of the cluster virial radius, while the large wavelength coverage allows a detection of at least one strong emission line over a wide redshift range. RPS tails of ionised gas have indeed been detected through the [OII]$\lambda$3727 \AA\ emission line in two cluster galaxies at $z$ $\sim$ 0.7 by \citet{Boselli19a}.

The H$\alpha$ ($\lambda$ = 6563 \AA) emission line, which is mostly used to detect extended ionised gas features associated to perturbed cluster galaxies, is due to the recombination of hydrogen 
which has an ionisation energy of 13.6 eV and thus traces the distribution of gas at a typical temperature of $T \simeq 10^4$ K \citep{Osterbrock06}. In the case that the tails are 
detected in other emission lines, the ionised gas temperature can be estimated from the ionisation energy of these lines. However, a robust estimate of the gas temperature hinges on the detection of faint lines to build temperature sensitive line ratios (e.g. [NII]$\lambda \lambda$ 6584,6548/[NII]$\lambda$5755, \citealt{Fossati16}).

Within the disc of late-type galaxies
hydrogen is mainly ionised by the UV radiation emitted by young (age $\leq$ 10 Myr) and massive ($M_{\rm star}$ $\geq$ 10 M$_{\odot}$) O-early B type stars and is thus an excellent tracer of star formation
\citep{Kennicutt98, Boselli09a}. In the diffuse tails, where star formation is not always present, the atomic hydrogen stripped during the interaction and mixed within the hot ICM can be ionised by
several other mechanisms such has shocks in the turbulent gas, heat conduction, and magneto hydrodynamic waves \citep[e.g.][]{Ferland09,Fossati16, Boselli16}. In massive galaxies which host an AGN,
such as NGC 4388 in Virgo, the diffuse gas can also be ionised by the powerful nuclear source \citep{Yoshida04}.
Filamentary structures such as those seen in the tails NGC 4330 in Virgo \citep{Fossati18}, reproduced by the most recent hydrodynamic simulations, suggests that magnetic field might also play an important role in keeping the stripped gas in unmixed filamentary structures in the tail \citep{Tonnesen14, Ruszkowski14}. The same simulations also indicate that magnetic fields contribute to keep the stripped gas confined within extended tails
without significantly affecting the stripping rate from the disc. Large scale magnetic fields aligned along the tail of stripped gas have been measured using polarised radio observations in the jellyfish galaxy JO206 by \citet{Muller21}. 

The large number of H$\alpha$ extended tails observed so far in nearby clusters allows us to have a clear view of the variety of possible configurations produced during the interaction
between the cold ISM of freshly-infalling gas-rich galaxies with the surrounding ICM. In most of the observed cases, the diffuse tails include compact regions where star formation is taking place.
Very deep narrow band imaging suggest that even in the most extreme cases dominated by compact \hii\ regions such as IC 3418 (VCC 1217) in Virgo \citep{Hester10, Fumagalli11, Kenney14}, a diffuse component is also present (Sun et al. in prep.).
There are, however, a few galaxies such as NGC 4569 \citep{Boselli16} in the Virgo cluster and CGCG 097-073 and 097-079 in A1367 \citep{Pedrini22} where star formation is not present in the stripped gas. 

The data derived from narrow-band imaging observations, once corrected for [NII] contamination ([NII]/H$\alpha$ $\sim$ 0.5, \citealt{Boselli16,Poggianti19}) and dust attenuation (expected to be low in these environments), and IFU 
data can be used to derive several physical properties of the ionised gas tails. First of all H$\alpha$ luminosities can be used to estimate mean ionised gas densities as in \citet{Fossati16} and \citet{Boselli16}
making simple assumption on the geometrical distribution of the gas and on the filling factor. Indeed, the H$\alpha$ luminosity of the gas is given by the relation \citep{Osterbrock06}:

\begin{equation}
L({\rm H\alpha}) = n_{\rm e} n_{\rm p} \alpha^{\rm eff}_{\rm H\alpha} V f h \nu_{\rm H\alpha}
\end{equation}

\noindent
where $n_{\rm e}$ and $n_{\rm p}$ are the number density of electrons and protons ($n_{\rm e}$ $\simeq$ $n_{\rm p}$), $\alpha^{\rm eff}_{\rm H\alpha}$
is the H$\alpha$ effective recombination coefficient ($\alpha^{\rm eff}_{\rm H\alpha}$ = 1.17 $\times$ 10$^{-13}$ cm$^3$ s$^{-1}$), $V$ is the volume of the emitting region, $f$ the filling factor, $h$ the Planck's constant, and $\nu_{\rm H\alpha}$ the frequency of the H$\alpha$ transition. This leads to:

\begin{equation}
n_{\rm e} = \sqrt{\frac{L({\rm H\alpha})}{\alpha^{\rm eff}_{\rm H\alpha}Vfh\nu_{\rm H\alpha}}}
\end{equation}

\noindent
with typical values of $n_e \simeq 10^{-2}$ cm$^{-3}$, and total masses of ionised gas in the tail of:

\begin{equation}
M_{\rm ionised} = V f n_e
\end{equation}

\noindent
of several $10^8$ M$_{\odot}$ for $f = 0.1$ 
%{\color{red}{MS: the X-ray/Ha correlation suggests an Ha filling factor of ~ 4000 smaller than 0.1. MF: CHANGES TO BE MADE TO THE TEXT?}}
and tails $\sim$ 100 kpc long, thus comparable to those of the cold atomic hydrogen which is lacking from the perturbed galaxies, as suggested by \hi\ observations and 
tuned models or simulations \citep[e.g.][]{Boselli16}. It is worth noticing that these gas densities, although very uncertain because of the huge uncertainties on the geometrical distribution of the gas and on the filling
factor, are very low, and can be hardly compared to those derived using standard optical line ratio diagnostics ([SII]$\lambda$6716/6731 \AA, [OII]$\lambda$3729/3726 \AA) since these tracers are degenerate when $n_e$ $\leq$ 10
cm$^{-3}$ \citep{Osterbrock06}. Infrared line diagnostics such as [NII]$\lambda$205/122 $\mu$m, sensitive to lower electron densities \citep{Kewley19}, should be used to confirm the low gas density within the tail.
The gas density can also be used to estimate the typical hydrogen recombination timescale:

\begin{equation}
\tau_{\rm rec} = \frac{1}{n_{\rm e} \alpha_{\rm A}}
\end{equation}

\noindent
where $\alpha_{\rm A}$ is the total recombination coefficient ($\alpha_{\rm A}$ = 4.2 $\times$ 10$^{-13}$ cm$^3$ s$^{-1}$; \citet{Osterbrock06}.
For a typical density of $n_{\rm e}$ $\simeq$ 10$^{-2}$ cm$^{-3}$, the recombination time is $\tau_{\rm rec}$ $\simeq$ 8 Myr, 
a short time when compared to the galaxy travel time that can produce a tail $\simeq$ 50 kpc long (30 Myr). Indeed, several observations \citep[e.g.][]{Chung07} suggest 
that the gas can even be stripped as \hi, being further heated up to the ionised phase when in the tail. While some of the energy input could come from the UV radiation produced by \hii\ regions in the tail, at the edges of the stripped disc, or from a powerful AGN radiation field, 
the fact that we detect ionised gas also in poorly star forming galaxies suggests that additional energy sources are shocks, ICM heat conduction or MHD waves \citep{Merluzzi13, Fossati16, Boselli16}, once again supporting the scenario where cold gas can change phase once stripped from the galaxy disc.

\subsubsection{Hot gas}
\label{subsec:hotgas}

%{\color{red}{MS (Mar. 5, 2021) --- I suggest we shorten the next two paragraphs significantly to save space, as the early-type galaxies are not much relevant here and some are stripping of the group X-ray emission. My writing will be in red and we can discuss.\\}}
%YES MING, SHORTEN IT AS YOU WISH TO FOCUS ON RP ON GALAXIES ONLY (AND MAYBE JUST MENTION THAT THIS CAN OCCUR ALSO ON GROUPS, SLOSHING)

%1. very briefly mention stripping in X-rays and X-ray tail in general (only mention a few papers) and summarize the main results and the constraints on the ICM microphysics.
%2. X-ray tails of late-type galaxies \& summary

Tails of ram pressure stripped gas have also been searched for and detected in the hot ($T \sim 10^{7}$ K) gas emitting in X-rays. As massive early-type galaxies are typically rich on X-ray gas, many examples of X-ray stripped tails behind cluster early-type galaxies have been found \citep[e.g.,][and references/citations therein]{Machacek06,Sun07a,Randall08,Wood17}. Even cluster early-type galaxies are beyond the scope of this paper, the physical processes related to their X-ray tails are often the same as or similar to those related to X-ray tails of cluster late-type galaxies. Since X-ray emission from cluster early-type galaxies and their tails is typically more luminous than that of cluster late-type galaxies, their studies offer important clues to our understanding of the X-ray tails behind cluster late-type galaxies, including constraints on the ICM transport processes (i.e., viscous diffusion and heat conduction).
% a brief summary on the importance of these tails,
%The recent results have shown:
It is now known that:
1) the most significant observational signature is the contact discontinuity (or cold front) separating the galactic medium and the ICM \citep[e.g.,][]{Zuhone16}. While a shock front is expected ahead of the contact discontinuity for galaxies moving supersonically, such a shock front has never been robustly detected in X-rays for galactic halos as the Mach number is typically not large (e.g., less than 1\% of cluster galaxies with a Mach number of $>$ 3, \citealt{Faltenbacher05}) and the emission enhancement on the galaxy scale ($\sim$ 10 kpc) has to be projected on the ICM emission on the cluster scale ($\sim$ 1 Mpc) along the line of sight. 
2) For cold fronts and X-ray halos of cluster early-type galaxies with the best data, it is concluded that both viscosity and heat conductivity have been suppressed by at least a factor of 20 - 100 relative to the isotropic Spitzer values \citep[e.g.,][]{Markevitch07,Sun07a,Su17,Ichinohe17,Kraft17}. Efficient mixing in stripped tails and at small scales of cold fronts, likely from Kelvin-Helmholtz instabilities, has been observed.
% NO very long X-ray tails
3) Magnetic field in both the ICM and the ISM can have a significant impact on the properties of the X-ray tails \citep[e.g.,][]{Shin14,Vijayaraghavan17}.
4) The actual galaxy trajectory and the ICM wind history can play important roles to shape the properties of the X-ray tails \citep[e.g.,][]{Sheardown19}.
% tail direction (slingshot)

%Most of them have been discovered associated to massive elliptical galaxies in nearby clusters such as Virgo (M86, Forman et al. 1979; Finoguenov et al. 2004; Randall et al. 2008; M49, Irwin \& Sarazin 1996; Kraft et al. 2011; M89, Machacek et al. 2006; M60, Randal et al. 2004, Wood et al. 2017), Coma (NGC 4839, Neumann et al. 2003; Lyskova et al. 2019), Fornax (NGC 1404, Jones et al. 1997; Machacek et al. 2005a, Su et al. 2017), Hydra (De Grandi et al. 2016), A2142 (Eckert et al. 2017), A85 (Ichinohe et al. 2015), and Z8338 (CGCG254-021, Schellenberger \& Reiprich 2015). Some of them are spectacular in terms of physical extension: in M86 the tail has a length of $>$ 380 kpc (Randall et al. 2008), in M49 60 kpc (Kraft et al. 2011), and in CGCG254-021 76 kpc (Schellenberger \& Reiprich 2015).

Compared with X-ray tails behind cluster early-type galaxies, X-ray tails behind cluster late-type galaxies bear several important differences:
1) Their X-ray tails typically co-exist with cooler gas while the tails of cluster early-type galaxies are single phase, hot gas removed from the galaxy.
2) While X-ray tails have been found behind late-type galaxies as faint as $\sim$ 0.15 $L_{*}$ galaxies (e.g., D100 and CGCG~097-079, \citealt{Sun22}), X-ray tails of cluster early-type galaxies are only observed in massive galaxies (e.g., $L > L_{*}$) with abundant X-ray emitting gas in the galaxy.
% need to check, N4552
% S07 sample, the only tail PGC 018313 in A3376 is a 1.44 L* galaxy
3) While X-ray tails of cluster late-type galaxies are typically fainter than their counterparts behind cluster early-type galaxies, they are often more luminous than their host galaxies in X-rays \citep{Sun22}, while that is seldom the case for cluster early-type galaxies.
% this is an important point for mixing and we can make a nice figure out of it.
4) With potentially abundant cold gas in the stripped tails of cluster late-type galaxies, the full evaporation of the stripped ISM can be slow, especially if the stripped cold clouds can grow when the radiative cooling in the mixing layers is strong \citep[e.g.][]{Gronke18,Sparre20}. On the other hand, radiative cooling is never strong in the X-ray tails behind cluster early-type galaxies.\footnote{For example, the radiative cooling time is 1.3 - 7.2 Gyr for several tail regions in NGC~4552's X-ray tail beyond the galaxy, with the X-ray gas properties derived by \citet{Kraft17}.}
Thus, one may expect fast evaporation in tails behind cluster early-type galaxies.
This would result in longer X-ray tails (relative to the galaxy size) behind cluster late-type galaxies than those behind cluster early-type objects.
%Thus, one may expect that the generally slower mixing in tails behind cluster late-type galaxies than tails behind cluster early-type galaxies can
% tail length vs. truncation radius
5) High-mass X-ray binaries and even ultraluminous X-ray sources can be found in tails of cluster late-type galaxies \citep{Sun06,Sun10,Poggianti19}, while only low-mass X-ray binaries are detected in early-type galaxies but none in their hot gaseous tails. 

Detections of X-ray tails behind cluster late-type galaxies require the sensitivity of \chandra{} and \xmm{}. The first possible candidate is the galaxy C153 in the cluster A2125 at $z$=0.253 \citep{Wang04}. While its X-ray tail is faint (for its distance) and not unambiguous, the UV trail and the \oii{} tail of the galaxy are significant \citep{Owen06}.
%but the low signal-to-noise makes this detection not unambiguous \citep{Wang04,Owen06}.
With the \chandra{} data, \cite{Sun05} presented the discovery of an X-ray tail behind UGC~6697, a starburst galaxy that was known to host tails in the radio continuum and H$\alpha$ \citep{Gavazzi78,Gavazzi01} in A1367. The \xmm{} data show that the X-ray tail extends to at least 120 kpc from the nucleus \citep{Ge19,Sun22}.
%The gas temperature within the 60 kpc long tail is $T$ $\sim$ 0.7 keV compared to that of the surrounding ICM of $T$ $\simeq$ 5-6 keV and has a metallicity 0.1-0.2 solar.
An X-ray tail was also observed behind the giant spiral NGC~6872 in the 0.5 keV Pavo group \citep{Machacek05a}, but it is uncertain how it was formed as the tail terminates on the dominant early-type galaxy of the group and the tail appears hotter than the group gas. Evidence of a short X-ray tail formed after an RPS process has also been found for the starburst galaxy NGC~2276 in the NGC~2300 group \citep{Rasmussen06}.

By far the brightest X-ray tail behind a cluster late-type galaxy is ESO~137-001 \citep{Sun06,Sun10} in the Norma cluster (A3627). Its X-ray tail, revealed from both \chandra{} and \xmm{}, is bifurcated and extended to at least 80 kpc in projected distance from the nucleus. Six X-ray point sources as candidates of intracluster ultra-luminous X-ray sources were also detected.
ESO~137-001 has then become the RPS galaxy with the most extensive multi-wavelength coverage, with the stripped tail detected in X-rays, H$\alpha$, FIR emission of dust, warm H$_{2}$ (from MIR), cold molecular gas, \hi\ and radio continuum \citep[][Ramatsoku et al. private communication]{Sun10,Fumagalli14,Fossati16,Sivanandam10,Jachym14,Jachym19}. It is also the only RPS galaxy in the GTO program of {JWST}.
%galaxy disc, with a mass of hot gas of $\sim$ 10$^9$ M$_{\odot}$ and a temperature $T$ $\sim$ 0.8 keV, thus as in UGC 6697 significantly colder than that of the surrounding ICM (6 keV; Sun et al. 2010).
%Particularly interesting is the fact that, with respect to those observed in ellipticals, this tail is detected also in H$\alpha$ and H$_2$, confirming the existence of a multiphase medium embedded within the hot cluster ICM (Sun et al. 2010). Thanks to the excellent X-rays data for the Norma cluster, Sun et al. (2010) have shown that the hot gas is well confined within the two parallel structures of the two tails, which have an almost constant width and a constant temperature along their length. 
%They also detected 19 X-rays point sources, out of which six have been identified as candidates of ultraluminous intracluster point sources with luminosities $L_{0.3-10 keV}$ up to 2.5 $\times$ 10$^{40}$ erg s$^{-1}$.
The Norma cluster also hosts another narrow X-ray tail behind ESO~137-002 \citep{Sun10,Zhang13}, with the X-ray emission detected to at least 40 kpc from the nucleus.

Candidates of X-ray tails in the closest Virgo cluster were reported behind NGC~4388 and NGC~4501 \citep{Wezgowiec11}, while the analysis by \cite{Gu13} also suggested a soft X-ray tail behind NGC~4388, positionally coincident with its \hi{} tail discovered by \cite{Oosterloo05}. \cite{Gu13} also presented the evidence of excess X-ray absorption in the NGC~4388's tail on the background bright M86 emission, likely originated from the extra \hi{} absorption column in NGC~4388's tail.
% X-ray shadowing also seen in X-ray cool cores (e.g., Centaurus, A2029)
A wide X-ray tail was also detected behind the most massive spiral in the Virgo cluster, NGC~4569 \citep{Boselli16,Sun22}. 

Another spectacular example of X-ray tails in a rich cluster is the galaxy D100 in Coma \citep{Sanders14}, which was first discovered in H$\alpha$ \citep{Yagi07,Yagi10}.
It has quickly become another RPS tail with rich multi-wavelength data for detailed studies \citep{Jachym17,Cramer19}.
%Making some assumptions on the geometrical distribution of the gas, the authors estimated that the electron density of the X-rays emitting gas is $n_e$ $\simeq$ 8$\times$ 10$^{-3}$ cm$^{-3}$, and that the total mass of hot gas ($\simeq$ 10$8$ M$_{\odot}$ equals that of the ionised gas detected in H$\alpha$. They also claimed that in this object radiative cooling is not important (Sanders et al. 2014). 
The GASP  survey has reported two X-ray tails, IC~5337 in A2626 \citep{Poggianti19} and KAZ~364 (or JO201) in A85 \citep{Campitiello21}. IC~5337 is a massive RPS galaxy and its short X-ray tail is luminous, also with multi-phase medium and SF detected. The X-ray tail of KAZ~364 is however very short.

%Despite the unsuccess of several targeted surveys (e.g. Jachym et al. 2013; Boselli et al. 2018c), we expect that other tails could be detected in X-rays provided that they are searched for with sufficiently deep X-rays observations. 
%Another interesting use of X-rays data in the study of the multiphase properties of the stripped medium has been done by Gu et al. (2013) in the M86 - NGC 4388 complex. By measuring the X-rays absorption along the line of sight in the diffuse tail of M86 due to the overlapping \hi\ tail of NGC 4388, Gu et al. (2013) measured the column density of the gas in the tail located in front of the diffuse hot halo component of M86. They also detected a X-ray counterpart of the \hi\ tail of NGC 4388, and estimated its mass which exceeds that of the \hi\ in the same region, concluding that most of the gas in the tail changed phase because of heat conduction once in contact with the surrounding hot ICM.

More X-ray tails behind cluster late-type galaxies were recently presented in \cite{Sun22}, including CGCG~097-079, 2MASX~J11443212+2006238 and CGCG~097-092 in A1367, NGC~4848, GMP~3816 (or NGC~4858) and GMP~4555 in Coma, all benefiting from recent deep \chandra{} and \xmm{} observations.
\cite{Sun22} also presented a tight, linear correlation between the X-ray surface brightness and the H$\alpha$ surface brightness in the RPS tails at scales of $\sim$ 10 - 40 kpc. 

From all the studies so far, we now know the following properties of X-ray tails behind cluster late-type galaxies:
1) The gas temperature is typically 0.8 - 1.2 keV, if fitted with a single thermal component (e.g., APEC). However, this should only be taken as an effective temperature as the tail is multi-phase with a broad temperature distribution \citep[e.g.,][]{Sun10}.
2) the best-fit abundance is low (0.1 - 0.4 solar typically) from the single-$T$ fits \citep{Sun22}, which again suggests the multi-$T$ components in the X-ray tails. This is similar to the ``iron bias'' first identified in X-ray cool cores. Non-equilibrium ionization may also play a role on the low abundance \citep{Sun10}.
3) a strong H$\alpha$ - X-ray linear correlation for diffuse gas beyond the star-forming regions has been established in tails \citep{Sun22} at 10 - 40 kpc scales, which suggest mixing of the cold ISM and the hot ICM as the origin of multi-phase tails behind cluster late-type galaxies. Both X-ray and H$\alpha$ surface brightness may be tied to the external pressure of the surrounding ICM \citep{Sun10,Tonnesen11}.
4) star formation in stripped tails can produce high-mass X-ray binaries and even ultraluminous X-ray sources \citep{Sun10,Poggianti19}.
%5) some narrow and bifurcated; face-on stripping and edge-on stripping
%5) more in clusters than in groups groups are on average poorer in the hot gas content than clusters \citep[e.g.,][]{Sun12}.

Soft X-ray emission in the stripped tails provides an important constraint on mixing and serves as a key link between the external heat pool of the ICM and the stripped cold ISM. Looking forward, more X-ray tails should be discovered from the {eRosita} survey in 5 - 10 years and eventually future X-ray observatories like {Athena} and {Lynx} will allow X-ray kinematics and line diagnostics studies in the tails, besides adding many more detections. Trajectories of cluster galaxies can no longer be hidden in X-rays!

\subsubsection{Dust}
\label{subsec:dust}

As a major constituent of the ISM, the dust component is also expected to be removed during the hydrodynamic interaction with the surrounding ICM. 
Stripping, however, should be less efficient than for the diffuse \hi\ since dust is mainly associated to the molecular gas component deeply 
embedded into giant molecular clouds (see Sec. \ref{subsec:molecular}). Considering that the typical gas-to-dust ratio ($G/D$) in spiral galaxies is of the order of $\sim$ 100
we can roughly estimate that for a gas loss of $M_{\rm gas} \simeq 10^8-10^9$ M$_{\odot}$ in a typical late-type galaxy of stellar mass $M_{\rm star} \simeq 10^8-10^9$ M$_{\odot}$
the amount of dust removed during the interaction is $M_{\rm dust}$ $\simeq$ 10$^6$-10$^7$ M$_{\odot}$. If, once removed, this dust is 
distributed on a diffuse tail of $\sim$ 50 kpc length and $\sim$ 10 kpc width, its typical column density would be $\Sigma_{\rm dust}$ $\simeq$ 2 $\times$ 10$^3$ - 2 $\times$ 10$^4$ M$_{\odot}$ kpc$^{-2}$. This estimate can be 
refined considering that the gas-to-dust ratio is metallicity dependent as \citep{Remy-Ruyer14}\footnote{Different recipes for gas-to-dust ratio vs. metallicity relation are given in this reference.}; 
see also \citet{Sandstrom13}):

\begin{equation}
\log(G/D) = 2.21 + (8.69 - 12 \rm{\log O/H)}
\end{equation}

\noindent
where the solar metallicity is 12 + log O/H = 8.69 \citep{Asplund09}, which leads to $G/D_{\odot}$ = 164 \citep{Zubko04}.

Dust masses and column densities can be transformed into flux densities $S_{\lambda}$ and surface brightnesses in different photometric far-IR bands using the relation \citep[e.g.][]{Spitzer78, Boselli11}:

\begin{equation}
M_{\rm dust} = \frac{S_{\nu} D^2}{K_{\nu}B(\nu,T)}
\end{equation}

\noindent
where $D$ is the distance of the galaxy, $K_{\nu}$ the grain opacity at a given frequency and $B(\nu,T)$ the Planck function for dust at a temperature $T$.  
Assuming a dust temperature of $T$ = 20 K, a grain opacity of $K_{250}$ = 4.00 cm$^2$ g$^{-1}$ \citep{Draine03}, the typical far-IR surface brightnesses of these tails would be 
$\Sigma S_{250 \mu m}$ $\simeq$ 5-50 $\mu$Jy arcsec$^{-2}$.
These surface brightness values are comparable to the confusion limit of SPIRE on \her{} (5.8, 6.3, 6.8 mJy/beam at 250, 350, and 500 $\mu$m respectively, which for a beam size
of FWHM 18.1, 24.9, and 36.6 arcsec correspond to 22.6, 12.9, 6.5 $\mu$Jy arcsec$^{-2}$, \citet{Nguyen10} and are thus hardly detectable with this 
instrument even using very deep observations. Thanks to their resolving power, interferometric observations such as those available with ALMA or NOEMA at the Plateau du Bure 
would thus be ideal to resolve the background point sources from the diffuse and extended emission of the dust tails possibly formed during an RPS event.

At present the only study dedicated to the detection of the diffuse dust component in the tail of ram pressure stripped galaxies is the work of \citet{Longobardi20a}. 
Using the far-IR data gathered during the \her{} Reference Survey (HRS; \citep{Boselli10}, a complete survey of a $K$-band selected, volume-limited sample of nearby galaxies
with available data in the SPIRE (250, 350, 500 $\mu$m, \citep{Ciesla12} and PACS (100, 160 $\mu$m, \citep{Cortese14}) bands, \citet{Longobardi20a} looked for and detected with a typical $S/N$ of $\sim$ 5
a diffuse dust component associated to the \hi\ and H$\alpha$ tails of three Virgo cluster galaxies (NGC 4330, 4522, 4654) undergoing an RPS event \citep{Chung07}. The gas-to-dust ratio 
derived in these tails for reasonable values of $T$ are consistent with those observed in the outer disc of spiral galaxies, making the detection trustworthy, and confirming that during an RPS event dust is removed with the other components of the ISM \citep{Longobardi20a}. This also indicates that dust stripped from gas-rich late-type systems can contribute 
to the pollution of the ICM \citep{Domainko06, Longobardi20}. The results of \citet{Longobardi20a} are in line with the evidence that cluster galaxies have, on average, a dust-to-stellar mass ratio lower than that of similar objects in the field, as observed by \citet{Bianconi20} in the cluster A1758. They are also consistent with the presence of kpc scale dusty filaments partly decoupled from the surrounding ISM perpendicular to the plane of the disc seen in absorption in the spectacular \hst{} images of a few ram pressure stripped galaxies in the Virgo cluster \citep{Abramson16}.
For completeness, it is also worth mentioning that dust has been detected in the tails of ram pressure stripped galaxies using absorption line measurements through the Balmer decrement \citep{Fossati16, Poggianti19}. This dust, however, has a clumpy structure since it has been formed within the star forming regions present in some of the perturbed galaxies. This observational technique is hardly applicable to the diffuse ionised gas since it requires that the gas is photoionised by the young stellar component (case B), which is not always the case in these particular environments (see Sec. \ref{subsec:ionised}).

\subsection{Indirect effects on the stellar spatial distribution}
\label{subsec:stars}

While the ICM ram pressure cannot move stars, it can affect the stellar distribution of the galaxy indirectly, if the stripped ISM can form stars. Star formation in the stripped ISM was suggested in the early simulations by \cite{Schulz01}. Isolated star formation has been known in the Virgo cluster since 2002 \citep{Gerhard02,Cortese04a} but lacked direct evidence to link with RPS (also see discussion in \citealt{Yoshida04}). The first observational data to reveal star formation in the stripped ISM come from \cite{Owen06} on C153 in Abell~2125, \cite{Cortese07} on 235144-260358 in Abell~2667 and 131124-012040 in Abell~1689, \cite{Sun06,Sun07} on ESO~137-001 in Norma, \cite{Yoshida08} on RB199 and \citet{Smith10}, \citet{Hester10}, \citet{Yagi10} in several other galaxies in Coma. As tidal effect always exists for cluster galaxies, it was important to have this ``early burst'' of observational evidence to firmly establish the action of star formation in the stripped ISM by ram pressure. Since then, many more examples have been revealed.
At intermediate redshift, where the full extent of clusters has been easier to cover with deep UV \footnote{Diffuse filaments in the FUV and NUV bands must be interpreted with care since the UV emission could be originating from hydrogen 2-photon continuum or resonant lines of CIV and MgII generated by shocks rather than from stellar continuum emission \citep[e.g.,][]{Bracco20}.} and optical imaging, galaxies with blue knots of star formation outside of the stellar disk have been found by \citet{Cortese07}. More recently \citet{Ebeling14} presented a large survey of such objects in X-Ray selected clusters at $z>0.3$, directly linking star forming knots to RPS.

These objects, commonly named ``jellyfish galaxies'' because of their peculiar morphology with tentacles of materials that appears to be stripped from the main body of the galaxy \citep[e.g.][]{Chung09, Ebeling14, Poggianti16}, have been more recently defined by \citet{Ebeling14} whenever they exhibit in optical broad band images the following features:

\noindent
1) a strongly disturbed morphology indicative of unilateral external forces.  

\noindent
2) a pronounced brightness and colour gradient suggesting extensive triggered star formation.

\noindent
3) a compelling evidence of a debris trail. 

\noindent
4) the directions of motion implied by each of these three criteria have to be consistent with each other.

%%%%\citet{Poggianti16}, using the \citet{Ebeling14} criteria, further defined five morphological classes according to the visual evidence of stripping signatures in the optical bands, from JClass 1 for the weakest to JClass 5 for the clearest cases. 
Since their discovery in intermediate redshift  massive clusters ($z \sim 0.2$) by \citet{Owen06} and \citet{Cortese07} who did not yet use the term ``jellyfish'', these objects are now becoming common in different environments in the local \citep{Smith10, Poggianti16, Poggianti17} 
and intermediate redshift Universe up to $z \simeq 0.7$
\citep{Owers12, Ebeling14, Rawle14, McPartland16, Ebeling19, Roman-Oliveira19, Durret21}. From the relative position of these galaxies with respect to that of the cluster (they are principally located in the outskirts), and form the analysis of their velocity distribution 
it has been concluded that jellyfish galaxies are infalling within the high-density region as a group along trajectories with high impact parameters rather than infalling into the cluster centre along filaments (small impact parameters) as individual objects \citep{McPartland16, Ebeling19}. %They have also shown that the stripping event is generally short lived ($\lesssim$ 500 Myr). They can thus be used to trace the dynamical state of the density region and to find ongoing merging clusters.  

We want to stress, however that jellyfish galaxies defined solely per the criteria above can not be \textit{a priori} considered as systems undergoing an RPS event
because the stellar distribution
has a too small cross-section to be perturbed by the external gas flow\footnote{The case of Mira indicates that the external pressure can perturb the wind produced by the mass loss of evolved stars as AGB stars \citep[e.g.,][]{Li19b}. In general, stellar winds are subject to ram pressure.}. Galaxies with this morphological aspect can be the results of an RPS event only if all stars in the tail have been formed outside the stellar disc after the stripping event. Indeed, as largely discussed in \citet{Cortese07}, these asymmetric morphologies with strong colour (age) gradients in the stellar populations and stellar tails can also be
produced in long-lived gravitational interactions with nearby companions \citep{Merritt83}, high velocity encounters with other cluster members (harassment, \citet{Moore98, Moore99}), or 
by the perturbation induced by the gravitational potential well of the cluster \citep{Byrd90, Valluri93}.
%A typical example is the galaxy NGC1427A in the Fornax cluster, characterised by a peculiar stellar jellyfish morphology and a long tail of \hi\ gas \citep{Lee-Waddell18}. Recent hydrodynamic simulations indicate that while the gas has been stripped by an RPS event, the stellar distribution has been perturbed by the gravitational potential well of the cluster \citep{Mastropietro21}.
RPS is a relatively short-lived process ($\lesssim$ 500 Myr, see Sec. \ref{subsec:timescale}),
it is thus conceivable that stars with ages older than these typical timescales have been formed within the disc of the galaxy and later removed during the interaction. 
For this reason optical selections, which are sensitive to the
distribution of stellar populations of ages of several Gyrs, can be easily contaminated by galaxies undergoing gravitational perturbations. 
%Furthermore, it has to be proven that the stars newly formed in the tail, which are insensitive to the ongoing hydrodynamic pressure and thus get decoupled from the gaseous component which keep being decelerated during the interaction (see Sec. \ref{subsec:fate}), keep being gravitationally bound to the main body of the parent galaxy for such long timescales, which might exceed the crossing time of clusters ($\tau_{\rm cross}$ $\simeq$ 2 Gyr, \citealt{Boselli06}).

We can quantify the importance of high velocity gravitational interactions with other cluster members or with the potential well of the cluster and compare them to those of RPS following \citet{Henriksen96} and \citet{Cortese07}. An infalling galaxy suffering a gravitational perturbation is subject to two different accelerations, 
a radial tidal one ($a_{rad}$), which is pulling matter along the direction connecting the galaxy and the perturber on either sides of the disc:

\begin{equation}
a_{\rm rad} = GM_{\rm pert}\left[\frac{1}{r^2} - \frac{1}{\left(R+r\right)^2}\right]
\end{equation}

\noindent
and a transverse acceleration ($a_{\rm tr}$), which is pushing matter versus the inner regions on the disc and in the nucleus along the direction perpendicular to the
line connecting the galaxy and the perturber \citep{Henriksen96, Cortese07}:

\begin{equation}
a_{\rm tr} = GM_{\rm pert} \frac{R}{\left[R^2+(R+r)^2\right]^{1.5}}
\end{equation}
 
\noindent 
where $M_{\rm pert}$ is the mass of the perturber within $r$, $R$ is the radius of the perturbed galaxy, and $r$ is its distance from the perturber.
If the radial acceleration overcomes the internal galaxy acceleration keeping the matter linked to the gravitational potential well of the galaxy:

\begin{equation}
a_{\rm gal} = \frac{GM_{\rm dyn}}{R^2}
\end{equation}

\noindent
where $M_{\rm dyn}$ is the dynamical galaxy mass, the perturbation is able to remove matter from the galaxy disc. The perturber can be either another cluster galaxy or the cluster itself.
%For a given stellar mass, the total dynamical mass of a galaxy can be estimated following \citet{Moster10} or \citet{Behroozi13}. 
As shown in \citet{Cortese07}, 
for a non-interpenetrating galaxy-galaxy high velocity encounter the impact parameter is at least equal or greater than the galactic radius ($R \geq r$), which implies
that the interaction is able to remove mass only whenever $M_{\rm pert}$ $\geq$ 1.33$M_{\rm dyn}$, and the perturber should not be at a distance larger than the typical length of the tail.
This condition makes the formation of jellyfish galaxies quite unlikely for massive objects, whose number density in rich cluster remains limited, while it is still possible in low mass, low surface brightness systems \citep{Moore99, Mastropietro05}. 

\begin{figure}
\centering
\includegraphics[width=0.75\textwidth]{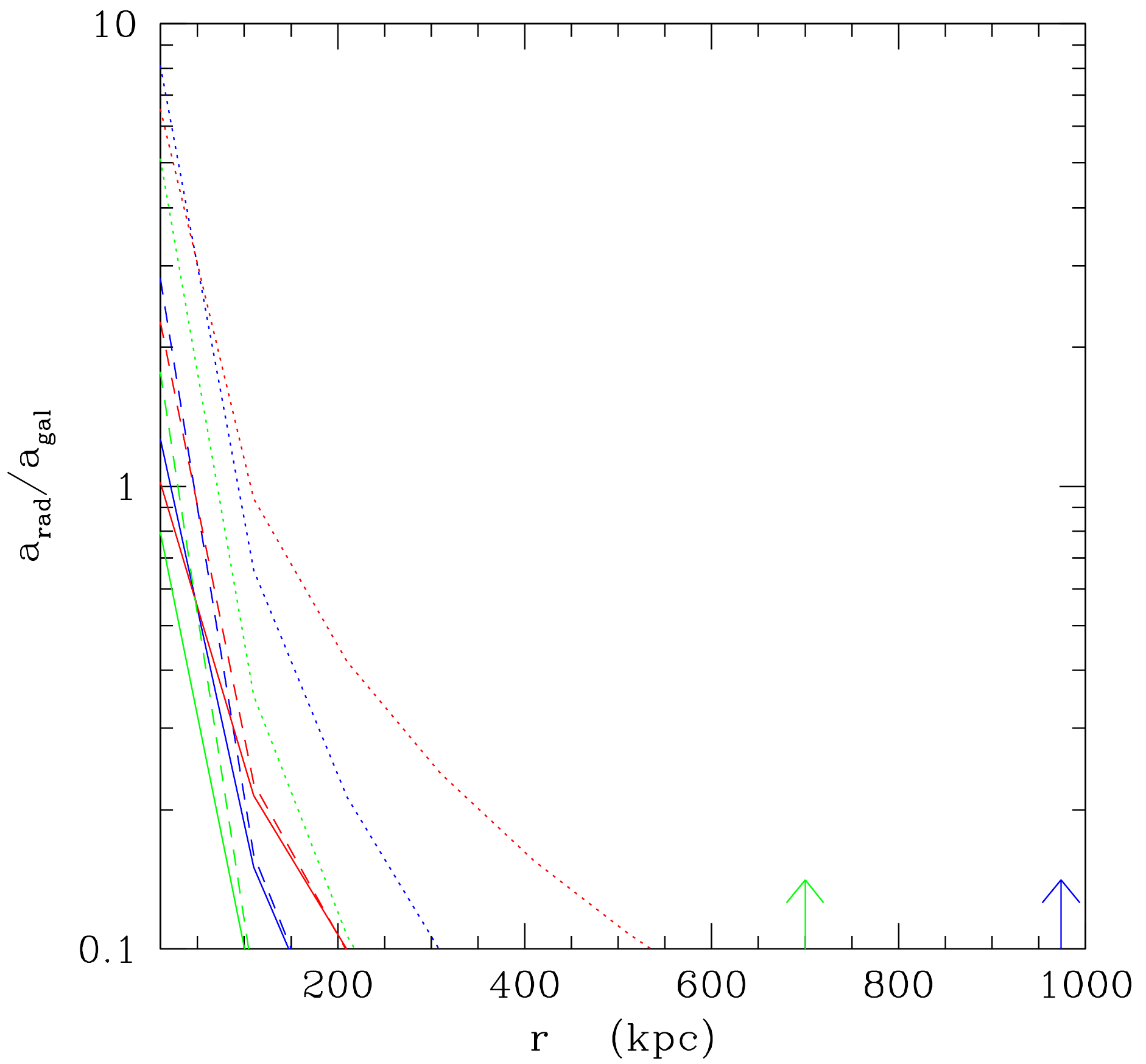}
\caption{Variation of the ratio between the radial acceleration on the galaxy disc induced by the gravitational potential well of the cluster ($a_{\rm rad}$) and the
acceleration due to the potential well of the galaxy ($a_{\rm gal}$) as a function of the distance from the cluster centre for four representative high density regions (Groups - green; Virgo - blue; 
Coma - red).
The different lines are for galaxies of different stellar mass: solid ($M_{\rm star}$ = 10$^{10}$ M$_{\odot}$), dotted ($M_{\rm star}$ = 10$^9$ M$_{\odot}$), and dashed ($M_{\rm star}$ = 10$^8$ M$_{\odot}$). The green and blue vertical arrows indicate $r_{200}$ for groups and for the Virgo cluster, respectively. The mass of the overdensity regions are derived using a NFW density profile assuming $c$ = 8.6, 4, and 8 for the Virgo, Coma, and Fornax (group) clusters, respectively, with $r_s$ = $r_{200}/c$.
}
\label{acc}       
\end{figure}

Both dwarfs and massive systems, however, can be perturbed by the gravitational potential well of the cluster itself. Following \citet{Cortese07}
we can quantify the effect of tidal forces induced by gravitational potential well of the cluster by assuming a NFW mass profile \citep{Navarro97} given in Sec. \ref{subsec:prophidens} where the mass of the perturber $M_{\rm pert}$ is derived using Eq. \ref{NFW2}.
We can than estimate under which conditions $a_{\rm rad}$ $\leq$ $a_{\rm gal}$
in different clusters such as Virgo, Coma, and A1689 and in representative galaxies of stellar mass $M_{\rm star} = 10^{10}$, $10^9$, and 10$^8$ M$_{\odot}$. 
For the three clusters we use the NFW parameters given in Table \ref{TabNFW}. 
%taken from \citet{McLaughlin99} (Virgo), \citet{Geller99} (Coma, similar to those of \citet{lokas03} and \citet{Kubo07} ), and \citet{Cortese07} (A1689). 
For the galaxies we calculate the total dynamical mass $M_{dyn}$ following
\citet{Behroozi13}, and assume as radii $r$ = 1.5 $R_{23.5}(i)$ derived from SDSS data of the HRS galaxies by \citet{Cortese12}, thus considering that the stellar disc extends outside
the $i$-band isophotal radius measured on the shallow SDSS images, as indeed deep UV images suggest (XUV discs, e.g. \citet{Thilker07}).
Figure \ref{acc} shows that, under some conditions, galaxies can be tidally perturbed by the gravitational potential well of the cluster up to $\simeq$ 250 kpc from the cluster centre.
This process is more efficient in massive clusters ($M_{\rm pert}$ $>$ 10$^{15}$ M$_{\odot}$) and in galaxies of intermediate stellar mass ($M_{\rm star}$ $\simeq$ 10$^9$ M$_{\odot}$)
because of their low $M_{\rm dyn}/M_{\rm star}$ ratio. We should recall that these estimates are probably lower limits because this effect is expected to be more efficient if the galaxies 
infall into the clusters within small groups, as indeed semi-analytic models and hydrodynamic cosmological simulations suggest: $\sim$ 40\%\ of the galaxies in massive clusters have been accreted as satellites of smaller groups \citep{Gnedin03, Gnedin03a, McGee09, DeLucia12, Han18}. 
In this case, high speed encounters with other members can rapidly change the gravitational field of the infalling galaxies, 
accelerating mass loss for tidal heating \citep{Taylor01}.

Using similar arguments we can also estimate which is the truncation radius outside which matter is removed from the perturbed galaxy
\citep{Binney08}:

\begin{equation}
R_{\rm trunc} \simeq R \left(\frac{M_{\rm dyn}}{M_{\rm pert}(<r)}\right)^{1/3}
\end{equation}

\begin{figure}
\centering
\includegraphics[width=0.75\textwidth]{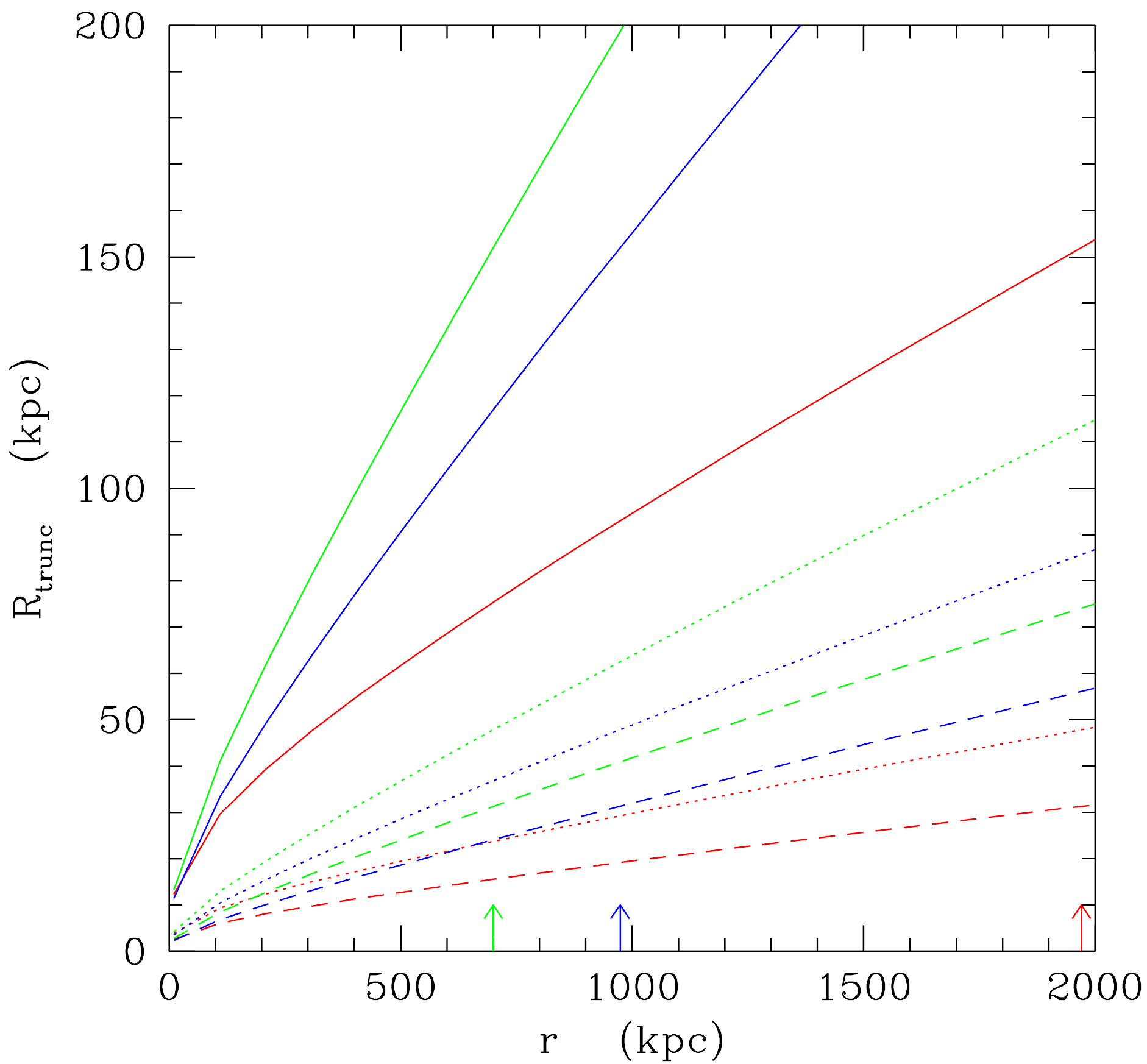}
\caption{Variation of the truncation radius as a function of the distance from the cluster centre for four representative high density regions (Group - green; Virgo - blue; Coma - red).
The different lines are for galaxies of different stellar mass: solid ($M_{\rm star}$ = 10$^{10}$ M$_{\odot}$), dotted ($M_{\rm star}$ = 10$^9$ M$_{\odot}$), and dashed ($M_{\rm star}$ = 10$^8$ M$_{\odot}$). The vertical arrows indicate $r_{200}$ for the different clusters.
}
\label{trunc}       
\end{figure}

\noindent
Figure \ref{trunc} shows how the truncation radius changes as a function of the distance from the cluster core for representative galaxies of three different stellar mass 
($M_{\rm star}$ = 10$^{10}$, 10$^9$, and 10$^8$ M$_{\odot}$) whose typical optical isophotal radius is $R_{23.5}(i)$ = 10, 5, 2 kpc \citep{Cortese12}, often ($\sim$ 30 \%) characterised by extended UV discs of twice the optical size and with \hi\ discs of size $\simeq$ $\times$ 1.8-1.9 the optical size \citep[][see Sec. \ref{subsec:truncdiscs}]{Cayatte94}. As mentioned above, in massive clusters ($M_\mathrm{cluster} \simeq 10^{15}$ M$_{\odot}$)
gravitational perturbations are able to remove gas and stars in the outer discs of infalling galaxies whenever they get close to the core of the cluster. 
%It is worth mentioning that, while observed in a variety of environments down to groups scales \citep{Poggianti16}, {and present in the field \citep{McPartland16} optically selected jellyfish galaxies are relatively rare in nearby clusters such as Virgo and Coma (see however \citep{Roberts20}). The galaxy NGC 4654 in Virgo has an asymmetric morphology similar to that of optically selected jellyfish galaxies. However, both gravitational and hydrodynamic perturbations have been invoked to explain its peculiar morphology \citep{Vollmer03}}. 
The observed by \citet{Cortese07} in two massive clusters at $z$ $\simeq$ 0.2 have been probably formed after the combined action of the gravitational interaction with the potential well of the clusters and ram pressure. 
%As extensively discussed in \citet{Cortese07}, the transverse acceleration pushes the gas towards the inner regions, favoring cloud-cloud collisions and thus increasing the density of the gas which can overcome the critical density for cloud collapse. This can lead to a rapid increase of the star formation activity in the disc and in the nucleus \citep{Henriksen96}, as indeed observed in several jellyfish galaxies \citep{Cortese07, Smith10, Vulcani18}.
To conclude, although the jellyfish morphology suggests an ongoing hydrodynamical interaction with the surrounding ICM, gravitational interactions can produce similar properties in the perturbed systems
thus this peculiar morphology cannot be taken by itself as an unambiguous sign of an ongoing RPS event.

\subsection{Truncated discs}
\label{subsec:truncdiscs}

Cluster galaxies are generally deficient in \hi\ gas (see Sec. \ref{subsec:integrated}).
Interferometric observations at 21 cm of late-type systems in nearby clusters have shown that the \hi\-deficiency\footnote{The \hi\-deficiency parameter, first defined by \citet{Haynes84}, 
is a measure of the logarithmic difference between the expected and the observed \hi\ mass of galaxies of different morphological type and size, where the estimated measure
is taken from standard scaling relations of unperturbed galaxies in the field. For an updated calibration of these relations, see \citet{Boselli09}.} is in most cases due to a reduced \hi\ discs. In Virgo, where
all objects are fully resolved thanks to the proximity of the cluster, it has been shown that the degree of truncation of the \hi\ disc
increases towards the core of the cluster where RPS is more efficient \citep{Warmels86, Cayatte90, Cayatte94, Chung09}. 
A similar trend has been observed also in the Coma \citep{Bravo-Alfaro00} and Fornax clusters \citep{Loni21}.
Given the radial exponentially declining gas and stellar surface densities of late-type galaxies, Eq. \ref{eq:rstrip} indicates that during a face-on interaction RPS 
removes the gas located in the outer disc producing truncated profiles while unperturbing the stellar distribution
(outside-in, \citep{Warmels88, Warmels88a, Warmels88b, Cayatte90, Cayatte94, Vollmer01}). It should be noted that, because of the conservation of the 
angular momentum including that of the spin vector, a late-type system entering a cluster will soon or late cross the ICM face-on although this might occur at different clustercentric distances if the infall is along a radial orbit. The observed outside-in truncation of the gaseous disc is in line with the prediction of simple models \citep{Hester06} and hydrodynamic simulations \citep{Abadi99, Quilis00, Roediger05, Roediger07}. 

\begin{figure}
\centering
\includegraphics[width=0.75\textwidth]{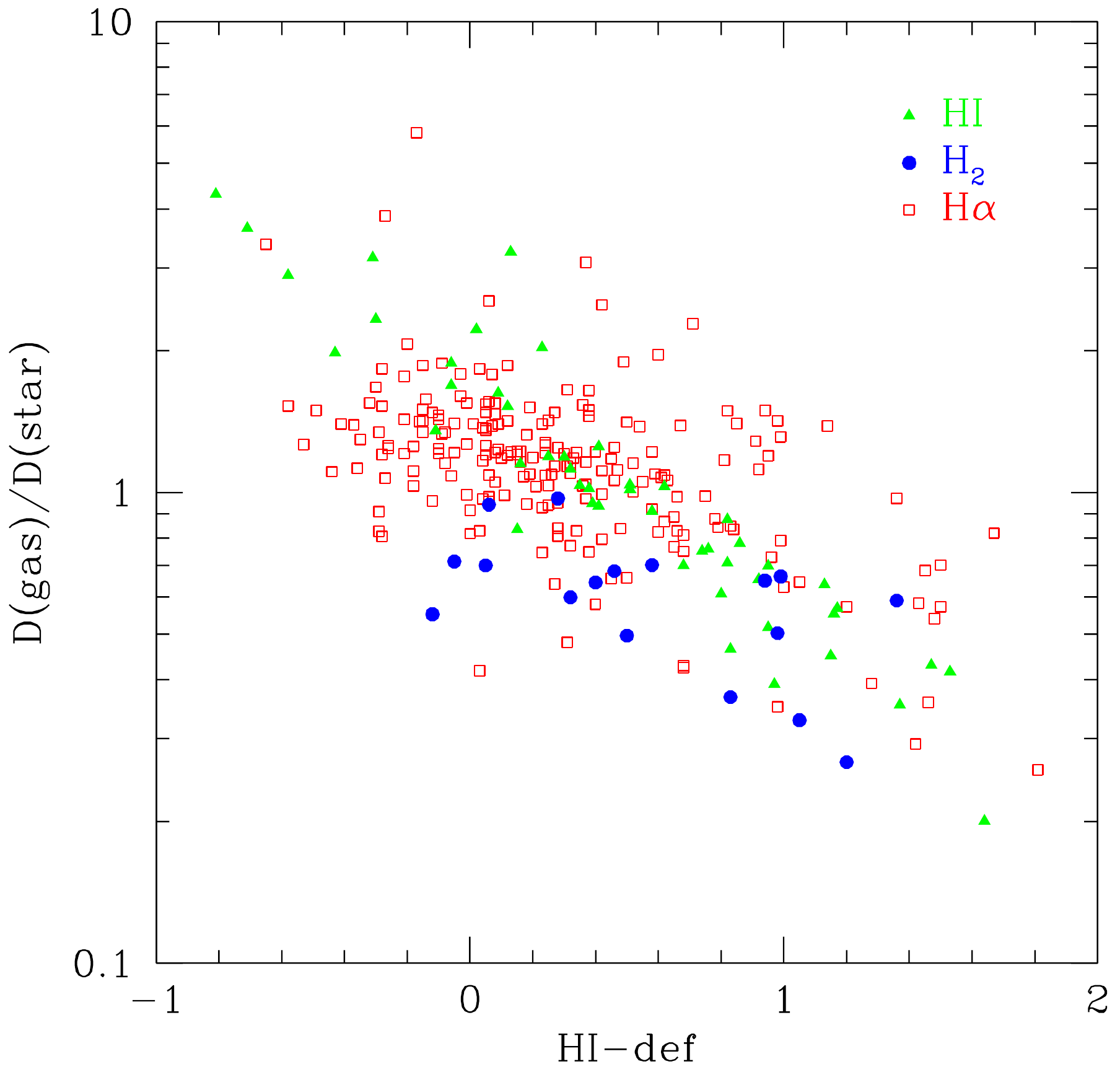}
\caption{Variation of the gas-to-stellar disc diameters of the atomic (green filled  triangles, with \hi\ isophotal diameters taken from \citet{Chung09}), the molecular 
(blue solid circles, with isophotal CO diameters taken from \citet{Chung09a}), and the ionised (red empty squares, with effective H$\alpha$ equivalent width diameters 
taken from \citet{Boselli15}) gas phases as a function of the HI-deficiency parameter for galaxies in the Virgo cluster (\hi, CO) and in the \her{} Reference Survey (H$\alpha$).
}
\label{profili}       
\end{figure}

Truncated discs in cluster galaxies have been also observed in other components of the ISM, from the molecular gas \citep{Fumagalli09, Boselli14a, Mok17} to the dust \citep{Cortese10, Cortese14},
with similar trends of the ratios of the CO-to-optical and dust-to-optical radii with the HI-deficiency parameter, suggesting that all the different components of the ISM are stripped away during the interaction.
Figure \ref{profili} shows the relationship between the ratio of the \hi\ and molecular gas isophotal radii to the optical radius (stars) 
as a function of the HI-deficiency parameter for the sample of Virgo cluster galaxies with interferometric \hi\ data gathered during the VIVA survey \citep{Chung09} and with available CO 
maps \citep{Chung09a}. Figure \ref{profili} clearly shows how the gaseous disc is reduced with respect to the stellar disc in galaxies deprived of their \hi\ content, indicating
that the stripping occurs outside-in. In the most deficient galaxies ($HI-def$ $\simeq$ 1) the gas disc is a factor of $\simeq$ 2 less extended than the optical disc, as expected in a ram
pressure stripping scenario where only the gaseous component is removed while leaving unaffected the stellar disc. It also shows that the molecular gas component, mainly located in the inner
disc within giant molecular clouds, is less subject to gas stripping than the diffuse \hi\ component (see Sec. \ref{subsec:molecular}).

\begin{figure}
\centering
\includegraphics[width=1.5\textwidth, angle=-90]{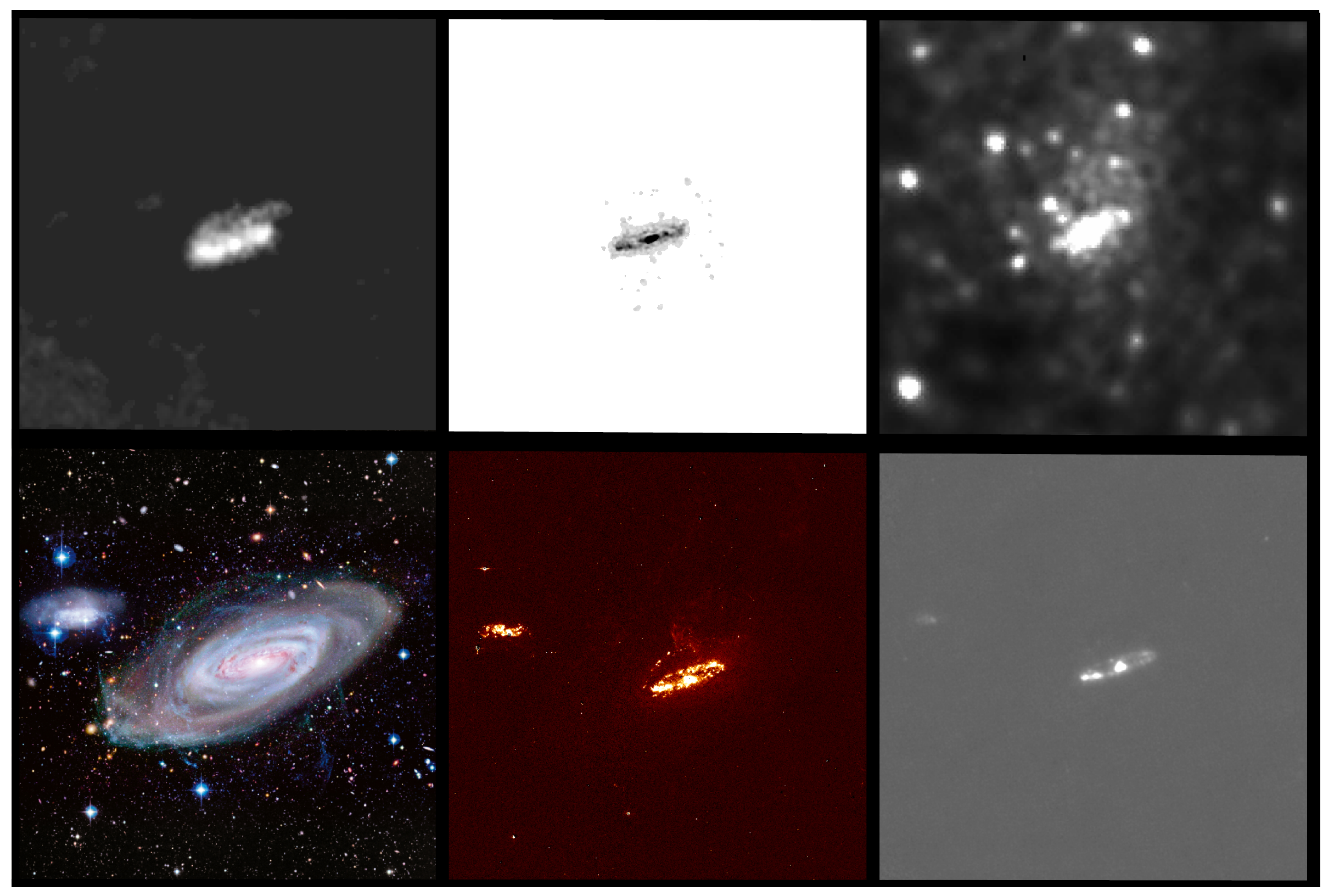}
\caption{Comparison of the 2D-distribution of the different components of the RPS galaxy NGC 4569. Left column, from top to bottom: colour image tracing the distribution of the stellar component (adapted from \citet{Boselli16}); continuum subtracted H$\alpha$ image tracing the distribution of the star forming regions (adapted from \citet{Boselli16}); \her/PACS image at 70 $\mu$m tracing the distribution of the dust component (from the KINGFISH survey, \citealt{Kennicutt11}). Right column, from top to bottom: atomic gas distribution (from the VIVA survey, \citet{Chung09}); molecular gas distribution derived from the $^{12}$CO(1-0) line transition (from the BIMA survey, \citealt{Helfer03}); hot gas distribution from \xmm. All panels are on the same spatial scale.
}
\label{N4569trunc}       
\end{figure}

The lack of gas, principal feeder of star formation, quenches the activity of star formation in the outer disc, producing truncated discs also in the youngest stellar populations
such as those traced by the H$\alpha$ line (O-early B stars of age $\leq$ 10 Myr, \citep{Kennicutt98, Boselli09a}), or by the UV emission 
(A-F stars of age $\leq$ 100 Myr), as depicted in Fig. \ref{profili}. Truncated discs in H$\alpha$ and UV have been observed in several HI-deficient Virgo cluster galaxies
by \citet{Koopmann04, Koopmann04a, Koopmann06}, \citet{Boselli06}, \citet{Cortese12}, and \citet{Fossati13}, and predicted by simulations \citep{Bekki09, Bekki14, Tonnesen12}. In dwarf galaxies, simulations indicate that the gas can be retained only in the nucleus, where the gravitational potential well is the deepest \citep{Steyrleithner20}. This gas can feed a nuclear activity of star formation, as indeed observed in several dwarf ellipticals in the Virgo cluster probably formed after an RPS event \citep{Boselli08}. The time elapsed since the first gas removal can be estimated by comparing models predictions
of radial gas stripping on the age-sensitive stellar populations observed in different photometric bands, as firstly done on the iconic galaxy NGC 4569 in the Virgo cluster \citep{Boselli06a}, illustrated in Fig. \ref{N4569trunc}. As shown in this figure, all the different  components of the ISM (atomic, molecular, ionised, and hot gas, dust) have been removed away from the outer disc during the stripping event. The lack of gas also reduced the activity of star formation, which is now limited only to the inner disc.
The truncation of the star forming disc has been observed in other environments down to massive groups \citep{Bretherton13, Schaefer17, Schaefer19, Owers19} and in clusters at redshift $0.3< z < 0.6$ \citep{Vaughan20}.

\subsection{Asymmetries}
\label{subse:asymmetries}

It has been suggested that galaxies suffering an external pressure can be characterised by an asymmetric distribution of the gas within the stellar disc,
sometime visible also in the shape of the two-horns \hi\ profile \citep{Gavazzi89, Roberts20}. In an edge-on
interaction, the gas is compressed on the leading side while it is stretched along the tail on the opposite side, as indeed observed in several representative
galaxies (CGCG 097-073 and CGCG 097-087 in A1367, \citep{Gavazzi95, Gavazzi01, Consolandi17, Pedrini22}; NGC 4501, NGC 4654 and IC 3476 in Virgo, \citep{Vollmer08a, Nehlig16, Vollmer03, Boselli21}; ESO 137-001 in Norma, \citep{Sun07, Fossati16}. At the leading side, the compressed gas can form a bow-shocked structure where star formation can take place,
sometime after a vigorous burst which leads to the formation of giant \hii\ regions as those observed in CGCG 097-073 in A1367 \citep{Gavazzi95, Gavazzi01} and IC 3476 in Virgo \citep{Boselli21}. The compression of the gas on the leading side occurs also in less inclined interactions, although in a less spectacular way \citep{Bekki14}. By comparing the properties of a sample of 11 ram pressure stripped galaxies extracted from the GASP  survey to those of a reference sample of unperturbed objects, \citet{Bellhouse21} have shown that the pitch angle of perturbed galaxies increases radially as predicted by tuned simulations. Ram pressure, thus, unwind the spiral arms of cluster galaxies. The presence of asymmetries can be taken as a proof of an ongoing RPS event only whenever the 
distribution of the old stellar population is unaffected. This is something difficult to demonstrate even when deep near-IR imaging is available
since red supergiants formed after the interaction can heavily contaminate these photometric bands. 
A less biased tracer is the asymmetry of the stellar and gas radial velocities, made accessible by large FoV IFU instruments. In these cases \citep{Fossati16, Fossati18} the asymmetry of the ionised gas distribution marks a strong contrast with symmetric stellar rotation fields in case of pure RPS, while in the case of gravitational interactions we see clear signs of the perturbation of the stellar rotation field \citep[][see also Sec. \ref{subsec:kinematics}]{Boselli21, Bellhouse21}.

Asymmetries in the different gas phases of the ISM have been systematically searched for and observed in several nearby clusters whenever the angular resolution of the data
was not prohibitive. Evidence of an asymmetric distribution of the \hi\ gas has been first observed by \citet{Warmels88b}, \citet{Cayatte90}, \citet{Chung09}, and \citet{Reynolds20} in the Virgo cluster, by \citet{Scott18} in A1367, and \citet{Bravo-Alfaro00} in Coma. Asymmetries have been also looked for in the molecular gas component
\citep{Scott15}, although systematic studies of complete samples of cluster galaxies are still lacking.

Thanks to its high angular resolution, narrow-band H$\alpha$ imaging data are perfectly suited to trace the distribution of the gas ionised by young ($\lesssim$ 10 Myr)
massive ($\gtrsim$ 10 M$_{\odot}$) stars \citep{Kennicutt98, Boselli09a} formed after the interaction. Given the stochasticity of the star formation process,
which results in a clumpy structure over the entire galactic disc, the interpretation of asymmetries as a product of an external perturbation can be reached only 
on statistical basis when confronted to a reference sample of unperturbed, field objects, or in the study of selected objects in conjunction with other indicators.
Asymmetries in the ionised gas distribution have been searched for using the concentration, asymmetry, and smoothness (CAS) parameters
\citep{Conselice00} in the nearby clusters Virgo, Coma, and A1367 by \citet{Fossati13} and \citet{Boselli15}. When plotted against the HI-deficiency parameter,
often used as a direct tracer of an ongoing or past external perturbation, the CAS parameters are fairly constant in the $r$-band, while the asymmetry parameter $A_{\rm H\alpha}$ and the clumpiness parameter $S_{\rm H\alpha}$
systematically decrease with increasing HI-deficiency. This trend, however, has been interpreted as a second order effect related to the decrease of the specific star
formation activity observed in cluster galaxies rather than an evidence of an ongoing RPS event. The presence of galaxies suffering an ongoing perturbation
in the Coma cluster has been also claimed after the analysis of the structural parameters of galaxies extracted from a set of high-resolution 
broad band optical images by \citet{Roberts20}. The presence of asymmetries has been correlated with their direction and the position of the galaxies 
within the cluster to identify objects undergoing an RPS event. Finally, it is worth mentioning that associated to these asymmetries in the stellar distribution, stars are also less concentrated in the inner regions as indicated by a lower Sercic index compared to unperturbed systems \citep{Roman-Oliveira21}.

\subsection{Integrated properties}
\label{subsec:integrated}

The tight relationship between the radial extent of the different components of the ISM and the total available reservoir of atomic gas shown in Fig. \ref{profili} 
naturally brought to the general assumption that the HI-deficiency parameter can be statistically taken as an indirect tracer of an RPS 
perturbation \citep[e.g.][]{Boselli06}. Since this parameter can be easily derived from single dish observations of unresolved galaxies, available for several thousands of objects 
in the nearby Universe, the HI-deficiency parameter has become one of the most used tracer of RPS in statistical analyses \citep[e.g.][]{Haynes84, Solanes01, Catinella13, Boselli14a, Boselli14b, Gavazzi05, Gavazzi06, Gavazzi13a, Loni21}. 

It has been shown that the HI-deficiency parameter systematically increases towards the centre of rich clusters of galaxies
consistently with the observed decrease of the HI-disc extension gathered from interferometric observations (see Sec. \ref{subsec:truncdiscs}).
The statistical analysis of large samples of galaxies in nearby clusters has indicated that the atomic hydrogen starts to be depleted at clustercentric distances as high as 
$\simeq$ 1-2 $\times$ $r_{200}$ reaching its maximum at the cluster centre where up to $\gtrsim$ 90\%\ of the atomic gas can be stripped \citep{Solanes01, Gavazzi05, Gavazzi06, Gavazzi13a, Healy21, Morokuma-Matsui21}.
Again, this statistical evidence matches the results obtained from the analysis
of representative objects with interferometric data showing the presence of extended \hi\ tails up to similar clustercentric distances (see Sec. \ref{subsec:atomic}).
It has been shown that the fraction of \hi-deficient galaxies increases as a function of the X-rays luminosity of the cluster \citep{Giovanelli85},
as expected considering that the X-ray luminosity is tightly connected to the density of the ICM. This relation, however,
has not been confirmed on a large sample by \citep{Solanes01}. The correlation between the HI-deficiency parameter 
and the X-rays luminosity, however, seems to be valid locally when the hot gas distribution is compared to the position of individual objects as done in the Virgo cluster 
\citep{Solanes01, Boselli14b}. Furthermore, HI-deficient galaxies are preferentially located on radial orbits \citep{Solanes01}.
Using a combination of direct observations and stacking 
techniques, it has been also shown that groups of galaxies are also populated by HI-deficient late-type systems \citep{Fabello12, Catinella13, Hess13, Brown17, Hu21}.
HI-deficiency has been also observed in compact groups by \citet{Verdes-Montenegro01} and interpreted as possibly/partly due to ram pressure in some of these objects by \citet{Rasmussen08}. Similarly, a decrease of the atomic gas content has been observed also in the filaments at the periphery of the Virgo cluster \citep{Castignani22}.
We recall, however, that this last observational evidence cannot be unequivocally interpreted as an evidence of RPS just because 
the atomic gas, being loosely bound to the gravitational potential well of the galaxy, can be removed also by gravitational perturbations.

By comparing the integrated properties of different samples of local galaxies \citet{Fumagalli09} and \citet{Boselli14a} have shown that the
molecular gas content also decreases in HI-deficient galaxies, suggesting that part of the molecular phase can be stripped away along with the atomic gas \cite[see however][]{Mok16, Chung17}.
In some galaxies such as NGC 4569 the stripping process can affect also the dense molecular gas component \citep{Wilson09}.
The stripping of the molecular gas component, which is deeply embedded in the gravitational potential well of the galaxies in large part inside giant molecular clouds
with a low cross section vs. the external pressure, is less efficient than that of the atomic phase, leading to an increase of the H$_2$/\hi\ gas ratio.
These results are in line with the most recent Illustris TNG100 hydrodynamical simulations \citep{Stevens21}, or with tuned simulations of RPS in local galaxies 
\citep[e.g.][]{Tonnesen09, Boselli21}.
Being closely connected to the star formation process, the lack of atomic and molecular gas induces a reduction of the star formation activity \citep[e.g.][]{Koyama17}
which is observed in late-type cluster galaxies at roughly similar clustercentric distances ($R/r_{200}\simeq 1-2 $; \citep{Lewis02, Gomez03, Gavazzi13a}, in groups \citep{Hu21}, or in the filaments surrounding rich structures such as Virgo \citep{Castignani22}. It is worth noticing that in the specific population of jellyfish galaxies (see Sec. \ref{subsec:molecular}) the molecular gas fraction is the dominant cold gas component over the disc of the perturbed systems \citep{Moretti20a}. Here the observed increase in the H$_2$/\hi\ ratio is probably related to the increase of the external pressure which favors the formation of molecular hydrogen \citep{Blitz06}. This molecular gas formation, however, seems to occur only under a specific geometrical configuration, where the interaction with the surrounding ICM is mainly edge-on. In this case, the gas is compressed over the disc of the galaxy on the leading side of the interaction, producing a bow-shocked region where the gas density increases and the star formation can be boosted, as indeed observed in the galaxy IC 3476 in the Virgo cluster \citep{Boselli21}, or predicted by tuned simulations \citep{Safarzadeh19}.

Given that along with the gaseous component also dust and metals can be stripped during the perturbation, and that ram pressure preferentially removes gas located in the outer disc, which is known to be metal poor, the stripping process can alter the typical metallicity and dust-to-gas ratio of the perturbed galaxies. Effects on the gas metallicity of ram pressure stripped galaxies have been first looked for by \citet{Skillman96} in nine Virgo cluster galaxies. Their analysis have shown that HI-deficient galaxies in the core of the cluster have, on average, higher gas metallicities than similar objects in the field. This result has been later confirmed on a large sample of 260 late-type galaxies in nearby clusters by \citet{Hughes13}, and interpreted as due to the removal of the gas poor component located in the outer regions (for a different interpretation see \citet{Peng15} and \citet{Gupta17}). By studying the properties of galaxies extracted from the \textit{Herschel} Reference Survey \citep{Boselli10}, a complete volume-limited, mass-selected 
sample of local galaxies with a complete set of multifrequency data, \citet{Cortese16} have shown that while the $M_{\rm gas}/M_{\rm dust}$ ratio varies by no more than a factor of two when moving from the core of the Virgo cluster to the field, the $M_{\rm HI}/M_{\rm dust}$ ratio increases and the $M_{\rm H_2}/M_{\rm dust}$ decreases with the HI-deficiency. They have explained this opposite behaviour as due to a different radial distribution of the different gas phases and of the dust component whenever the stripping process occurs outside-in, as indeed happens during an RPS event. These results are fairly consistent with the previous work of \citet{Corbelli12}. Evidence for dust stripping in rich environments comes also from the multifrequency study of the cluster A1758 which shows a decrease of the dust-to-stellar mass ratio and a lower level of infrared emission in the star forming galaxies belonging to the cluster with respect to that of similar and coeval objects in the field \citep{Bianconi20}.

\subsection{Galaxy distribution within high-density regions}
\label{subsec:galdist}

The radial decrease of the density of the ICM and that of the velocity of infalling galaxies
suggest that outside a given clustercentric radius, which might depend on the properties of the high-density regions and of the mass of the perturbed system, ram pressure is not sufficient to overcome
the gravitational forces keeping the gas anchored to the gravitational potential well of the galaxy. The observation of galaxies with evident signs of an ongoing interaction, 
such as extended tails of gas in its different phases, or of truncated discs, have been used to physically identify the region where RPS is active. The results of these works are summarised in Sec. \ref{subsec:galdist2}. Thanks to the advent of deep spectroscopic surveys on statistically significant samples we are now able to add the relative velocity of perturbed galaxies as a new parameter to trace their distribution within the gravitational potential well of high-density regions. Positions and velocities have been recently used to trace the distribution of galaxies within the phase-space diagrams. The results of these works,
which can be easily compared to the predictions of simulations, are summarised in Sec. \ref{subsec:phasespace}.

\subsubsection{Galaxy distribution as a function the angular distance}
\label{subsec:galdist2}

%{\color{red}{MS: a lot of description here can be replaced with a phase diagram for the local RPS galaxies, likely size scaled by mass and color-coded by SFR?}}

In the nearby Virgo cluster, galaxies with extended tails of gas undergoing RPS have been observed up to a projected distance of $R/r_{200}\simeq 1 - 1.3 $,
in A1367 and in Coma up to $R/r_{200} \simeq 0.6 $ and $R/r_{200} \simeq 1.4 $, respectively (see Table \ref{tab:RPSgalaxies}), although these numbers are likely lower limits given the limited extension of the blind narrow-band H$\alpha$ imaging survey carried out with Suprime Cam at the Subaru telescope \citep{Yagi10, Yagi17, Gavazzi18a}. Tails of gas have been observed
even at further distances, up to twice $r_{200}$ in A1367, but in objects where ram pressure combined with other mechanisms have been invoked to explain the origin of the 
perturbation \citep{Scott12}. \hi\ interferometric observations of galaxies in the Virgo cluster have also revealed that the truncation of the atomic gas disc decreases with increasing clustercentric 
distance, as expected in an RPS scenario \citep{Cayatte90, Chung09}. Galaxies with \hi\ discs less extended than their stellar counterparts have been observed at distances up to 1.5 $\times$ $r_{200}$, although this radial distance is quite uncertain given the complex structure of the Virgo cluster. On a more solid statistical basis, it is well known that
the HI-deficiency parameter starts to increase at clustercentric projected distances smaller than 1.5 $\times$ $r_{200}$ in Virgo \citep{Gavazzi13a, Boselli14b} 
and 2 $\times$ $r_{200}$ in Coma \citep{Gavazzi06, Boselli06} and in other nearby clusters \citep{Solanes01}. 
%WATCH OUT THAT THE NORMALISATION RADIUS OF SOLANES IS THE ABELL RADIUS... 
The presence of HI-deficient galaxies has been also observed along the filaments surrounding the Virgo cluster region \citep{Castignani22}, where pre-processing might have been acting.
These observational results are corroborated by the predictions of cosmological hydrodynamical simulations now able to resolve the \hi\ gas component of satellite galaxies of stellar mass $M_{\rm star}$ $\sim$ 10$^9$ M$_{\odot}$ \citep{Marasco16}. 
These trends observed in \hi\ are mirrored in H$\alpha$: in the Virgo cluster, for instance, the star formation activity of late-type galaxies starts to decrease at the same clustercentric 
distance where the \hi-deficiency increases ($R$ $\simeq$ 1.5 $\times$ $r_{200}$; \citep{Gavazzi13a, Boselli14b, Castignani22}.
The same decrease of the star formation activity is seen in several other well known nearby clusters \citep{Haines06, Haines15}, and on larger samples in all high density regions in the local Universe within 1-2 $\times$ $r_{200}$ from the central galaxy as derived from the analysis of SDSS \citep{Gomez03}, 2dF \citep{Lewis02}, and \galex{} \citep{Catinella13} surveys. In particular, \citet{Kauffmann04} have shown that the
decrease of star formation activity per unit stellar mass (specific star formation rate) is more pronounced in intermediate mass galaxies with respect to massive systems. This mass segregation effect
in the quenching mechanism typically occurs in nearby massive haloes $M_{\rm halo} \sim$ 10$^{13}$ - 10$^{14}$ M$_{\odot}$ \citep{Catinella13}, 
suggesting a relevant (although maybe not dominant) role of RPS in the quenching of a large population of galaxies orbiting these haloes. It is also worth mentioning that in two clusters at $z \sim 0.25$ the orientation of the ionised gas asymmetries, which are generally pointing away from the cluster centre, has been used to identify RPS as the dominant perturbing mechanism \citep{Liu21}. This observational evidence also indicates that RPS occurs principally during the first passage of an infalling galaxy within the cluster, and that the quenching episode is rapid \citep{Liu21}.

\subsubsection{Galaxy distribution within the phase-space diagram}
\label{subsec:phasespace}

\begin{figure}
\centering
\includegraphics[width=1.0\textwidth, angle=0]{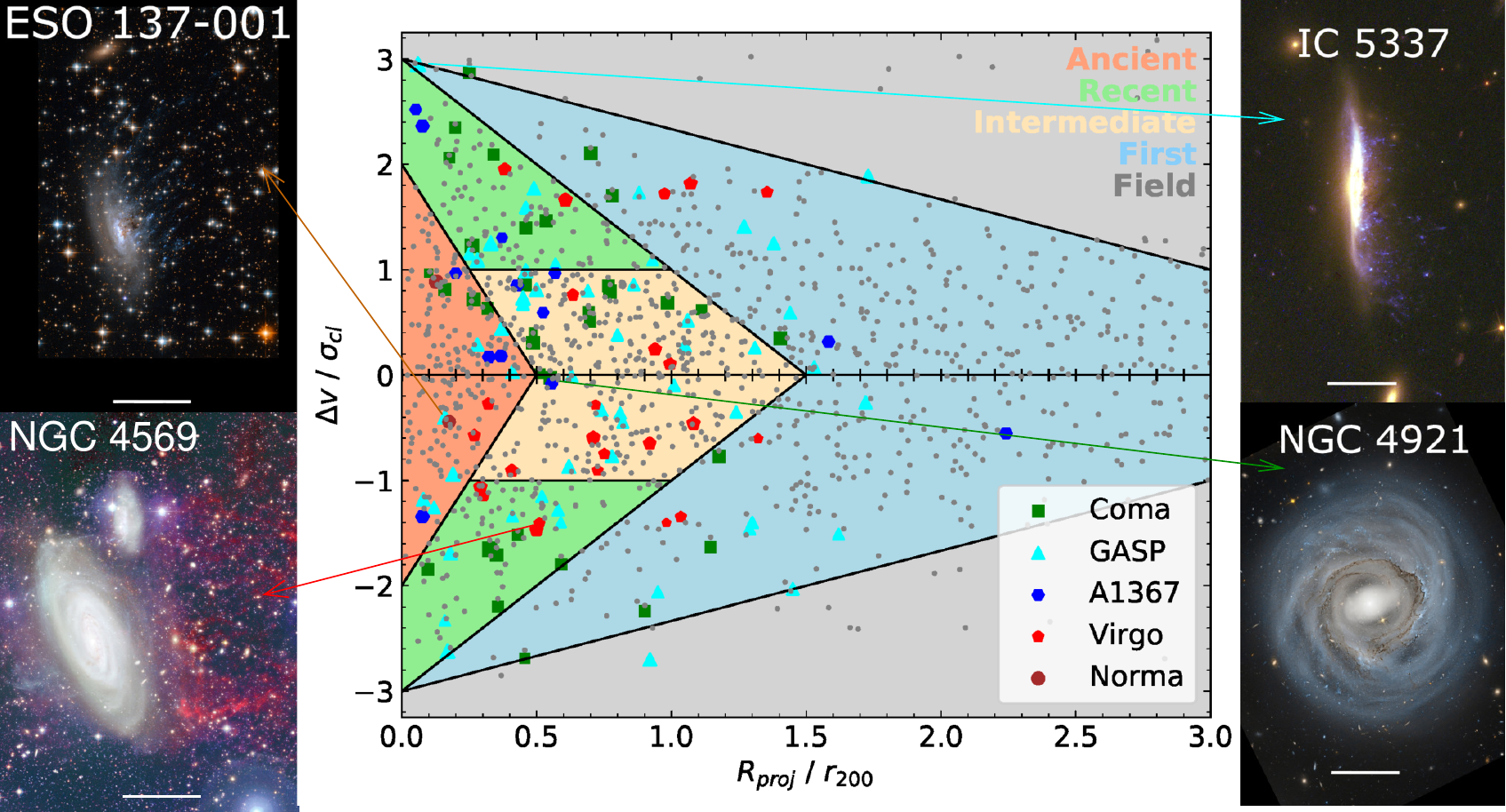}
\caption{Phase-space diagram of the Coma supercluster constructed using the sample of \citep{Gavazzi10} (gray filled dots) and including the galaxies suffering an RPS event in the Virgo (red pentagons), Coma (green squares), A1367 (blue hexagons), and Norma (brown circles) clusters listed in Table \ref{tab:RPSgalaxies}. The cyan triangles indicate the jellyfish galaxies of the GASP  survey extracted from \citet{Gullieuszik20}. The solid lines and the coloured segments delimit the different regions extracted from the simulations of \citet{Rhee17} to identify galaxies in different phases of their infall into the cluster: first (not fallen yet - light blue), recent (0$<$ $\tau_{\rm inf}$ $<$3.6 Gyr - light green), intermediate (3.6$<$ $\tau_{\rm inf}$ $<$6.5 Gyr - orange), and ancient (6.5$<$ $\tau_{\rm inf}$ $<$13.7 Gyr -red) infallers. Optical RGB images of four RPS galaxies at different positions on the phase diagram are also shown (image credits: ESO~137-001 and NGC~4921 --- STScI; NGC~4569 --- \citealt{Boselli16}; IC~5337 --- Sunil Laudari with the \hst{} data from the program 16223, PI: Gullieuszik).
The white scalebar shows 10 kpc.
}
\label{fig:PSDbox}       
\end{figure}

An alternative way of identifying the region of action of a given perturbing mechanism is that of plotting galaxies with evident signs of that ongoing perturbation in the phase-space diagram, i.e.
in a diagram showing the relationship between the line of sight velocity of the galaxy with respect to the mean velocity of the cluster as a function of the projected distance from the cluster centre. An example of phase-space diagram is shown in Fig. \ref{fig:PSDbox} constructed using the sample of galaxies in the Coma supercluster of \citet{Gavazzi10} combined with identified RPS galaxies in the Virgo, Coma, A1367, and Norma clusters given in Table \ref{tab:RPSgalaxies}, and the GASP  sample of jellyfish galaxies. This diagram has the
advantage of putting the studied galaxies extracted from different clusters in a normalized parameter space where both axis are tightly connected to the ram pressure force ($\rho_{\rm ICM}V^2$), the line of sight velocity within the cluster (Y-axis) and the clustercentric distance, which varies as the density of the ICM.  The position of galaxies in this diagram can be used to statistically identify samples at different phases of their journey within the cluster, from their first infall, to their long term presence in the central virialised region, and can easily be compared to the predictions of cosmological simulations in the same observed parameter space \citep{Oman13, Oman16, Sorce16, Sorce19, Sorce21, Rhee17, Arthur19}.

This exercise has been done in several nearby high-density regions 
\citep{Mahajan11, Hernandez-Fernandez14, Haines15, Jaffe15, Jaffe18, Gavazzi18a, Rhee20, Wang21}, including Virgo \citep{Vollmer01, Boselli14b, Yoon17, Morokuma-Matsui21} and Fornax \citep{Loni21}, and in 
high redshift clusters \citep{Muzzin14, Liu21}.  We recall that while early-type galaxies are mainly virialised within the inner regions of high-density environments and have isotropic orbits, spiral galaxies have radial orbits with no significant evolution with redshift \citep{Colless96, Biviano09}. The results of these works do not always agree. In the Virgo cluster, where galaxies suffering ram pressure have been clearly identified thanks to the 
excellent angular resolution of the data, the analysis of the phase space diagram suggests that the hydrodynamic interaction between the hot ICM and the cold ISM acts preferentially on fresh infalling systems 
rapidly stripping the gas and quenching the activity of star formation on short timescales. Here the most perturbed galaxies, which have suffered the external perturbation on longer timescales, 
are located deep within the potential well of the cluster \citep{Vollmer01, Boselli14b, Yoon17}. \citet{Jaffe18} also indicated that most of the jellyfish galaxies identified in the WINGS survey
and targets of the MUSE GASP  survey are located in the infalling regions, and thus have higher peculiar velocities than the other cluster members. 
A similar picture consistent with a rapid quenching of the star formation activity (see Sec. \ref{subsec:timescale})
in the infalling galaxies has been also claimed thanks to the statistical analysis of several nearby clusters by \citet{Mahajan11}, \citet{Hernandez-Fernandez14}, \citet{Jaffe15}, and \citet{Oman16}, again consistent with a dominant
RPS scenario at least within $1-2$ virial radii of massive clusters. 
Longer timescales ($\tau$ $\sim$ 1.7 Gyr), on the contrary, suggesting a smooth decrease of the star formation activity as predicted by a starvation scenario, have been suggested by the analysis of other nearby clusters by \citet{Haines15}. The analysis of $z \sim 1$ clusters extracted from the GCLASS survey by \citet{Muzzin14} indicates a delayed ($t_{\rm d}$ $\sim$ 2 Gyr) then rapid ($\tau$ $\sim$ 0.4 Gyr) quenching of the star formation activity. As we discuss in Sec. \ref{subsec:redshift} the delay time is shorter than the one derived for satellite galaxies in less dense environments (e.g. groups) suggesting some degree of dynamical gas removal in clusters also at these cosmic epochs.

\section{The impact of RPS on galaxy evolution}
\label{sec:impactgal}

\subsection{Induced effects on the star formation process}
\label{subsec:indsfr}

\begin{figure}
\centering
\includegraphics[width=0.75\textwidth]{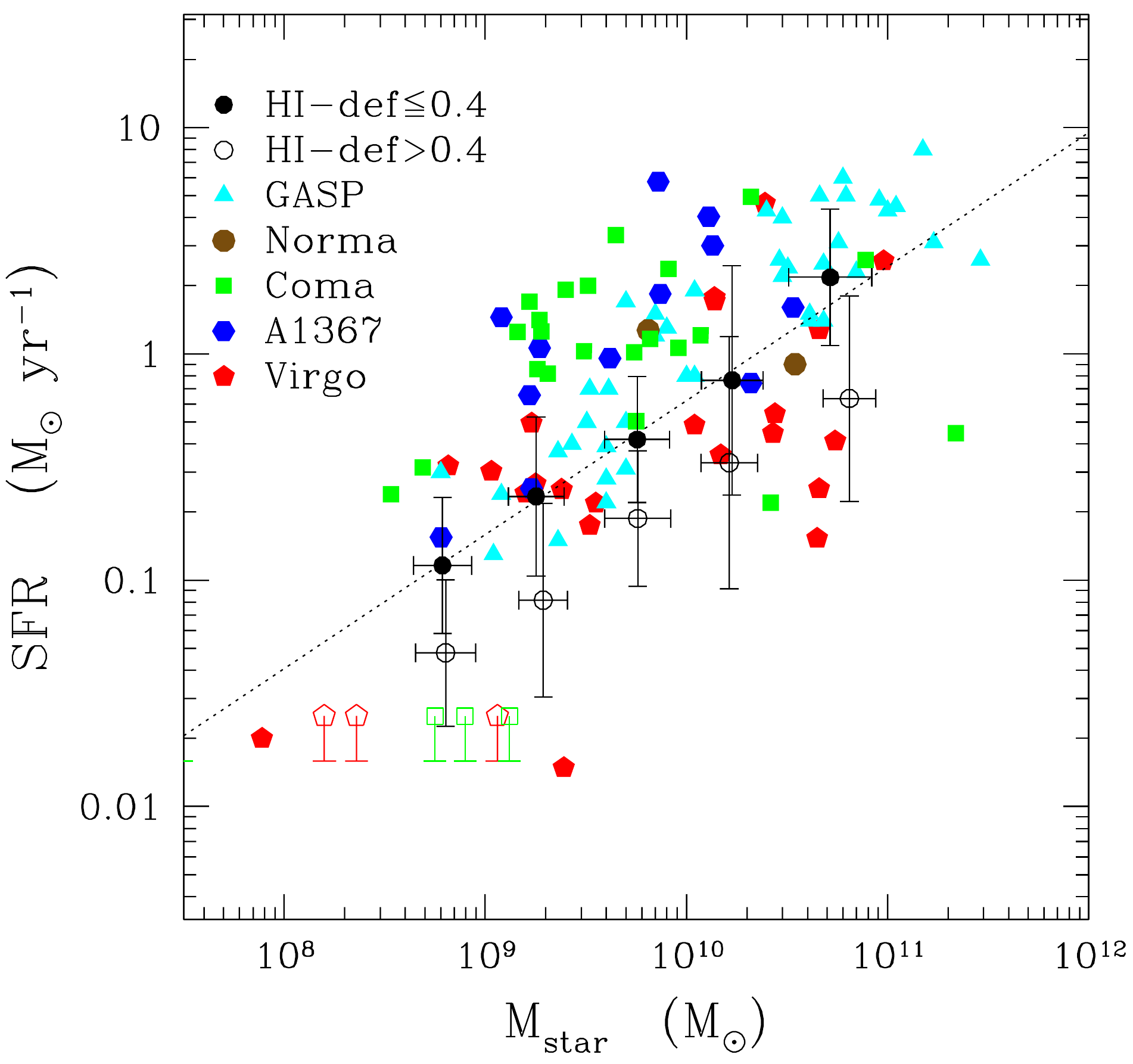}
\caption{Relationship between the star formation rate and the stellar mass (main sequence) for galaxies suffering an RPS event listed in Table \ref{tab:RPSgalaxies} 
or extracted from the GASP sample compared to the mean main sequence relation extracted from the \textit{Herschel} Reference Survey for HI-normal (black filled dots) and HI-deficient (black empty circles) galaxies \citep{Boselli15}. For the GASP sample (cyan filled triangles), data have been extracted from the compilation of \citet{Vulcani18},
while for the Virgo cluster (red pentagons) from \citet{Boselli15} or from the VESTIGE survey. Those for Coma (green squares), A1367 (blue exagons), and Norma (brown circles) have been measured from H$\alpha$ fluxes available in the literature (from dedicated MUSE papers when available, as indicated in Table \ref{tab:RPSgalaxies}, or from \citep{Gavazzi15} and references therein) corrected for dust attenuation and [NII] contamination (whenever necessary) consistently with \citet{Boselli15}. Empty red pentagons and green squares are for those galaxies with clear 
H$\alpha$ extended tails but with no emission in the disc as seen in the deep VESTIGE \citep{Boselli18} or Subaru \citep{Yagi10} narrow-band H$\alpha$ images. The black dotted line gives the best fit to the \textit{Herschel} Reference Survey for HI-normal galaxies (unperturbed reference sample). Stellar masses and star formation rates have been estimated assuming a Chabrier IMF. Adapted from Boselli et al. (in prep.).}
\label{main}       % Give a unique label
\end{figure}

The suppression of gas removed during the interaction has heavy consequences on the future evolution of the perturbed galaxies.
The main gas component removed during a RPS event is the diffuse atomic gas reservoir distributed on an extended disc. The one located within the stellar disc, which is the principal supplier of giant molecular clouds where star formation takes place, is only moderately or indirectly affected by the interaction\footnote{Infall of the \hi\ gas located in the outer regions to the inner disc has been invoked to explain the observed strong correlation between the total gas content (\hi\ plus H$_2$) of isolated 
late-type galaxies, a large fraction of which is located outside the stellar disc, and their overall star formation activity \citep{Boselli01}. The supply of this gas component to the inner star forming regions, however, would occur on timescales relatively long compared to the typical ones for gas stripping (see Sec. \ref{subsec:timescale}).}. Whenever this occurs, the decrease of the atomic and molecular gas components in the inner regions, (the former partly swept away during the interaction, the latter gradually exhausted by the lack of feeding from the atomic gas reservoir), induce a decrease of the star formation activity, as indeed observed in the large majority of the HI-deficient galaxies in nearby clusters \citep[e.g.][]{Kennicutt83, Gavazzi91, Gavazzi98, Gavazzi02, Gavazzi06a, Gavazzi13a, Donas90, Donas95, Moss93, Haines07, Haines09, Vulcani10, Boselli14b, Cybulski14} and groups \citep{Hu21}. 
Within the Virgo cluster, the radial dependence of the specific star formation activity with clustercentric distance perfectly matches that of the increase of the HI-deficiency parameter, which starts to systematically change from that of unperturbed field objects at projected $R/r_{200} \lesssim 1-2$ \citep{Gavazzi13a}. We recall that this clustercentric distance measured in Virgo also corresponds to that of the decrease of the star formation activity observed in large statistical samples extracted from the 2dF \citep{Lewis02} and SDSS \citep{Gomez03} surveys, as described in Sec. \ref{subsec:galdist}.
The overall decrease of the star formation activity put HI-deficient galaxies below the main sequence relation traced by unperturbed field objects (see Fig. \ref{main}; \citep{Boselli15, Boselli16a, Ciesla16}, or in the green valley when colour indices including a UV band, sensitive to the youngest stellar populations, are used in the definition of the colour-magnitude or colour-stellar mass relation \citep[e.g.][]{Hughes09, Cortese09, Boselli14a, Boselli14b}. In the Virgo cluster, the spatial distribution of HI-deficient late-type galaxies with reduced star formation activity (red $NUV-i$ colours) seems fairly correlated with the distribution of the hot gas emitting in X-rays \citep{Cayatte90, Chung09, Boselli14b}.

As described in Sec. \ref{subsec:truncdiscs}, ram pressure removes the gas outside-in, forming truncated gas discs
whose radial extension decreases at the increasing of the HI-deficiency parameter, thus in those galaxies located closer to the cluster core. 
%As a consequence, the quenching of the star formation activity occurs outside-in, as clearly seen in the prototypical galaxy NGC 4569, a giant late-type system undergoing an RPS event as indicated by its extended cometary tails of ionised gas \citep[e.g.][]{Vollmer04a, Boselli06a, Boselli16}. Here, all the components of the ISM (\hi, H$_2$, dust) are removed outside $\sim$ 0.5 $r_{25}$ ($r_{25}$ being the isophotal radius), where the star formation activity is totally quenched as deduced from the lack of any ionised gas (see Fig. \ref{N4569trunc} in Sec. \ref{subsec:truncdiscs}). 
As a consequence, the star formation activity of the perturbed galaxies mainly decreases in the outer disc, but there are some cases where it is also reduced in the inner disc as deduced by a lower H$\alpha$ surface brightness with respect to similar objects in the field \citep{Koopmann04}. The decrease of the star formation activity might be drastic in dwarf systems, where the shallow gravitational potential well is not sufficient to keep the cold gas anchored to the stellar disc. Gas rich, low mass systems can thus be transformed in quiescent dwarf ellipticals in relatively short timescales \citep{Haines07, Boselli08, Boselli08a, Boselli14c, Toloba09, Toloba11, Toloba12, Toloba15, Smith12, Mahajan11a, Janz21, Junais21}. In a few objects, some ISM is retained only in the nucleus, where the gravitational
potential well is at its maximum \citep{Boselli08, De-Looze10, Boselli14b}. Here star formation can continue at small rates \citep{Boselli08} and possible be at the origin of the blue nuclei observed in the most massive dwarf ellipticals within the Virgo cluster \citep{Lisker06}.

Under some conditions, however, the RPS process can boost the overall activity of star formation and bring galaxies above the main sequence, as indeed suggested by several simulations \citep{Fujita99, Bekki03, Bekki14, Steinhauser12, Steinhauser16, Henderson16, Steyrleithner20, Lee20, Troncoso-Iribarren20} (see Fig. \ref{main}). A systematic enhancement of the star formation activity of local cluster galaxies has been looked for using data gathered during several blind surveys, but never convincingly shown so far \citep{Donas90, Donas95, Moss93, Mahajan12}. There are, however, clear examples of galaxies with an enhanced activity such as the jellyfish galaxies observed during the GASP survey \citep{Vulcani18} (cyan filled triangles in Fig \ref{main}). Other examples are the blue cluster galaxies discovered by \citet{Bothun86}, whose spectroscopic properties (presence of strong emission and absorption Balmer lines) suggest a recent enhanced star formation activity (some of these are indicated as green filled squares in Fig. \ref{main}). 
%These actively star forming galaxies could be the analogues of the blue systems dominating in intermediate redshift ($z$ $\sim$ 0.5) clusters (Butcher-Oemler effect, \citealt{Butcher84}), although some of them are merging systems progenitors of local lenticulars \citep{Dressler97, Poggianti99}.  
Representative RPS galaxies with enhanced star formation activities are also UGC 6697, CGCG 097-073 and CGCG 097-079 at the periphery of the cluster A1367 \citep{Gavazzi95, Gavazzi01, Boselli94a, Scott15, Consolandi17} (blue filled pentagons in Fig \ref{main}). On the leading-edge opposite to the observed radio continuum tail these galaxies have a bow-shocked structure of star forming complexes probably formed after the interaction. Another spectacular case is IC 3476 in the Virgo cluster \citep{Boselli21}. Despite a marked HI-deficiency ($HI-def$ = 0.67), the galaxy is located in the upper envelop of the main sequence drown from unperturbed isolated systems (see Fig. \ref{main}). The exquisite quality of the H$\alpha$ narrow-band imaging gathered during the VESTIGE survey indicates that giant \hii\ regions are formed on the leading side of the disc, at the interface between the stripped ISM and the surrounding ICM. Here, the observed increase of the star formation seems due to the compression of the gas which induces an increase of its mean electron density \citep{Boselli21} and probably of the molecular gas phase, as indeed observed in a few other Virgo galaxies \citep{Nehlig16, Lizee21}. Tuned simulations suggest that the overall increase of the star formation activity of these objects is mainly localised in this front region, but also show that any bursting phase generally lasts only a few tens of million years \citep{Weinberg14, Troncoso-Iribarren20}. The origin of this burst, as those observed in the galaxies mentioned above, can be due to the inclination angle of the infalling galaxies, all entering the cluster mainly edge-on, as indeed indicated by tuned simulations \citep{Safarzadeh19}. In this particular configuration, most of the stripped gas has to cross the stellar disc before leaving the galaxy, creating turbulences and instabilities which favour gas collapse and star formation \citep{Lee20, Boselli21}. 
The perturbed galaxies with an enhanced star formation activity located above the main sequence might thus be perturbed systems representative of a particular short living phase of the ram pressure stripped population, while the quiescent population associated to the gas deficiency the following long living phase, as suggested by models \citep{Boselli06a, Boselli08, Boselli14b}. 
The excellent quality of the VESTIGE data in terms of angular resolution ($\simeq$ 50 pc) and sensitivity ($L({\rm H\alpha})$ $\sim$ 10$^{36}$ erg s$^{-1}$) also allowed to extend the study the effects of RPS down to the scale of individual \hii\ regions \citep{Boselli20}. This analysis has shown that the HI-deficient galaxies have, on average a lower number of \hii\ regions with respect to similar gas-rich objects. It has also shown that \hii\ regions are mainly lacking in the outer disc, thus consistent with an outside-in truncation of the star formation activity as expected in an RPS scenario \citep{Koopmann04a, Boselli20}.

\subsection{Timescales for stripping and quenching}
\label{subsec:timescale}

The different perturbing mechanisms affecting galaxies in high density environments act on different timescales for the gas removal, and thus have different effects on the star formation activity of the perturbed objects. 
The typical timescale for RPS to be efficient has been derived using semi-analytical models and full or zoomed-in hydrodynamic cosmological simulations of rich clusters of galaxies \citep{Abadi99, Balogh00, Quilis00, Mori00, Schulz01, Marcolini03, Tonnesen07, Roediger07, Marasco16, Trayford16, Lotz19, Pallero22} or tuned simulations of selected objects with multifrequency data consistently indicating that they are undergoing a stripping event \citep{Vollmer03, Vollmer03a, Vollmer04a, Vollmer05, Vollmer06, Vollmer08, Vollmer08a, Vollmer09, Vollmer12, Vollmer21, Boselli21}.
3-D hydrodynamic simulations suggest that a large fraction of the total gas reservoir \citep{Abadi99, Quilis00, Schulz01, Tonnesen07, Roediger07, Lotz19} can be efficiently stripped on relatively short timescales ($\lesssim$ 1 Gyr) compared to the typical crossing time within a cluster ($\sim$ 2 Gyr; see however \citet{Bahe15}). This is particularly true in dwarf systems, where the shallow gravitational potential well hardly keeps the gas anchored to the stellar disc \citep{Mori00, Marcolini03}. For these low mass objects the typical timescale for gas stripping can be of $\sim$ 100-200 Myr. Simulations based on a single and homogeneous gas phase such as these, however, generally underestimate the efficiency of RPS, suggesting that these timescales could even be overestimated \citep{Tonnesen09}. 

Observationally, these timescales can be estimated by reconstructing the star formation history of the perturbed galaxies through the analysis of their stellar population (quenching episode). Statistically, this has been done by comparing the star formation history of cluster galaxies (reconstructed from multifrequency datasets) with that predicted by cosmological simulations or semi-analytic models of galaxy evolution. The outside-in truncation of the gaseous disc of RPS galaxies gradually quenches the activity of star formation in the outer regions, inverting the typical colour gradients of spiral systems which generally gets bluer in the outer disc. The UV photometric bands, most sensitive to the youngest stellar populations, are those which become redder first after the stripping of the gas has been completed \citep{Crowl06}. The comparison of the light profiles in different photometric bands with the predictions of spatially resolved models of gas stripping can thus be used to estimate the typical timescale for the quenching of the star formation activity. This experiment has been successfully done for the first time on the spiral galaxy NGC 4569 in Virgo \citep{Boselli06a}, and led to the conclusion that the cold gas content and the star formation activity of this ram pressure stripped galaxy, as indicated by the presence of a long cometary tail of ionised gas \citep{Boselli16}, have been reduced by $\sim$ 95\%\ in about 100 Myr \citep{Boselli06a}. This short timescale is consistent with the one derived by \citet{Vollmer04a} ($\sim$ 300 Myr) by comparing the kinematical properties of the perturbed gas with those derived from tuned hydrodynamic simulations. Since then, several works have tried to estimate the typical timescale for star formation quenching in representative RPS galaxies \citep{Pappalardo10, Abramson11, Fossati18, Boselli18a, Boselli18b, Boselli21, Vollmer18, Vulcani20} or in larger statistical samples \citep{Crowl08, Boselli08, Boselli14b, Boselli16a, Smith12, Wetzel13, Haines15, Ciesla16, Owers19, Gallazzi21}. The most recent used SED fitting techniques on FUV-to-FIR photometric data with the purpose of removing any possible degeneracy related to the reddening of the stellar population, which might result from either an ageing of the stellar population or an increasing dust attenuation \citep{Ciesla16, Boselli16a}. The most recent SED fitting codes are now able to combine photometric with spectroscopic data, these last important since they include age-sensitive Balmer absorption lines \citep{Poggianti97}. Whenever the star formation quenching episode needs to be dated with a time resolution $<$ 100 Myr, the use of the Balmer emission lines such as H$\alpha$ ($\lambda$ 6563 \AA) or H$\beta$ ($\lambda$ 4861 \AA) becomes critical \citep{Boselli16a, Fossati18}, if combined with UV photometric bands. Indeed, these lines are due to the emission of the gas ionised in \hii\ regions by O-early B stars, whose typical age is $\lesssim$ 10 Myr \citep{Kennicutt98, Boselli09a}. The short quenching timescales derived from the analysis of the stellar population is consistent with the observed orientation of the ionised gas tails of perturbed galaxies in local and $z$ $\sim$ 0.25 clusters, which are mainly pointing away from the cluster centres, suggesting a rapid stripping and quenching process occurring during the first infall within the cluster potential \citep{Yagi10, Yagi17, Gavazzi18a, Liu21}. 

\citet{Donnari21} using data from the IllustrisTNG simulations revealed a great diversity of quenching pathways for satellite galaxies found in $z=0$ clusters. Satellites can quench before infall in dense environments or after being accreted into groups (pre-processing) or in the host where they currently reside. The frequency of each quenching track depends on the satellite stellar mass and the host halo mass. As a result, the timescales derived using SED fitting analysis of a large sample of galaxies in the Virgo cluster, which indicate a rapid evolution, suggest that RPS is the dominant perturbing mechanism in clusters \citep{Boselli16a, Oman16} but its effects become less prominent when larger populations of satellites are considered. A consistent result is given by the analysis of a sample of 11 massive early-type galaxies in the Coma cluster with high signal-to-noise spectra carried on by \citet{Upadhyay21}. By studying the variation of their star formation activity as a function of their orbital parameters as derived from tuned simulations, the authors concluded that in these objects the activity of star formation ceased $\sim$ 1 Gyr after their first pericentre passage. They ascribed this rapid quenching to RPS possibly combined with tidal interactions.

\subsection{Nuclear activity}
\label{subsec:nuclear}

The recent discovery by the GASP survey that a large fraction of jellyfish galaxies host an AGN \citep[optical definition based on emission line diagnostics][]{Poggianti17a} raised a new interesting question on the interplay between RPS and the triggering of nuclear activity. In this scenario, the AGN feedback could efficiently combine with ram pressure to expel the gas from the galaxy disc and quench the activity of star formation \citep{George19, Radovich19}. Theoretical considerations and simulations consistently suggest that during the stripping process the external perturbation can induce a loss of angular momentum of the gas in the disc that could fall in the inner regions and feed the nuclear activity \citep{Schulz01, Tonnesen09, Ramos-Martinez18}. 
Other cosmological hydrodynamical simulations also show no signs of enhanced AGN activity in massive ($M_{\rm star}$ $>$ 10$^{10.7}$ M$_{\odot}$) galaxies infalling in high mass clusters, where the activity of star formation is also significantly reduced during an RPS event \citep{Lotz21}.

\begin{figure}
    \centering
    \includegraphics[width=0.95\columnwidth]{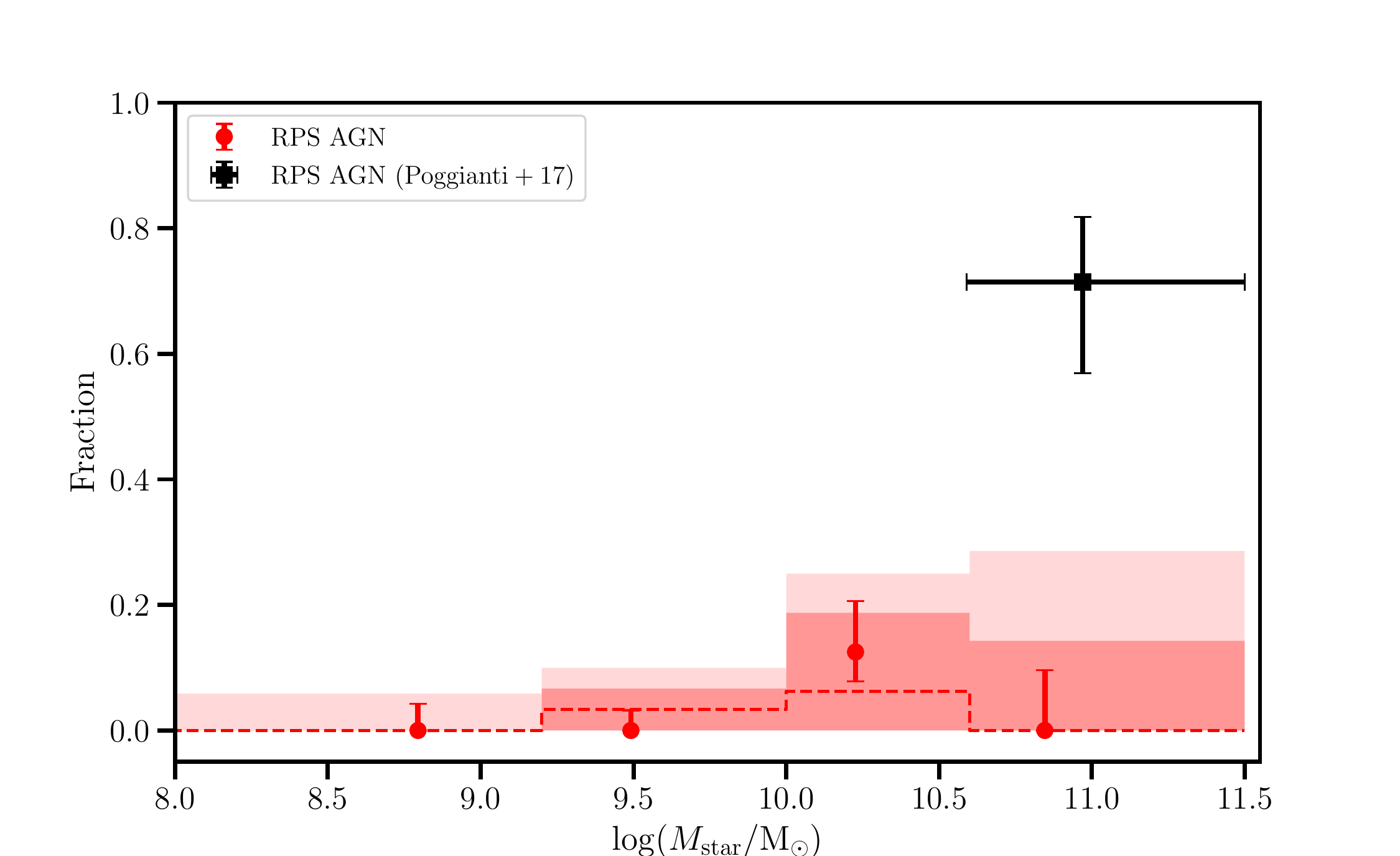}
    \caption{Fraction of RPS galaxies from Table \ref{tab:RPSgalaxies} with an AGN. The fraction of AGN in the RPS sample from \citet{Poggianti17a} is also shown. The dashed lines show the median fraction of galaxies with AGNs from a bootstrap mass matched sample of a local galaxy catalogue (G.Gavazzi private comm.). The dark and light shaded areas show the boundaries of the 1$\sigma$ and 2$\sigma$ confidence levels, respectively. }
    \label{fig:rpsfraction}
\end{figure}

In \citet{Poggianti17a}, the evidence of an increase of the fraction of galaxies hosting an AGN in jellyfish galaxies is based on a very limited sample of galaxies (5 objects out of 7 observed) and thus might be uncertain due to the low number statistics. 
\citet{Peluso21} extended this analysis in sample size and stellar mass range finding that the AGN fraction drops to 24\% if a sample of 82 objects at $M_{\rm star} > 10^9 \rm M_\odot$ is considered while it is 80\% in a sample of 15 high mass galaxies ($M_{\rm star} > 10^{10.5} \rm M_\odot$). These authors, however, include as AGNs also LINERs, which are known not to always correspond to accreting super massive black holes in the galaxy centres \citep[see e.g.][]{Belfiore16}.
However, several observational results do not seem to confirm this result. The analysis of the cluster A901/2 at $z$=0.165 carried on with H$\alpha$ GTC and ACS HST imaging by \citet{Roman-Oliveira19} has clearly shown that, among the 70 jellyfish candidates with a stellar mass 10$^9$ $\leq$ $M_{\rm star}$ $\leq$ 10$^{11.5}$ M$_{\odot}$, the majority are star forming systems (53) and only 5 probably host an AGN (defined using optical line diagnostics). In these five AGN candidates the nuclear activity does not seem to be correlated with the presence of extended tails. In local clusters such as Coma, A1367, and Virgo, where the large sets of multifrequency data acquired during several complete, blind surveys allowed the identification of relatively complete sets of galaxies currently undergoing an RPS event. Indeed, as summarised in Table \ref{tab:RPSgalaxies}, the fraction of late-type systems suffering an RPS event and at the same time hosting an AGN is 3\% (2/70) and increases to 14\% (10/70) when LINERs are included. The BPT classification adopted is the same used in \citet{Poggianti17a} and is based on the \citet{Kewley01} and \citet{Kauffmann03} thresholds to divide star formation from AGN ionisation. The sample from Table \ref{tab:RPSgalaxies} is divided into four stellar mass bins with the highest mass bin corresponding to the range probed by \citet{Poggianti17a}, and the AGN fraction is shown by the red points with Binomial uncertainties in Figure \ref{fig:rpsfraction}. To accurately compare to the general galaxy population we randomly select a mass matched sample from a complete local Universe catalogue (G.Gavazzi private comm.), with RPS galaxies removed.  Bootstrap resampling statistics shows that across all the four stellar mass bins there is no excess of AGNs in RPS galaxies compared to the general galaxy population.

An increase of the AGN fraction in ram pressure stripped galaxies seems also at odd with the analysis of large statistical samples extracted from the SDSS survey which rather indicate a decrease of the AGN fraction with galaxy density \citep{Kauffmann04} and that while Seyfert galaxies are less clustered than normal galaxies, LINERs  follow a similar distribution \citep{Constantin06}. More recent works aimed at studying the incidence of AGNs in different environments, despite the uncertainty introduced by different selection criteria (optical emission lines, radio emission, or X-rays data) or by a different definition of the environment, appear to agree that the fraction of AGNs decreases towards the inner parts of high density regions, from clusters to groups \citep[AGN selected from optical line ratios, and excluding LINERs][]{Sabater13, Sabater15, Lopes17, Li19}, \citep[X-rays selected AGNs][]{Ehlert14, Ehlert15,Koulouridis18, Mishra20}.
The observed decrease of the AGN fraction in star forming systems towards the centre of rich clusters of galaxies is interpreted as due to a lack of cold gas available to feed the nuclear activity \citep{Sabater13}. A statistical reduction of the cold gas content in samples of these objects is likely due to RPS, leading to the conclusion that ram pressure generally inhibits rather than increases the AGN activity. 
A possible increase of the AGN activity is present in the infalling regions of low mass clusters and massive groups \citep[spectroscopically selected AGN,][]{Pimbblet13, Gordon18};  \citep[X-rays selected but spectroscopically confirmed AGN,][]{Koulouridis18}, or in low mass groups \citep[spectroscopically selected AGN,][]{Manzer14}.
Such an increase is rather interpreted as due to the gravitational perturbations acting on the infalling systems (pre-processing). The relative close position within infalling groups, combined with their low velocity dispersion, give enough time to the infalling galaxies to be tidally perturbed, fuelling the gas towards the inner regions, and thus feeding the nuclear activity \citep{Manzer14, Gordon18, Koulouridis18}.

How can the GASP results be then explained? The first possible interpretation is that the jellyfish galaxies observed by the GASP survey are only a small fraction of the galaxies undergoing an RPS event, possibly those where the interaction with the surrounding ICM is mainly edge-on. In this geometrical configuration, the gas located in the outer regions is pushed along the disc before being totally stripped from the parent galaxy \citep{Consolandi17, Boselli21}. It can thus be partly accreted by the nucleus. 
A second possible interpretation is that the peculiar morphology of these jellyfish galaxies is not (only) due to an RPS event, but also to gravitational perturbations (see Sec. \ref{subsec:stars}), which could create strong instabilities able to fuel the cold gas of the disc towards the centre \citep[e.g.][]{Lake98}. 

An interesting related question is the possible contribution of the nuclear activity via feedback to the quenching episode that follows the stripping of the gas. A detailed analysis of NGC 4569, a massive late-type galaxy in the Virgo cluster hosting a nuclear starburst or a weak AGN as derived from X-rays observations, and optically classified as a LINER, revealed that the quantity of gas expelled by the nuclear outflow is only a very small fraction of that removed by the RPS phenomenon \citep{Boselli16}. On the contrary, the lack of molecular gas observed in the central regions of the massive jellyfish galaxy JO201 suggests that the suppression of the star formation activity in these regions could mainly be driven by the feedback from a relatively strong AGN \citep{George19}. 

\subsection{Gas and stellar kinematics}
\label{subsec:kinematics}

RPS acts on the gaseous component of the late-type galaxies infalling for the first time into rich clusters. It is thus conceivable that, along with its distribution, the kinematical properties of the gas are also affected during the perturbation. This is indeed predicted by hydrodynamic simulations of galaxies suffering an RPS event \citep{Schulz01, Hidaka02, Kronberger08a, Smith12a, Haan14}. The interaction is expected to displace the kinematical centre of the rotating gas with respect to that of the unperturbed stars, change the shape of its rotation curve and modify the velocity field mainly in the outer regions where the gas is only weakly bound to the gravitational potential well of the galaxy \citep{Kronberger08a, Bellhouse21}, while producing on long terms warps \citep{Haan14}. The degree of the perturbations and the effects on the rotation curve depend on the impact angle and are more important in nearly edge-on interactions, but are generally less pronounced than in those due to tidal interactions \citep{Kronberger08a}.
The first systematic study of the kinematical properties of the ionised gas of cluster perturbed galaxies is the work of \citet{Rubin88} suggesting modified rotation curves in RPS galaxies. Since then, a major improvement in the 
observational data has been gathered thanks to IFU spectroscopy able to probe at the same time the gas and stellar 2D-kinematics whenever the stellar continuum is sampled.
Optical IFU spectroscopy, which now has the required sensitivity, angular, and spectral resolution to reconstruct the detailed velocity field of nearby perturbed galaxies, confirmed this picture. 
The first clear evidence of gas perturbed velocity fields in a statistically representative sample of ram pressure stripped galaxies comes from the Fabry-Perot spectroscopic observations of a large sample of late-type galaxies in the Virgo cluster \citep{Chemin06}. This work confirmed previous claims of a modified velocity field discovered using radio interferometry in NGC 4522 \citep{Kenney04} and later 
detected in another Virgo cluster object NGC 4330 still using \hi\ data \citep{Abramson11}. Systematic differences between the gas and stellar velocity field in a galaxy suffering an RPS event in the cluster A3558 in the Shapley supercluster have been shown by \citet{Merluzzi13}. This object has a complex gas velocity field, with the extraplanar gas showing signs of rotation up to 13 kpc from the stellar disc (projected distance), while the stellar velocity field is uniform and symmetric. Another interesting example of galaxy suffering an RPS event with a peculiar kinematics is an object in Abell 2670. This object, which has an elliptical morphology, is characterised by an extended ionised gas tail with a significant rotation, without a stellar component \citep{Sheen17}. The IFU observations of ESO 137-001 \citep{Fumagalli14}, UGC 6697 \citep{Consolandi17}, and NGC 4330 \citep{Sardaneta22}, suggest that the gas conserves its rotation close the galactic disc. The rotation velocity decreases with increasing distance from the galaxy disc where the gas becomes more turbulent. 
Further evidence for a perturbed velocity field of the gaseous component comes from the MUSE and Fabry-Perot observations of the galaxy IC 3476 in the nearby Virgo cluster, where the angular resolution of the instruments and the low distance of the cluster allow to resolve the different components down to scales of $\sim$ 100 pc \citep{Boselli21}. The perturbation, which here is acting almost edge-on ($\simeq$ 70 deg. from face-on), has displaced the gas kinematical centre from that of the unperturbed stars by $\sim$ 500 pc along the direction of the motion of the galaxy within the cluster on timescales of the order of 50 Myr. It has also changed the mean velocity of the gas with respect to that of the stars in each position along the disc by $\sim$ 20 km s$^{-1}$. All this observational evidence has been reproduced by tuned hydrodynamic simulations for a galaxy with similar properties undergoing an RPS event \citep{Boselli21}. Another interesting case is the edge-on late-type galaxy NGC 4330 in the Virgo cluster, where the stripping process, which acts mainly face-on, allows us to study the effects of the perturbation along the $z$-axis. Fabry-Perot observations have shown a decrease of the rotational velocity of the stripped gas with increasing distance from the disc plane, suggesting a gradual but not linear loss of angular momentum \citep{Sardaneta22}.  It is also worth mentioning that tuned simulations of RPS events indicates that the velocity of the stripped gas can be lowered close to the stellar disc and increased far away in the presence of magnetic fields \citep{Tonnesen14}.
By reconstructing the kinematical properties of the gas in perturbed systems using tuned N-body simulations, Vollmer and collaborators were able to estimate the epoch of the first interaction of these objects with the surrounding ICM and thus pose strong constraints on the time required by an RPS event to remove the cold gas component \citep{Vollmer99, Vollmer00, Vollmer01a, Vollmer04a, Vollmer05, Vollmer06, Vollmer08, Vollmer08a, Vollmer09, Vollmer12, Vollmer18, Vollmer21}. These works also showed for the first time that part of the stripped gas can fall back on the stellar disc once the galaxy has passed the peak of the interaction \citep{Vollmer01, Steyrleithner20}.

Other simulations, however, indicate that while ram pressure pushes the gas outside the stellar disc, the drag force exerted by this gas on the dark matter component is able to modify the gravitational potential well, moving the position of the dynamical centre of the perturbed galaxy by several kpc \citep{Smith12a}. For this reason, also the stellar velocity field would be modified during the interaction. For the same reason, the perturbation would also thicken the stellar disc by a factor of $\sim$ 2 mainly in the outer regions where most of the gas can be stripped, consistently with simulations \citep{Farouki80, Clarke17, Safarzadeh17, Steyrleithner20}. A boxy shape in the outer disc probably resulting from the induced perturbation in the stellar orbits has been observed in the edge-on spiral galaxy NGC 4330 in the Virgo cluster which is suffering a RPS event as suggested by the presence of extended tails of ionised gas \citep{Sardaneta22}. Using different arguments, a thickening of the stellar disc has been also invoked by \citet{Boselli08, Boselli08a} to explain the formation of dwarf ellipticals in clusters. Indeed, the lack of cold gas which in dwarfs is rapidly stripped during the interaction \citep{Mori00, Marcolini03} prevents the formation of new stars on the stellar disc. Since young stars are formed within a thin layer of cold gas, they have a low velocity dispersion. The lack of this newly formed stellar population induces an instability in the stellar disc which heats up, dumping spiral waves on a few revolutions \citep{Sellwood84, Fuchs98, Elmegreen02, Bekki02}. This effect is expected to be more important in dwarf systems, where most of the gas can be easily removed during the interaction and where the velocity rotation of the disc is relatively low. Indeed, in low-luminosity systems the scale-height of the disc increases by a factor of $\sim$ 2 in $\sim$ 3 Gyr \citep{Seth05}. This mechanism could explain the properties of dwarf ellipticals in clusters if they were formed after an RPS event able to remove on short timescales their gas content from the disc. In this picture, the flattening of their disc could be anti-correlated to the age of the perturbation. It is worth mentioning that this process would increase the velocity dispersion of the stellar component while conserving its angular momentum \citep{Steyrleithner20}. This picture is also consistent with the evidence that several dE galaxies in the Virgo cluster are rotating systems, and that their $v$/$\sigma$ ratio or $\lambda_{\rm R}$ is higher in those systems located at the periphery of the cluster and characterised by a young stellar population \citep{Toloba09, Toloba11, Toloba12, Toloba15, Boselli14b, Bidaran20}. This result is in agreement with the evidence that satellite galaxies have, at fixed stellar mass and ellipticity, lower values of $v$/$\sigma$ than similar main sequence objects in the field \citep{Cortese19}. This observed decrease of the spin parameter of satellites is mainly due to satellite galaxies which have already reached the red sequence.

\section{The impact of RPS on the surrounding environment}
\label{sec:impactenv}

\subsection{The fate of the stripped gas}
\label{subsec:fate}

As soon as a galaxy becomes satellite of a larger structure, its halo, which is principally composed of hot gas, gets mixed with the hot gas trapped within the gravitational potential well of the central system. The hot halo of the galaxy is not available any more for supplying gas to the disc, reducing at long timescales the activity of star formation in the satellite. This mild phenomenon is generally called starvation or strangulation \citep[e.g.][]{Larson80, Bekki02, van-den-Bosch08}. Ram pressure, which is a more violent process, adds up to this effects and removes the cold gas from the disc. As mentioned in Sec. \ref{subsec:mainmech}, the gas phase which is most efficiently removed is the atomic one, since located on a disc extending far out from the stellar disc, where the gravitational potential well of the galaxy is shallow. It is thus interesting to understand what is the fate of the stripped gas.

We can first notice that the perturbed gas can definitely leave the galaxy only if it overcomes the gravitational forces keeping it bound to the disc, i.e. if the velocity of the gas overcomes the escape velocity from the galaxy, which for a spherically symmetric solid body is given by the relation:

\begin{equation}
v_{\rm e} = \sqrt\frac{2GM_{\rm DM}}{r_{\rm gas}}    
\end{equation}

\noindent
where $M_{\rm DM}$ is the total mass of dark matter of the perturbed galaxy and $r_{\rm gas}$ is the distance of the gas from the galaxy. Whenever this condition is not satisfied, once the galaxy gets to the apocentre of its orbit, where the external pressure on its ISM is significantly reduced, the stripped gas can fall back on the disc \citep[e.g.][]{Vollmer01, Roediger05, Boselli18a, Steyrleithner20, Cramer21}.
If removed from the galaxy, the gas gets mixed with the surrounding hot gas, where it can be heated and change phase, than becoming part of the ICM. The change of phase in the stripped gas is now commonly observed in ram pressure stripped tails, as mentioned in Sec.~\ref{subsec:ionised}. Indeed these tails have been seen in cold atomic gas, in molecular gas through different molecular transitions (see Sec.~\ref{subsec:molecular}),
witnessing thus a different gas density and temperature, in ionised and hot gas, this last with temperatures generally lower than those typically observed in the surrounding ICM (see Sec.~\ref{subsec:hotgas}). 

Different mechanisms have been proposed to explain the change of phase of the stripped gas. These includes shocks in the instabilities at the interface of the cold ISM and the hot ICM formed during the journey of the galaxy within the cluster
\citep[e.g.][]{Nulsen82, Roediger06a, Roediger08, Tonnesen09, Tonnesen10, Tonnesen11, Poggianti19a},
thermal evaporation and mixing \citep[e.g.][]{Cowie77}, magneto-hydrodynamic waves \citep[e.g.][]{Tonnesen14, Ruszkowski14, Vijayaraghavan15}, and photoionisation by young stars formed in the stripped tails (see Sec.~\ref{subsec:stars}).
All these physical mechanisms are now explored in the most recent hydrodynamic simulations which began to reproduce some observed properties of the different gas phases in the tails.
There are, however, several observational results which remains unexplained. For instance, it is still unclear why the tails are rarely seen in all the different gas phases at the same time, or why star formation occurs in some of them why not in others. The reasons behind these still unexplained results are manifold. First, they can depend on the properties of the surrounding medium (gas temperature and pressure \citep{Tonnesen11}, two radially dependent variables which might affect the typical timescales for the gas change of phase), on the impact parameter
of the galaxies with the hot ICM (which might differently favor the formation of turbulent regions where the gas can collapse to form giant molecular clouds), or on the progression with time of the gas stripping process (which might act on different components of the ISM working this mechanism outside-in). Moreover, other complex physical processes can shape the morphology and physical conditions of the tails, including but not limited to the presence of magnetic fields, shocks, and instabilities.

\subsubsection{Gas evaporation}
\label{subsec:evaporation}

To understand whether RPS is the dominant mechanism affecting the late-type population in these environments we have to quantify the fraction of galaxies which are now undergoing or underwent this perturbation.
Truncated gaseous discs combined with unaffected stellar discs are a clear sign of a past RPS event. If the typical timescale for gas evaporation is short, as suggested by the low fraction of \hi\ tails observed in nearby clusters, these tails are short lived phenomena and rather indicate an ongoing perturbation. This evidence also suggests that the \hi\ gas, once stripped, changes of phase while mixing with the surrounding ICM.
How does this gas phase transformation occurs?
The typical timescale for gas evaporation can be measured using different approaches. The total \hi\ mass in the tail can be first compared to the lacking mass of \hi\ over the disc of the galaxy by mean of the \hi-deficiency parameter \citep{Haynes84}. Derived from different mean scaling relations in field objects, the \hi-deficiency parameters gives the difference in logarithmic scale 
between the expected and the observed total mass of \hi\ that a galaxy of given type, size, or luminosity has with respect to similar objects in the field. Recent calibrations of this relation are given in \citep{Boselli09}. The lacking mass of \hi\ can be compared to that observed in the tail to see whether all the stripped \hi\ gas is still there (see for an example \citet{Boselli16}).
The mass, the column density, and the volume density of the \hi\ gas can be directly measured from the \hi\ observations, the latter making simple assumptions on the 3D geometrical distribution of the gas within the tail. The largest uncertainty in these estimates is the clumpiness of the gas leading to a very uncertain volume filling factor.

In the case that the time since the beginning of the interaction $\tau_{RPS}$ is known, as for instance derived from simulations, models,
or from SED fitting analysis as described in the next sections, these data can be used to estimate an evaporation rate \citep{Vollmer07}:

\begin{equation}
\dot M_{\rm evap} \sim \frac{(M_{\rm expected} - M_{\rm observed})}{\tau_{\rm RPS}} ~~~~ \rm{[M_{\odot} yr^{-1}]}
\end{equation}

\noindent
As suggested by \citet{Vollmer07}, this estimate can be compared to that derived from the evaporation rate of a spherical gas cloud \citep{Cowie77a}:

\begin{equation}
\dot M_{\rm evap}^{\rm sphere} = 1.74 \times 10^{-3} (\frac{T_{\rm ICM}}{\rm 3\times10^7 K})^{5/2} (\frac{r}{\rm pc}) (\frac{37}{\ln\Lambda}) ~~~~ \rm{[M_{\odot} yr^{-1}]}
\end{equation}

%\begin{equation}
%\dot M_{\rm evap}^{\rm sphere} = 4.34 \times 10^{-22} T^{5/2}_{\rm ICM} r_{\rm pc} (30 \ln\Lambda) ~~~~ \rm{[M_{\odot} yr^{-1}]}
%\end{equation}

\noindent
where $r$ is the cloud radius and $T_{\rm ICM}$ is the ICM temperature and $\ln\Lambda$ is the Coulomb logarithm. Noticing this mass loss rate is proportional to the cloud radius, one can analytically solve the evaporation timescale of this spherical cloud:

\begin{equation}
\tau_{\rm evap}^{\rm sphere} = 2.5 \times 10^{6} (\frac{\overline{n_{\rm H}}}{\rm 1 \,cm^{-3}}) (\frac{T_{\rm ICM}}{\rm 3\times10^7 \,K})^{-5/2} (\frac{r}{\rm 100 \,pc})^{2}  (\frac{\ln\Lambda}{37}) ~~~~ \rm{[yr]}
\end{equation}

\noindent
where $\overline{n_{\rm H}}$ is the average hydrogen number density of the cloud. We also include a factor of 1.4 on the total cloud mass from He and other heavy elements. This timescale\footnote{Note that the classical evaporation is assumed here. If instead saturated evaporation \citep{Cowie77} is assumed, the evaporation timescale typically increases by a factor of a few. On the other hand, galactic clouds are clumpy so the actual contact surface between the hot ICM and the cold clouds can be much larger than the surface of a single spherical cloud, which will reduce the evaporation timescale, along with the likely turbulence around the mixing layers.} is typically less than 100 Myr for galactic clouds, e.g., an \hi{} cloud with a radius of 500 pc (corresponding to the scale height of a galaxy disc) and an average density of 1 cm$^{-3}$, or a molecular cloud with a radius of 50 pc and an average density of 10$^{2}$ cm$^{-3}$. Similarly,
%, and for the number of clouds in the tail (accurate recipes for prolate and oblate gas distributions are given in \citet{Cowie77}). For a typical  size of $r_{\rm pc}$ = 500 pc (corresponding to the scale height of a galaxy disc), an ICM temperature of $T_{\rm ICM}$ = 3 $\times$ 10$^7$ K, and a Coulomb logarithm of ln $\Lambda$ = 30,
\citet{Vollmer07} estimated that the evaporation rate $\dot M_{\rm evap}^{\rm sphere}$ $\sim$ 1 M$_{\odot}$ yr$^{-1}$, which for a tail composed of $\sim$ 1000 similar clouds
with a given geometric form where 10\%\ of the surface of the spheres is surrounded by the hot ICM, the total evaporation rate of the tail is $\dot M_{\rm evap}^{\rm tail}$ $\sim$ 100 M$_{\odot}$ yr$^{-1}$.
If the mass of the gas in the tail $M_{\rm tail}$ is known (measured as described above from the HI-deficiency parameter, 
from a direct measure, or simply assuming that the mean density of the gas within the spheres is $n$ = 1 cm$^{-3}$), this quantity can be used to estimate the typical 
timescale for evaporation once the cold gas in the tail get in contact with the hot ICM.
%\begin{equation}
%\tau_{\rm evap} = \frac{M_{\rm tail}}{\dot M_{\rm %evap}^{\rm tail}}    ~~~~ \rm{[yr^{-1}]}
%\end{equation} 
%\noindent
For a typical tail of mass $M_{\rm tail}$ $\sim$ 10$^9$ M$_{\odot}$, this leads to an evaporation timescale of $\tau_{\rm evap}$ $\sim$ 10 Myr, a relatively short timescale compared to the 
typical crossing time of a cluster ($\sim$ 2 Gyr, \citet{Boselli06}). With a typical velocity within the cluster of $\sim$ 1000 km s$^{-1}$, during this time a galaxy travels $\sim$ 10 kpc, 
a length comparable to that of the tails observed in \hi{}. This means that \hi\ tails are short living structures, thus their relatively low number in blind surveys does not 
rule out RPS as a dominant mechanism affecting cluster galaxy evolution. It is worth remembering that this result is in agreement with what found from the analysis of H$\alpha$ narrow-band imaging data: indeed, the large amount of ionised gas measured within the tails of ram pressure stripped galaxies, which is fairly comparable to that of the \hi\ removed from the galaxy disc, combined with its short recombination time ($\lesssim$ 5 Myr), and the length of the tails (up to 100 kpc) suggests that the gas has been stripped as cold atomic hydrogen and later ionised and kept ionised in the tail once in contact with the hot ICM (see Sec. \ref{subsec:ionised}). 

On the other hand, such a steady, non-radiative evaporation model is certainly too simplified. As stated in Sec.~\ref{subsec:hotgas}, heat conductivity is typically largely suppressed in the ICM and the interface between the ICM and the hot ISM, which is typically attributed to magnetic field. Even in the absence of magnetic field, hydrodynamical instabilities and turbulence enhance mixing of the stripped cold ISM and the hot ICM by drastically increasing the contact surface area between two phases. However, radiative cooling can be significant as the mixed gas (with a temperature of about the geometric mean of the cold and hot gas temperatures) can cool quickly. Clouds with sufficiently short cooling time in the mixed gas can even grow with time \citep[e.g.,][]{Gronke22}.
Such kind of turbulent radiative mixing layer has received a lot of attention recently \citep[e.g.,][]{Gronke18,Fielding20,Sparre20,Tan21,Gronke22}, with the focus on galactic winds ad circumgalactic medium. As hydrodynamical instabilities, turbulence and magnetic field are inevitable in stripped tails, more works are required to better understand evaporation and mixing of the stripped ISM in the surrounding hot ICM.

\subsubsection{Gas cooling into giant molecular clouds}
\label{subsec:cooling}

The detection of large amounts of molecular gas at large distances from the galaxy opened the question whether this molecular gas has been stripped from the inner disc,
or it has been formed in-situ after the cooling of the atomic phase. Indeed, as clearly explained in \citet{Vollmer01}, two major physical processes are competing within the tail of stripped cold gas embedded in a hot
diffuse ICM: cooling and heating. As explained in previous sections, once in contact with the hot ICM, the diffuse molecular gas component can be evaporated by the hot ICM. 
%destroyed by the harsh X-rays radiation {\color{red}{MS: collisional heating should be much more important here, I will check more. ACTION ITEM MING}}, 
%becoming atomic hydrogen and evaporating into ionised gas and hot plasma.
If the density of the stripped gas is sufficiently high, however, self shielding in the outer layers can protect the inner regions from heating and allow the cooling of the gas 
into giant molecular clouds, where star formation can take place. The presence of diffuse dust, now confirmed by recent far-infrared observations \citep{Longobardi20a}, could contribute to the cooling of the gas within the tails (see Sec. \ref{subsec:dust}). The formation of molecular gas generally occurs whenever the column density of \hi\ exceeds 3$\times$ 10$^{20}$-10$^{21}$ cm$^{-2}$ \citep{Schaye04}
and becomes more efficient in the presence of dust, which can act as a catalyst in the formation of molecular gas and can
contribute to the cooling of the gas via absorption of the incoming radiation \citep{Maloney96, Krumholz09}. The typical timescale for the formation of molecular hydrogen 
is $\tau_{\rm H_2}$ $\simeq$ 10$^9$/$n_{\rm e}$ yr \citep{Hollenbach97}.
\citet{Vollmer01} also noticed that, if stripped from the inner regions where star formation occurs, 
the molecular gas is not further heated by the UV radiation produced by the newly formed stars and thus can collapse if not supported by another source of energy. 
Furthermore, in the lack of magnetic fields molecular clouds can be destroyed by Kelvin-Helmholtz instabilities before they collapse to form new stars \citep{Yamagami11}.
Within a typical environment with characteristics similar to those encountered in the Virgo cluster ($n_{\rm e}$ $\simeq$ 10$^{-3}$ cm$^{-3}$, $T$ $\simeq$ 10$^7$ K),
most of the gas in the tail is expected to be either hot ($T$ $\geq$ 10$^6$ K), ionised, or low density ($n_{\rm e}$ $\geq$ $n_{\rm ICM}$), or molecular, cold ($T$ $\sim$ 10-100 K), and high-density ($n_{\rm e}$ $\geq$ 100 cm$^{-3}$) \citep{Vollmer01}. It is thus clear that under specific conditions, 
when the density of the gas is sufficiently high, new stars can be formed in the stripped gas. It is worth recalling that the coexistence of different gas phases, from the cold molecular gas, through the ionised phase, up to
the hot plasma, has been also observed in other characteristic regions on the edges of galaxies and their surrounding ICM such as in cool cores in NGC 1275 in Perseus  
\citep{Conselice01, Salome06, Salome11} and M87 in Virgo \citep[e.g.][]{Forman07, Sparks09, Boselli19}. These observations confirm that different gas phases at different temperature and density can cohabit the hostile intracluster environment.
It has also been stressed that, once star formation has occurred, the newly formed stars, which do not suffer the ram pressure exerted by the ICM because of their very low cross-section, get decoupled from the molecular gas cloud which is still decelerated by the external pressure during the travel of the galaxy within the high-density region,
as indeed observed in ESO137-001 in the Norma cluster \citep{Tonnesen12, Jachym19} or in IC3418 in Virgo \citep{Kenney14}.

\subsubsection{Star formation within the tail}
\label{subsec:sfrtail}

%{\color{red}{What is the difference between 3.2.7 and 5.1.3? IN MY OPINION IN 3.2.7 WE JUST EXPLAIN HOW AN ASYMMETRIC STELLAR DISTRIBUTION IS USED TO (OFTEN ERRONEOUSLY) IDENTIFY GALAXIES SUFFERING an RPS EVENT, IN 5.1.3 HOW THE STRIPPED GAS CAN OR CANNOT COLLAPSE TO FORM STARS}}
A key question in the study of the fate of the stripped gas is understanding whether it can contribute to the intracluster light by forming new stars outside the stellar discs of stripped galaxies, or whether it just pollutes the diffuse ICM. Indeed, blind surveys such as those carried out in \hi\ or in narrow-band H$\alpha$ imaging clearly indicate that star formation in the tails of stripped galaxies is not ubiquitous \citep{Boissier12, Boselli16, Boselli18a, Fossati16, Yagi17, Gavazzi18a, Laudari22}
%\footnote{Star formation is present in all the jellyfish galaxies of the GASP survey (although most lack a long H$\alpha$ tail), as indeed expected given that these objects have been selected to have young stars in their tails \citep[e.g.][]{Poggianti17, Poggianti19}.}. 
There is also evidence that whenever star formation is present, this occurs with an efficiency reduced by a factor of $\gtrsim$ 5-10 with respect to that of star forming discs \citep{Boissier12, Jachym14, Jachym17, Moretti18, Moretti20, Gullieuszik20}.

IFU observations indicate that, within the compact regions in the tails, the velocity
dispersion of the gas is lower than in the diffuse component ($\sigma$ $\simeq$ 25-50 km s$^{-1}$; \citealt{Fossati16}). The density of the gas here is sufficiently high, 
as the one encountered in typical GMC in the Milky Way (gas mass 5 $\times$ 10$^4$ $\leq$ $M_{\rm GMC}$ $\leq$ 5 $\times$ 10$^6$ M$_{\odot}$, size 5 $\leq$ $r_{\rm GMC}$ $\leq$ 30 pc, 
volume densities 50 $\lesssim$ $n(\mathrm{H_2})$ $\lesssim$ 500 cm$^{-3}$, \citealt{Solomon87,Engargiola03}). At these densities the typical collapse time (free fall time) for
a cloud with a spherically symmetric distribution of mass is:

\begin{equation}
t_{\rm ff} = \sqrt\frac{3\pi}{32G\rho} \simeq 5 \rm{Myr}
\end{equation}

\noindent 
where $\rho$ is the mean density of the gas. On relatively short timescales compared to the typical time necessary for a galaxy travelling at
$\simeq$ 1000 km s$^{-1}$ to form 50 kpc tails ($\simeq$ 30 Myr) stars are formed within the tail. Thanks to the spectacular angular resolution in the multifrequency data available for perturbed galaxies in nearby clusters, and in particular in Virgo, the physical, kinematic, and spectro-photometric properties of the \hii\ regions formed within the tails have been studied in detail. 
These compact \hii\ regions are characterised by star clusters of mass $M_{\rm star}$ $\simeq$ 10$^3$-10$^5$ M$_{\odot}$ and luminosities 
$L({\rm H\alpha})$ $\simeq$ 10$^{36}$-10$^{38}$ erg s$^{-1}$, corresponding to star formation rates $SFR$ $\simeq$ 10$^{-3}$-10$^{-4}$ M$_{\odot}$ yr$^{-1}$ \citep{Sun07,Hester10,Fumagalli11,Boselli21, Junais21}. Slightly larger values have been found for the star forming blobs detected in the tails of the jellyfish galaxies analysed during the GASP survey \citep{Poggianti19} probably because these blobs are unresolved in these higher distance galaxies.
The extraplanar \hii\ regions have thus characteristics similar to those observed in galaxies suffering a combined effect of
harassment and ram pressure (NGC 4254, \citealt{Boselli18a}; VCC1249, \citealt{Arrigoni-Battaia12}). 
The mean age of the stellar population of
these regions is of $\lesssim$ 100 Myr, or even younger ($\lesssim$ 30 Myr) whenever selected in narrow-band H$\alpha$ imaging, consistent with a picture where they were formed only recently as expected in a short-lived interaction. The metallicity of the gas in these regions is
generally similar to that of the parent galaxy in its outer disc \citep{Sun07,Fossati16}. Similar values of stellar masses, H$\alpha$ luminosities, and $SFR$ derived using \hst{} data to resolve the individual star forming regions
are observed in the tail of one ram pressure stripped galaxy in the Coma cluster (D100), where the mean age population is $\leq$ 50 Myr \citep{Cramer19}.

It is still unclear why \hii\ regions are formed within the tail in some objects and why not in others. These condensed regions are formed 
within the turbulences and instabilities created in the friction layers between the cold ISM stripped during the interaction and the hot ICM \citep[][G. Henseler priv.comm.]{Roediger14, Tonnesen21}.
The efficiency in producing high density regions might depend on the impact parameters and on the physical properties of the surrounding medium.
The lack of star forming regions far from the galaxy in the tails of NGC 4569 \citep{Boselli16}, NGC 4330 \citep{Fossati18} and ESO~137-002 \citep{Laudari22},
%which are suffering an almost face-on stripping event, 
%might indicate that this configuration is less efficient in producing turbulent regions where GMC can be formed than almost edge-on stripping events
contrasts with the large number of young \hii\ regions in the tails of ESO 137-001 \citep{Fossati16}, UGC 6697 \citet{Consolandi17}, and IC 3476 \citep{Boselli21}. Our samples are still to small to identify the main physical variable (if one exists) driving the presence of these features. Different scenarios should be further tested with tuned simulations, which explore differences in the quantity of stripped gas according to realistic wind angles experienced on different possible orbits of the galaxy within the cluster, and on the galaxy structural parameters. Furthermore, simulations also suggest that the presence of magnetic fields does not strongly affect the quantity of stripped material, but rather indicate that the gas is better confined in filamentary structures or bifurcated tails such as those observed in ESO 137-001, and thus might favor the formation of giant molecular clouds \citep{Roediger06a, Tonnesen14,Ruszkowski14, Ramos-Martinez18, Vijayaraghavan17}.
Some of these simulations also predict the formation of new stars in the tails, but assuming quite uncertain (and different) recipes for the star formation efficiency within the stripped material.

Early GADGET simulations \citep[e.g.,][]{Kronberger08,Kapferer09} cannot model mixing sufficiently well \citep[e.g.,][]{Sijacki12}, overpredicting the star formation activity in the stripped ISM.
%Those of \citet{Kronberger08} and \citet{Kapferer09} predict a large activity of star formation in the tails up to 100 kpc from the galaxy disc, with structures up to 1 kpc in diameter and stellar masses $\sim$ 10$^7$ M$_{\odot}$. 
%{\color{red}{MS: these early GADGET simulations, including Puchwein+10, used the code suppressing mixing so SF is enhanced --- well known in cosmological simulations now. I can revise this part. Some suggested writing made}}
%This activity of star formation is comparable to that observed within the disc of the perturbed galaxy and is able to produce up to 10$^8$ M$_{\odot}$.
On the contrary, the simulations of \citet{Tonnesen12} or \citet{Lee20} rather suggest a weak star formation activity in the turbulent wakes formed during the interaction ($\sim$ 1\%\ with respect to that formed within the disc). This activity is mainly
limited to clumps where the density of newly formed stars is $\sim$ 3 $\times$ 10$^4$ M$_{\odot}$ kpc$^{-2}$. The star formation in the tail can last $\sim$ 1 Gyr and significantly drop on longer timescales \citep{Calura20}. The star formation in the tail seems principally modulated by the external pressure of the ICM rather than by the strength of the ram pressure. Low rates of star formation in the tails of dwarf galaxies are also predicted by the simulations of \citet{Steyrleithner20}.

%- gas can fall back on the disc

%- gas can be transformed into stars (under which conditions): Sand et al 2017, stripping remnant in Virgo

%Dasyra et al 2012 says that whenever the density is sufficiently high \hi\ can collapse into H2 (nice discussion); if it makes stars, less efficiently than in the disc

%GASP Gullieuszik et al. 2020 do not find strong relations of SFR in the tail (which ranges from 10-4 a 1 Mo yr-1) with other parameters, just slightly higher in more massive galaxies (more gas, non normalised result). Those with higher SFR in the tail are in the inner regions of the phase diagram but with high velocities. ANd they are mainly located in clusters with sigma $<$ 900 km/s

%Fossati et al 2016: only in regions where the velocity dispersion of the gas is low

%Poggianti et al 2019 gives mean values of SFR, AV, sigma, Mstar ionising source etc

%- change phase and get first ionised and then hot: we could make a calculation using all galaxies with Halpha tails, estimate the mass of ionised gas, and
%see whether it is correlated to the HI-def parameter, and with the length of the tail to have an estimate of the timescale

\subsection{ICM clumping}
\label{subsec:clumping}

Recent X-ray observations have suggested that the ICM distribution is not smooth but instead inhomogenous (or clumpy), especially at the cluster outskirts \citep[e.g.,][]{Walker19}. The ICM clumping factor,
$C \equiv \langle n_{\rm e}^2 \rangle / \langle n_{\rm e} \rangle^{2}$ \citep{Mathiesen99}, can be derived from the X-ray data and typically measures both the azimuthal variations and local clumping at scales of below 100 kpc.
%$C = \frac{(\Sigma n_{\rm e})^2}{\Sigma n_{\rm e}^2}$
The clumping factor measured in clusters is small, from close to 1 within $R_{500}$ to typically $< 1.5$ at $\sim r_{200}$ \citep[e.g.,][]{Walker19}.
Since the X-ray emissivity of the ICM is proportional to $n_{\rm e}^2$, the ICM clumping can bias the measured ICM density, which can further bias the measured ICM gas mass \citep[e.g.,][]{Walker19}. The ICM clumping is typically induced by bulk motion and turbulence in the ICM. This deviation from the hydrostatic equilibrium suggests that the cluster mass from the X-ray data by assuming hydrostatic equilibrium is biased, which has become a key question in cluster cosmology \citep[e.g.,][]{Kravtsov12}.

Many of the ICM clumps can originate from cluster galaxies, detached via RPS or tidal interaction/stripping \citep[e.g.,][]{Dolag09,Vazza13} and now evaporating in or mixing with the ICM.
%The stripped gas cloud induce inhomogeneity or clumpiness in the ICM.
%3) ISM/ICM mass ratio about a few percent but can still be significant, along with other disturbance
The ISM-to-ICM mass ratio in local clusters is small. The ICM-to-stellar mass ratio in $M_{500} > 10^{14}$ M$_{\odot}$ clusters is $\sim$ 3 - 12 \citep[e.g.,][]{Leauthaud12,Gonzalez13}. The average ISM-to-stellar mass ratio in cluster galaxies can be estimated by combining the cold gas scaling relation in galaxies \citep[e.g.][]{Blanton09,Boselli14b} and galaxy mass function in clusters \cite[e.g.,][]{Ahad21}. Such an analysis gives an average ISM-to-stellar mass ratio of $\sim$ 0.3 in clusters. Thus, the average ISM-to-ICM mass ratio is only $\sim$ 5\% in nearby clusters. Nevertheless, the evaporation time for the stripped ISM can be substantial so the induced ICM inhomogenities can be long-lasting. The mixing between the stripped ISM and the ICM is poorly understood. The MHD simulations also suggest that the stripped ISM clouds can grow via mixing under certain conditions, e.g., when the cooling time of the mixed gas is shorter than the mixing time \citep[e.g.,][]{Gronke18,Fielding20,Sparre20}.
% more on mixing and growth
Moreover, the motion of ISM-less galaxies (or of their dark matter halos only) in clusters can still induce ICM inhomogeneities and turbulence in the wakes behind the galaxy orbits \citep[e.g.,][]{Kim07,Ruszkowski11}.
% Naked galaxies can do some work

There are isolated \hi{} clouds detected in the Virgo cluster \citep[e.g.,][]{Taylor12}, although no X-ray enhancement was detected from any of the \hi{} clouds from the existing data. Recently, the first example of an X-ray/H$\alpha$ isolated cloud in the intracluster space was discovered in A1367 \citep{Ge21} (the H$\alpha$ cloud was first discovered by \citealt{Yagi17}), presenting a robust example connecting the ICM clumps and the stripped ISM clouds. 
Follow-up observations with the IRAM 30m telescope also revealed molecular gas in the cloud \citep{Jachym22}.
This first example also suggests that at least some ICM clumps are multiphase in nature and implies that the ICM clumps can also be traced in H$\alpha$ from future deep and wide-field H$\alpha$ surveys.

\subsection{ICM metal enrichment}
\label{subsec:enrichment}

The ICM is enriched with metals with an average abundance of $\sim$ 0.3 solar
% or use 0.3-0.5?
\citep[e.g.,][]{Yates17,Mernier18}.
% Baumgartner+2005 on ASCA
The ICM is composed of pristine gas and ejected ISM from cluster galaxies over the formation and evolution of clusters. The ICM metals should mainly come from stars in cluster galaxies, ejected through AGN feedback (quasar and radio modes), stellar feedback, RPS, galaxy mergers and tidal interactions.
The properties of the ICM metals (abundance, distribution and evolution) present important constraints to reconstruct the SF history in cluster galaxies, the evolution of AGN and stellar feedback and other processes to transport metals from galaxies to the ICM.
Recent X-ray constraints on the ICM metallicity evolution generally support the ``early enrichment'' model, giving the little evolution of the ICM metal content since $z \sim 1$ and the typically constant metallicity at the cluster outskirts \citep[e.g.,][]{Mantz17,Liu20}. The strong mass dependency on the ratio between the ICM metallicity and the cluster stellar mass fraction from groups to clusters may also require early enrichment \citep[e.g.][]{Bregman10,Renzini14,Biffi18}.
% also https://ui.adsabs.harvard.edu/abs/2021arXiv210504638B/abstract
The ``early enrichment'' model (typically related to strong winds during the formation and rapid growth of galaxies, and sometime their SMBHs) was also supported by the recent simulations \citep[e.g.,][]{Vogelsberger18,Pearce21}, although the detail (e.g., IMF, SN type/rate/yields, transport efficiency) is not yet agreed.
Thus, the ICM is mostly enriched beyond $z$ of 1 and the effect of the late enrichment (including RPS) since $z \sim 1$ is minor and mainly shown in the inner regions of clusters.
%If RPS enriches the ICM, the ICM abundance evolution should be aligned with the galaxy ISM evolution.

How much can RPS contribute to the ICM enrichment? In section~\ref{subsec:clumping}, we estimated the average ISM-to-stellar-mass ratio. We can also estimate the average metallicity of cluster galaxies, weighted by the galaxy mass function and the ISM mass fraction. We do that by assuming the mass-metallicity relationship from \cite{Foster12}. Integrating between $M_{\rm star}$ of 10$^{8}$ - 10$^{11.5}$ M$_{\odot}$, the average metallicity of cluster galaxies is 1.2 solar. 
Thus, for an ISM-to-ICM mass fraction of 0.05 and an ISM metallicity of 1.2 solar, if 50\% of the ISM is stripped from galaxies, the expected ICM metallicity increase is only 10\%, for the original ICM metallicity of 0.3 solar.
%but stellar product can end in the ICM no matter whether the ISM still remain?.
This simple estimate is also consistent with the simulation results from \cite{Domainko06}, which suggests that RPS can only contribute to $\sim$ 10\% - 15\% of the ICM metal.
%The late-stage pollution would not add much metal in the ICM. 
Indeed, if RPS plays an important role on the ICM metal content, one may expect a strong correlation between the ICM metallicity and the cluster mass, as ram pressure increases with the cluster mass ($\propto M^{2/3}$ for the self-similarity). However, the observed correlation between the ICM metallicity and the cluster mass is very weak \citep[e.g.,][]{Truong19}.
% KEEP this in the comment: on the other hand, the roles of RPS in early enrichment is still unclear.

Nevertheless, RPS can still play a role to add on the ICM metal content and modify the ICM metal distribution. Simulations \citep[e.g.,][]{Schindler05,Domainko06,Kapferer07} suggest that RPS is more important around cluster centres where the ram pressure is high, while galactic winds are more important at the outskirts where winds are less confined by the surrounding ICM.
RPS also leaves trails and plumes of high-metallicity gas behind galaxies undergoing stripping, which can be detected with the high-resolution ICM abundance maps. Cluster mergers can also enhance RPS and the subsequent metal enrichment \citep{Ebeling19}.

%
%- are we able to make a first order calculation assuming a galaxy infall rate in the cluster of 200 gal/Gyr (Boselli et al 2008 Virgo, consistent with SORCE ET AL 2021), a typical \hi\ mass content, a gas-to-dust ratio, and quantify what is the content of gas and dust left over in the ICM after a stripping process? we have done it in Longobardi et al. i guess. But here i would add what is expected from the ellipticals after SN feedback. Is it possible to quantify what could be the contribution of Elliptical vs Spirals? See also Domainko et al 2006. The idea is the same than Domainko, but with an analytical prescription.

\subsection{Intracluster light and dust contribution}
\label{subsec:ICL}

The presence of diffuse stellar light in the intracluster space, dynamically hot and bound to the cluster, has been known for many years \citep[e.g.,][]{Gonzalez05,Mihos05,Montes19,Contini21}. The intracluster light (ICL) carries the imprint of the growth history of the cluster and contributes to a significant fraction of the stellar mass in clusters \citep[e.g.,][]{Pillepich18,Montes19,Contini21}.
While the bulk of the ICL is believed to be evolved stellar population removed from cluster galaxies by tidal interactions and mergers \citep[e.g.,][]{Contini21}, the discovery of star formation in stripped tails as discussed in Sec.~\ref{subsec:stars} suggests an {\em in situ}, young component of the ICL. There have also been reports of blue intracluster light \citep[e.g.,][]{Starikova20}. The cosmological simulation by \citet{Puchwein10} also attributes up to 30\% of mass in intracluster stars to young stars
formed in the intracluster space, inside cold ISM that was stripped from infalling galaxies by ram pressure, although one has to be aware of the uncertainties of transport processes, cooling and star formation processes in the simulation.
%The contribution of the stripped ISM to the ICL (through in-situ SF) has been explored in \cite{Sun07,Sun10}. It depends on the SF efficiency in the tail.

What is the fraction of the ICL that comes from this {\em in situ} channel?
%The importance of this channel depends on SF efficiency in the tail \citep[e.g.,][]{Sun10}.
One can write its contribution as: 

\begin{equation}
{\frac{M_{\rm ICL, SF}}{M_{\rm ICL}} = (\frac{M_{\rm ICL, SF}}{M_{\rm ISM, RPS}}) \times (\frac{M_{\rm ISM, RPS}}{M_{\rm ISM}}) \times (\frac{M_{\rm ISM}}{M_{\rm star}}) / (\frac{M_{\rm ICL}}{M_{\rm star}})}
\end{equation}

\noindent
where $M_{\rm ICL, SF}$ is the total mass of ICL formed via the {\em in situ} channel in the stripped ISM, $M_{\rm ICL}$ is the total mass of the ICL, $M_{\rm ISM, RPS}$ is the total mass of the stripped ISM, $M_{\rm ISM}$ is the total mass of the ISM and $M_{\rm star}$ is the total stellar mass of the cluster. $M_{\rm ISM} / M_{\rm star} \sim$ 0.3 as discussed in Sec.~\ref{subsec:clumping}. $M_{\rm ICL} / M_{\rm star}$ is uncertain, 0.05 - 0.5 from \cite{Montes19} (also depending on the detailed definition of the ICL). We can assume an average 
$M_{\rm ISM, RPS} / M_{\rm ISM} \sim$ 0.5, which is also consistent with the average estimates from \cite{Gullieuszik20}.
Existing observations generally suggest that the star formation efficiency in the stripped tails is not as high as that in the galactic disc \citep[e.g.,][]{Boissier12,Kenney14,Jachym14,Jachym17,Gullieuszik20}.
\citet{Gullieuszik20} concluded that the star formation efficiency in the tails of GASP galaxies is $\sim$ 5 times lower than in the galactic disks. In the galactic disks, $\sim$ 6\% of gas is converted to stars in 100 Myr \citep[e.g.][]{Leroy08}. Thus, we can take $M_{\rm ICL, SF} / M_{\rm ISM, RPS} \sim 0.02$, which is also consistent with the typically long gas depletion timescales in stripped ISM ($\sim 10^{10.5}$ yrs from e.g., \citealt{Cramer19}). With all these rough estimates, $M_{\rm ICL, SF} / M_{\rm ICL} = 0.006 - 0.06$. Here the star formation efficiency in the stripped tails is the biggest uncertainty, as already pointed out by \cite{Sun10}. Nevertheless, the current data do not suggest a significant contribution of the ICL from this {\em in situ} channel.

The contribution to the ICL by stars formed within the tails of perturbed galaxies has been studied recently using the data gathered during the GASP survey \citep{Gullieuszik20}.
An average SFR in the stripped tail is estimated to be $\sim$ 0.22 M$_{\odot}$ yr$^{-1}$ per cluster, for typical GASP clusters with a velocity dispersion of 750 km s$^{-1}$ at $z \sim $ 0. Simply extrapolating this estimate to $z = 1$, they concluded that an integrated average value per cluster of $\sim$ 4 $\times$ 10$^{9}$ M$_{\odot}$ of stars formed in the tails of RPS galaxies since $z \sim 1$. For this typical GASP cluster, $M_{500} \sim 2\times10^{14}$ M$_{\odot}$. With a stellar mass fraction of $\sim$ 3\%, the GASP result essentially suggests $M_{\rm ICL, SF} / M_{\rm ICL} \sim$ 0.0007. The GASP results should only be taken as a lower limit, giving the vast majority of the cluster volume not probed (only the small MUSE fields around each GASP galaxy studied) and the incomplete sample of the cluster galaxies.
%Nevertheless, the estimate is uncertain and one would expect the fraction should have a mass dependency.

Recent studies have also begun to reveal the presence of diffuse dust in the intracluster space (or intracluster dust) \citep[e.g.,][and references therein]{Planck16a,Gutierrez17,Gjergo20,Longobardi20a}. Two methods have been applied, direct detection of the FIR emission from the dust and the reddening of background sources. Combining the Planck and the {IRAS} data, \cite{Planck16a} constrained the average spectral properties of dust in clusters and derived an average dust-to-gas mass ratio ($D/G$) of $\sim 2\times10^{-4}$. However, most of the dust in clusters resides in galaxies so the contribution from the intracluster dust is small \citep[e.g.,][]{Gjergo20}. \cite{Gutierrez14,Gutierrez17} applied both methods and put an upper limit of $D/G$ of $\sim 10^{-4}$ for the intracluster dust.
% note they quoted ratio to the total mass of the cluster; assuming a hot gas mass fraction og 0.1
\cite{Gjergo20} attempted to model the dust content of cluster galaxies and suggested a rather small $D/G$ of $\sim 10^{-6}$ for the intracluster dust.
They also concluded that cluster spiral galaxies contribute to the bulk of the intracluster dust.
\cite{Longobardi20a} studied the reddening of background sources behind the Virgo cluster and derived a total dust mass of $\sim 2.5\times10^{9}$ M$_{\odot}$ within 430 kpc (or $\sim 0.43 \times r_{200}$), which suggests a rather high $D/G$ of $\sim 3\times10^{-4}$.
%- gas to dust ratio of the ICM vs galaxies and tails
The dust sputtering time in the hot ICM is on the order of 10 Myr in the cluster core and 1000 Myr at 1 Mpc from the cluster centre \citep{Draine79, Vogelsberger19} so the intracluster dust needs to be continuously replenished. We recall that the intracluster dust plays a major role in the energy balance of the ICM enhancing and can even dominating the cooling of the hot gas responsible for the X-ray emission \citep[e.g.,][]{Montier04}.
%{\color{red}{MS: probably a general point in conclusions that RPS and other processes populate cold gas/dust in the intracluster space, modifying the energy balance and affect ICM evolution --- the studies of the ICM evolution is incomplete with the X-ray data only!}}

\section{The importance of RPS at different epochs and in different environments}
\label{sec:importance}

\subsection{Clusters vs. groups and other environments}
\label{subsec:halomass}

In the previous Sections we have described the role of several environmental processes and most notably RPS on the transformations occurring to satellite galaxies in massive haloes (mostly galaxy clusters in the local Universe). We now describe the statistical relevance of these processes on the overall galaxy population as a function of galaxy stellar mass and host halo mass. By using the semi-analytic model of \citet{Henriques15} we estimate that the fraction of galaxies with $M_{\rm star} > 10^9 \rm{M_\odot}$ living in haloes more massive than $10^{14} \rm{M_\odot}$ is $25\%$ at $z=0$. This substantial fraction means that environmental processes acting in clusters cannot be neglected when a complete population of relatively massive galaxies is studied in the local Universe. However, this fraction drops to $5\%$ and $2\%$ at $z=1$ and $z=1.5$, respectively. It therefore seems that the most favorable environment for RPS to occur becomes increasingly rare as redshift increases. Moreover, the most massive haloes at $z>1.5$, the so-called protoclusters, are mostly unrelaxed structures which makes them difficult to identify in galaxy surveys \citep{Shattow13} and could imply that these forming structures do not have dense hot gas for RPS to be effective. 
A possible partial exception to this are the dense cores of high-redshift protoclusters (SPT-CLJ0459, \citealt{Strazzullo19} and JKCS041, \citealt{Andreon14}) where RPS could potentially be effective. The authors of these studies indeed find a high quiescent fraction ($> 80\%$) for galaxies spectroscopically confirmed in the clusters suggesting that environmental quenching was already active in these systems at these epochs. However, the high stellar mass limit of their observations coupled with the small statistics hampers a clear identification of the quenching mechanism.
The simulations presented by \citet{Quilis17} suggest that the overall effect of stripping is subdominant at $z=2$ even in clusters with a hot ICM, because a large fraction of this gas is accreted and cools onto the galaxies. These authors also found different behaviours for galaxies with stellar masses below 10$^{10} {\rm M_\odot}$, which have eccentric orbits at the outskirts of the cluster and are thus only mildly perturbed, and the most massive ones which have more radial orbits and therefore can reach the dense cluster cores. Even for the latter population the effect of stripping is marginal in these simulations which points to other processes to explain the high passive fractions found in the aforementioned high-$z$ clusters.

Another, less extreme, environment in which galaxies live are galaxy groups. Several definitions for what is a group exists, but in general those environments have a host halo mass in the range $10^{13}- 10^{14} \rm{M_\odot}$. The fraction of galaxies with $M_{\rm star} > 10^9 \rm{M_\odot}$ living in this halo mass range is $47\%$ at $z=0$ and stays constant (within $5\%$) up to $z=1.5$. This intermediate mass range is very important because halos that assemble into massive clusters are often in the group range and are known to pre-process the galaxies \citep{Dressler97} before their final arrival into the clusters. \citet{Oman21} studied the stripping and quenching times in a local sample of galaxies using models derived and calibrated on a cosmological simulation of massive clusters. These authors found that while in massive clusters the stripping of \hi\ is almost complete before the first pericenter passage, in groups it takes a significantly longer time ($\sim 3$ Gyr), likely due to the reduced ram pressure force exerted by the intragroup medium. In both cases a complete suppression of star formation occurs $2-3$ Gyr later. This is not inconsistent with the short quenching times derived for some RPS galaxies in Virgo and Coma. Indeed, this analysis of the full galaxy population includes a significant fraction of objects whose molecular gas reservoir is not stripped and therefore the quenching follows the H$_2$ consumption timescale of a couple of Gyr. Lastly, these authors found that the lowest mass groups in their sample ($M_{\rm{halo}} < 10^{13} \rm{M_\odot}$) strip and quench their satellites extremely inefficiently with typical time-scales that approach the age of the Universe. These results agree with the observations of nearby groups: \citet{Roberts21a} using a broad-band optical selection found that the frequency of RPS candidates in groups ($M_{\rm halo} < 10^{14} \rm{M_\odot}$) is lower than in clusters by a factor of $\sim 2$. Consistent with this result is also the presence of HI-deficient galaxies (\citep[e.g.][]{Fabello12, Brown17}) and of radio continuum tails \citet{Roberts21a} recently discovered in nearby groups.

A particularly favorable environment for RPS to be efficient might be merging clusters of galaxies \citep[e.g.][]{Roediger14, Stroe15, Stroe17, Ebeling19}. In these environments, galaxies belonging to a given structure might suffer an increased external pressure due to their non-isotropic motion once entering into the other massive halo. The compression of their ISM can trigger star formation, as indeed observed in several merging structures at redshift $z$ $\sim$ 0.15-0.3 \citep[e.g.][]{Stroe15, Stroe17, Ebeling19}. Clear examples of triggered RPS in merging systems in the very local Universe are rare because of the limited sampled volume.
There are, however, a few examples which might suggest an increased RPS efficiency in local merging structures. The first one is the observed concentration of young and blue low-mass poststarburst galaxies at the edges of the two infalling substructures within the Coma cluster observed by \cite{Poggianti04}. These objects might be the remnants of RPS galaxies where the external pressure exerted by the ICM of the main body of the cluster while crossing it as members of the infalling substructures enhanced their star formation activity and than totally stopped it on a short timescale. Another possible example is NGC 4522 in the Virgo cluster, a late-type galaxies with asymmetric atomic, molecular, and ionised gas distributions suggesting an ongoing RPS event \citep{Kenney99, Vollmer00, Vollmer04, Vollmer08, Kenney04, Abramson14, Abramson16, Stein17, Minchin19, Longobardi20a}. Tuned simulations indicate that the properties of this object can be hardly explained by the external pressure exerted by a smooth and static ICM, but would rather require large bulk motions and local density enhancements as those occurring in a shock-filled ICM experiencing subcluster merging \citep{Kenney04}.
Another intriguing local example is at the northwest of A1367 where a merger shock and a radio relic were detected \citep{Ge19}, around at least five RPS galaxies in proximity (at least in projection) \citep{Yagi17}.

\citet{Roberts19} studied the role of the ICM density as a function of satellite galaxy stellar mass from a sample of local galaxy clusters with \chandra{} X-ray data. These authors found that for low-mass galaxies the quenched fraction mildly increases in the cluster outskirts at ICM densities below $\rho_{\rm ICM}$ = 10$^{-29} {\rm g~cm^{-3}}$ before increasing sharply beyond this threshold. Their results are consistent with a picture where low-mass cluster galaxies experience an initial, slow-quenching mode driven by gas depletion via star formation in absence of accretion, followed by an accelerated quenching associated with RPS of the cold gas within a quarter of the virial radius. More massive galaxies, instead, might deplete their gas while still moving at larger clustercentric radii and their deeper gravitational potential wells makes them more resistant to RPS until they reach the most dense regions of the ICM. This study, while limited to clusters, might suggest that low-mass galaxies are stripped even in group-like environments while massive ones are more efficiently stripped in massive clusters.
A complete consensus on the physical conditions leading to effective RPS is still missing, for instance \citet{Mostoghiu21}, using more than 300 simulated clusters, found that galaxies lose most of their gas when crossing the virial radius (in projected space), while \citet{Oman16} suggest that the stripping occurs on average further-in, near pericenter.

We have seen that low mass galaxies are easily subject to stripping phenomena due to their shallow gravitational potential well. Indeed the stripping of these objects does not always require a massive halo and can take place in other types of environment. \citet{Benitez-Llambay13} proposed that dwarf galaxies can be stripped of their gas as they cross the network of filaments that builds the so-called cosmic web. The stripping has a rapid and dramatic influence on the star formation activity of these galaxies and could even explain the scarcity of dwarf galaxies compared with the numerous low-mass halos expected from the cosmological model in the Local Group. On larger scales \citet{Lee21} studied nearly 70 galaxy filaments around the Virgo cluster finding that galaxies in the filaments have a normal \hi\ content. This could imply that galaxies travelling along the filaments are experiencing a much milder RPS compared to those travelling across the filaments. This result, however, contrasts with the work of \citet{Castignani22} who found that the star formation, gas content and morphological properties of galaxies in cosmic filaments are at intermediate levels between galaxies in isolation and in clusters suggesting that filaments are dense enough to produce measurable environmental effects \citep[see also][]{Winkel21}). In a study of 24 jellyfish galaxies extracted from the GASP survey that do not reside in galaxy clusters, \citet{Vulcani21} found two star forming objects ($M_{\rm star} <10^{9.1} \rm{M_\odot}$) with ionised gas tails that could be associated to cosmic web stripping. The fact that these objects are actively forming stars, as opposed to the passive systems analysed by \citet{Benitez-Llambay13}, suggests that they are experiencing the early phases of the stripping or that the stripping is not rapid and efficient. 

RPS can also occur in the gaseous halo of individual galaxies, as shown by \citet{Mayer06} and \citet{Gatto13}. These authors coupled observations of dwarf quenched galaxies orbiting near the Milky Way (MW) with a stripping model to constrain the properties of the hot corona around the MW. A similar stripping event of a dwarf galaxy in the hot halo of a massive elliptical is reported by \citet{Arrigoni-Battaia12} that studied the interaction of VCC 1249 with M 49 in the Virgo cluster, concluding that both ram-pressure stripping and tidal interaction occurred. The joint action of the two mechanisms led to the removal of the \hi\ gas from the ISM of VCC 1249, while the gravitational tides perturbed the disc generating stellar tails. 

In conclusion, RPS can occur in a variety of environments provided that a galaxy travels with sufficiently high velocity through a medium that is dense enough (irrespective of its temperature) to overcome the gravitational potential well that keeps its ISM bound. The lower the mass of the target galaxy, the lower the RPS force needs to be to efficiently strip the gas and quench the star formation activity.

\subsection{Evolution with cosmic time}
\label{subsec:redshift}

%{\color{red}{MS: redshift evolution here}}
We can also examine the evolution of the average ram pressure in clusters with redshift.
As discussed in Sec.~\ref{subsec:prophidens}, for fixed cluster mass, one expects the ram pressure to increase with redshift, as $E(z)^{8/3}$, up to the time of cluster virialization and formation at $z \sim 3$. We can also follow the cluster growth track. The self-similar relation, as shown in the Eq. \ref{eq:Mdelta} is clearly incorrect as it does not account for the hierarchical mergers and growth. The cluster growth history presented by \cite{McBride09} can be roughly approximated by $M(z) \sim E(z)^{-3}$ at $z < 3$, which would suggest the ram pressure to evolve as $\sim E(z)^{2/3}$ (here we assume that RP scales as Eq. \ref{eq:P_ICM} that should apply for clusters). 
% yes, if P_RP ~ M^~1.2, RP not high in high-z if high-z groups are also gas poor in the inner regions
Therefore the ram pressure decreases with decreasing $z$ as the cluster assembles its mass. The above simple estimates are consistent with the results from the analytic work by \cite{Fujita01}, which shows that for a given cluster mass, RPS has more influence in high-$z$ clusters to the extent that most of galaxies in rich clusters at $z > 1$ are affected by RPS. As non-gravitational processes reduce the ICM content within $R_{500}$, the strength of RPS is also tied to the actual contribution of non-gravitational heating with redshift \citep{Fujita01}, which is however poorly constrained from the current X-ray data at $z > 1$. Nevertheless, RPS is expected to be substantial in high-$z$ clusters to at least $z \leq 2$ (also see \citealt{Singh19}).

The detection of direct evidences of RPS beyond the local Universe are made difficult by the surface brightness dimming of the cometary tails. Due to this observational limitation, direct evidence for ongoing RPS at higher redshift is still relatively poor. \citet{Cortese07} reported the detection of two cluster galaxies at $z\approx 0.2$ with asymmetric and extended features in $B$ band images obtained with \hst. Their modelling suggests that the morphologies of these galaxies have been influenced by the combined action of tidal interaction and RPS (see Sec. \ref{subsec:mainmech}). Since then, a few other tens of similar objects have been found in deep \hst{} images of clusters at $0.3<z<0.7$ \citep{Owers12,Ebeling14,McPartland16}. However, as discussed in Sec. \ref{subsec:stars}, the optical band selection is sensitive to intermediate age stars and cannot be taken as a direct proof of an ongoing or recent RPS event. While \hi\ and X-Ray tails are undetectable at high-redshift with current facilities, significant advancements on the optical spectroscopic instrumentation has made the detection of H$\alpha$ tails possible up to $z\sim 0.7$. \citet{Vulcani16} using \hst{} slit-less spectroscopy analysed the H$\alpha$ morphology of  galaxies in 10 clusters at $0.3<z<0.7$. Based on visual classification, these authors found that roughly 30\% of the galaxies have a regular H$\alpha$ morphology, a 20\% show sign of RPS and a 40\% show signatures of a recent merger or tidal interaction. By comparing to a field sample, this work demonstrates that the cluster environment shapes galaxy evolution at these redshifts, however the morphological transformations detected can be caused by a variety of mechanisms. Another indirect evidence of RPS is reported by \citet{Vaughan20} who found more compact H$\alpha$ radii compared to the stellar continuum in cluster galaxies at $0.3 < z < 0.6$ when contrasted to a field sample. 

\citet{Boselli19a} exploited the unique capabilities of MUSE at the Very Large Telescope to obtain deep IFU spectroscopy of the core of a $z=0.73$ cluster. These observations revealed the first direct evidence of ongoing RPS in intermediate redshift galaxies. These authors reported the presence of two galaxies with extended tails of diffuse gas detected in the [OII] line doublet. The tails extend up to 100 kpc from the galaxy discs, and are not associated with any stellar component, a typical signature of an RPS event.
The surface brightness of these tails, once accounting for the cosmological dimming and the [OII] to H$\alpha$ ratio, is $\sim 10$ times higher than those measured in the local Universe \citep[see also][]{Moretti22}. A possible explanation for this evidence can be found in the (molecular) gas fractions of main-sequence star forming galaxies at $z\sim 0.7$ which are $\sim 10$ times higher than their local counterparts \citep{Tacconi18}. A larger sample is however required to confirm whether the increased surface brightness is a common feature of intermediate-$z$ RPS tails or if it is limited to these objects. \citet{Boselli19a} also found that these RPS galaxies exhibit a significant star formation activity, consistent with normal galaxies of the same mass and redshift. This could be explained by a recent onset of the ram-pressure that could have produced the tails without significantly reducing the gas content in the star forming disc or by the initial higher gas fraction in these objects which makes the quenching of star formation more difficult.

At even higher redshift, direct evidence of RPS become even more challenging to be seen. As the H$\alpha$ line is redshifted in the near-IR, deep observations from the ground have to overcome the increasing brightness of the night sky, requiring extremely long integrations. Upcoming NIR space telescopes (e.g. the James Webb Space Telescope) will provide the IFU capabilities required for this task. At present, however, we can only rely on indirect evidences of the role of RPS at $z>1$. Many studies have exploited the exquisite quality of multiwavelength surveys in the deep fields to identify satellite galaxies and to derive the average time it takes for these galaxies to be quenched by environmental processes. \citet{Nantais17} reported that the quenching efficiency (i.e. the fraction of galaxies quenched by the environment) sharply decreases from $z\sim 0$ to $z \sim 1.5$, as first proposed by \citet{McGee09}. This result implies that at $z \sim 1.5$ the satellite galaxies population is indistinguishable from isolated galaxies in terms of their star formation activity, implying that any environmental process operates on scales comparable to the Hubble time at that epoch. At lower redshift, however, a passive population of galaxies made passive by the environment builds up. \citet{Balogh16}, \citet{Fossati17}, and \citep{Foltz18} found that the quenching times are long ($\sim 2-5$ Gyr in the stellar mass range $M_{\rm star} = 10^9-10^{11} {\rm M_\odot}$ and shorter for more massive objects) which points at the lack of infall of fresh gas from the cosmic web as the dominant environmental quenching mechanisms in galaxy groups. However, when those results are compared to cluster samples at the same redshift, \citet{Balogh16} and \citet{Foltz18} found that the average quenching times are shorter in clusters ($M_{\rm halo} > 10^{14} {\rm M_\odot}$) suggesting that in these massive haloes the dynamical removal of gas, possibly associated to RPS, is occurring. Since the results are obtained by studying the galaxy population as a whole it remains unclear if RPS is occurring only on some galaxies with high relative velocities with respect to the cluster or if it happens on a large fraction of galaxies although with a smaller efficiency than in the local Universe. Lastly, despite having shown analytically that the ram pressure should increase with redshift, we need to consider that high-$z$ galaxies are usually more compact \citep{van-der-Wel14} and therefore have a higher density of stars and gas, making the effects of  a stronger ram pressure less appreciable. Another consequence of the galaxy properties at high-$z$ is that their larger initial gas reservoir requires a significant depletion of the gaseous disc before an effect on the star formation activity can be observed.

%\subsection{The importance of RPS as a function of z, $\rho$, and $M_{\rm star}$ AB, MS, MF}
%\label{subsec:relativecont}

%- it would be nice to see from models of galaxy formation, hydrodynamic simulations, or other kind of models if we can estimate the number of infalling galaxies of different mass, in different environments, as a function of z and make a statistical analysis to estimate which would be the typical timescale for quenching they would have to match the observed population of gas rich and quiescent galaxies in local environments, MF:WORTH SOME TEXT IN THE DISCUSSION BUT HARD TO DO MORE THAN THAT

%\begin{figure}
%\centering
%\includegraphics[width=0.75\textwidth]{density.pdf}
%\caption{Radial profile of the density of the ICM within the Virgo cluster derived from X-rays observations %by Simionescu et al. (2017).
%Different symbols give the projected electron density for each galaxy. Empty symbols are for HI-deficient %galaxies, filled ones for HI-normal
%objects. Pentagons indicate galaxies undergoing an RPS event, triangles objects suffering %RPS combined
%with a gravitational perturbation.
%{\color{red}{MS: sorry what is the purpose of this Figure (giving the projection uncertainty)?
%AGAIN I AM HAPPY TO REMOVE IT, IT WAS JUST TO TEST WHERE OUR VIRGO GALAXIES ARE LOCATED (WILL PUT IT IN A %VESTIGE PAPER)}}}
%\end{figure}

\section{Comparison with other mechanisms}
\label{sec:comparemech}

The large set of results reviewed in the previous Sections can be used to derive some general considerations on the importance of RPS in shaping galaxy evolution in high density regions also in comparison with other perturbing mechanisms such as gravitational interactions (galaxy harassment, e.g. \citealt{Moore96,Moore98}) or the suppression of gas supply on the disc from the cosmic web once a galaxy becomes satellite of a larger halo (starvation, e.g. \citealt{Larson80, Bekki02, Kawata08}). In the following Sections we will show that the effects of RPS on galaxy evolution are significantly different than those due to starvation and harassment. We have, however, to caution that these processes are often acting together on individual galaxies and therefore it is not always possible to identify a single dominant mechanism responsible for specific observations.

%The former, with the suppression of gas infall on the disc, is expected to reduce the star formation activity uniformly at all galactocentric distances \citep[e.g.][]{Boselli06a}, thus producing fainter but non truncated stellar discs. The latter is able to truncate or perturb the gaseous and the stellar disc, in some instances producing disc instabilities such as bars driving gas infall into the nucleus possibly triggering nuclear star formation or AGN activity \citep{Moore96,Moore98,Lake98,Mastropietro05,Ellison11}. Gravitational interactions can easily increase the overall activity of star formation of the perturbed systems and move them over the main sequence \citep[e.g.][]{Ellison08,Scudder12}.

\subsection{Effects on galaxies and timescales}
%\subsection{Which is the typical timescale for gas stripping and star-formation quenching?}

In an RPS event gas removal occurs outside in, suppressing the star formation activity of the perturbed galaxies starting from the outer discs and moving to the inner regions until the ram pressure force cannot overcome the gravitational restoring force (see Sec. \ref{sec:physicalproc}). In dwarf systems, where the gravitational potential well is shallower, the stripping process can quench the star formation activity over the entire galaxy.
%The stripping process is more efficient when occurring face-on. In edge-on interactions, the ISM at the interface with the hot ICM can be compressed on the disc, triggering on short timescales the activity of star formation, producing giant \hii\ regions clearly observed in several nearby galaxies (e.g. CGCG097-073 in A1367, \citealt{Gavazzi95}, IC3476 in Virgo, \citealt{Boselli21}). Thus, at an early phase of the RPS interaction, when they are still star-forming and gas-rich, some galaxies may move above the main sequence relation, inevitably dropping below it once the stripping process has proceeded further.% This effect might be more pronounced when galaxies are accreted onto the cluster as isolated. Indeed, if they are accreted as part of a group they might have already lost or consumed part of their gas while in the group.
As detailed in Sec. \ref{subsec:timescale}, the RPS episode and the star formation quenching are relatively rapid ($\lesssim$ 0.5-1.0 Gyr) and typically shorter than the crossing time of the cluster ($\tau_{\rm crossing}$ $\simeq$ 2 Gyr, \citep{Boselli06}), with only a small fraction of the most bound gas being stripped on longer timescales. These timescales are valid for intermediate mass galaxies and can be even shorter for dwarf systems ($M_{\rm star} \lesssim 10^9 \mathrm{M_\odot}$). 
%With the cluster self-similar relations, the crossing time, $\sim R_{\rm vir} / \sigma_{\rm V}$, is independent of the cluster mass. 
In lower density regions the local conditions might not efficiently remove a significant fraction of the gas content of the perturbed galaxies, thus only partially reducing their star formation activity. As described in Sec. \ref{subsec:redshift}, RPS is expected to be more efficient at higher $z$ due to the evolution of the ICM parameters. However, at early epochs, the galaxies were more compact and gas-rich than their local analogues, making their gas more gravitationally bound. The gas removal due to ram pressure might thus have less critical effects on the star formation activity, possibly leading to longer RPS timescales compared to the local Universe.

These RPS timescales are shorter than those expected for other processes like starvation and harassment. Starvation or strangulation, i.e. the stop of gas infall on the stellar disc, quenches the activity of star formation on very long timescales. Although the molecular gas depletion timescale is $\tau_{\rm H_2}$ $\simeq$ 0.6-2.3 Gyr \citep{Leroy08, Leroy13,Bigiel11,Boselli14}, the total amount of cold gas (HI+H$_2$) available on the disc of unperturbed spiral galaxies can sustain star formation at a constant rate for $\simeq$ 4 Gyrs \citep{Boselli14}, or even longer considering that $\sim$ 30\%\ of the stellar mass is re-injected into the ISM through stellar winds as recycled gas \citep{Kennicutt94}. Indeed, tuned models of galaxies suffering a starvation scenario, where the infall of fresh gas has been stopped, show that a significant quenching of the star formation activity occurs only after $\gtrsim$ 7 Gyr \citep{Boselli06a, Boselli14b}. These long timescales can be slightly reduced if associated to efficient outflows \citep{McGee14, Balogh16, Trussler20}. The same models also indicate that the decrease of the star formation activity is uniform at all galactocentric distances, producing shallow but extended stellar discs \citep{Boselli06a, Boselli14b}. Starvation cannot produce truncated discs on short timescales as those observed in HI-deficient galaxies, nor thick stellar discs as those of typical bright lenticular galaxies in nearby clusters \citep[e.g.][]{Dressler04, Boselli06}. 
Some works based on complete samples of galaxies from large surveys (e.g. SDSS, GAMA, \galex) suggest long quenching timescales typical of starvation \citep{McGee09, Wolf09, von-der-Linden10, DeLucia12, Wheeler14, Taranu14, Haines15, Paccagnella16}. Indeed, \citet{Wetzel13}, deriving the quenching timescales of satellite galaxies in SDSS, found that for the complete satellite population the total quenching times are relatively long $\sim 4-6$ Gyr. Moreover these authors found that the most likely quenching path for satellite galaxies is the so-called ``delayed+rapid'' quenching where the SFR of galaxies remain largely unaffected during the delay time (at the expenses of \hi\ being converted into molecular gas) and then the SF activity fades rapidly when the molecular gas reservoir runs out. 
%It was also found that the delay times become shorter with higher halo mass and lower stellar mass suggesting an increasing contribution of faster quenched ram pressure stripped galaxies in the average population. These results are further complicated by the fact that satellite galaxies already experience environmental effects (at least a loss of fresh gas infall) when they enter group-like environments (pre-processing; e.g. \cite{Wetzel12, Wetzel13, Wijesinghe12}), while RPS becomes effective only in more massive haloes and possibly only for galaxies with high relative velocities with respect to the ICM.
Long quenching timescales appear to be common at $z \gtrsim 0.5$ indicating a larger and larger dominance of starvation (or lack of gas accretion onto satellite galaxies) compared to RPS as redshift increases \citep{Balogh16, Fossati17, Foltz18}.

Harassment also requires relatively long timescales ($\simeq$ 3 Gyr) to have significant effects on the perturbed galaxies \citep[e.g.][]{Moore96, Moore98, Gnedin03a, Mastropietro05, Bialas15}. This is due to the fact that the relaxation time, i.e. the time necessary for a galaxy to have its orbit perturbed by a tidal interaction, is very long ($\tau_{\rm relax}$ $\geq$ 10$^{10}$ yrs, \citep{Byrd90, Binney08}) in rich clusters such as Coma \citep{Boselli06}.
Indeed, the probability that an object has to undertake multiple encounters, or a close interaction with another cluster member, is relatively low and might require several crossing of the cluster \citep{Moore98, Mastropietro05}. Harassment is able to truncate the gaseous (and stellar) disc, but also favors dynamical instabilities able to form bars and eventually induce gas infall in the inner regions, thus triggering the nuclear star formation or AGN activity of the perturbed systems \citep{Henriksen96,Moore96,Moore98,Lake98,Mastropietro05,Ellison11}. Gravitational interactions can easily increase the overall activity of star formation of the perturbed systems and move them over the main sequence \citep[e.g.][]{Ellison08,Scudder12}. Evidence for an increased nuclear activity of cluster galaxies has been searched for in nearby clusters, but never found with statistically convincing results \citep[e.g.][]{Moss93,Moss00,Moss06}. An increased activity has been found in the infalling filaments, but far from the cluster, probably induced by tidal interactions \citep[e.g.][]{Porter08}.

\subsection{Region of influence}
%\subsection{In which environment is RPS the dominant perturbing mechanism?}

As presented in Sec. \ref{sec:obsevid}, there is a large amount of observational evidence corroborated by models and numerical simulations, indicating that RPS is the dominant perturbing mechanism in local clusters. In the Coma and A1367 clusters tails of ionised gas are observed in $\sim$ 50\%\ of the late-type galaxies \citep{Yagi10, Yagi17, Gavazzi18a} making this systems ideal targets for RPS studies. (see Sec. \ref{subsec:ionised}). This fraction can be even higher if we include tails recently detected in radio continuum \citep{Chen20, Roberts20} (see Sec. \ref{subsec:radcont}). 
If we consider that some tails might be missed for projection effects or for sensitivity, and 
that the tails observed in H$\alpha$ and radio continuum have a limited life time (10-40 Myr, see Sec. \ref{subsec:ionised} and \ref{subsec:radcont}), we can justify that ram pressure is active and probably dominant in local clusters of mass $M_{cluster}$ $\simeq$ 10$^{15}$ M$_{\odot}$.
%CHECK THE MASS OF A1367 THAT WE HAVE TO ADD IN TABLE 1, IN BOSELLI GAVAZZI 2006 IS 6.9 10$^{14}$. 
This result can also be extended to Virgo ($M_{\rm Virgo}$ $\simeq$ 10$^{14}$ M$_{\odot}$, see Sec. \ref{subsec:ionised}), where there is clear evidence that ram pressure is at place on some of the most massive spirals of the cluster (e.g. NGC 4501, NGC 4569, \citealt{Vollmer04a, Boselli06a, Boselli16}), and should be even more efficient on lower mass galaxies (see Sec. \ref{subsec:galpars}).%, making RPS effective also on lower mass clusters.
The recent extremely deep radio observations of the Fornax cluster with {MeerKAT} (see Sec. \ref{subsec:atomic}) identified a large number of late-type systems with extended tails of \hi\ gas proving that ram pressure can be efficient also in clusters/groups of mass $M_{\rm cluster} \simeq 5 \times 10^{13}$ M$_{\odot}$. 
At the lowest range of group haloes, galaxies with extended radio continuum tails have been observed by the LoTSS survey in groups of mass $M_{\rm group}$ $\simeq$ 3 $\times$ 10$^{12}$ M$_{\odot}$, showing that RPS can occur also in these haloes. However, in these environments RPS galaxies are rarer than in more massive clusters making the process subdominant and possibly less efficient \citep{Roberts21a}.

Having established that RPS is a very relevant (sometimes dominant) process in massive haloes, another question that arises is how its efficiency varies with the distance from the central galaxy. Tails of stripped gas have been observed mainly within the virial radius of local clusters. The most extreme known case being the galaxy CGCG097-026 in A1367, where an \hi\ tail has been detected at a projected $R/r_{200} \simeq$ 2  (\citet{Scott12, Scott22}, see Sec. \ref{subsec:atomic}). 
A similar evidence comes from galaxies with truncated gaseous or star forming discs which mostly live within the virial radius of nearby clusters and groups (see Sec. \ref{subsec:truncdiscs}). Furthermore, the distribution of HI-deficient galaxies (see Sec. \ref{subsec:integrated}) peaks in the core of the cluster becoming similar to that of isolated objects at $\simeq$ 1-2 $\times$ $r_{200}$. We can thus conclude that RPS shows its effects mainly within the halo virial radius as also shown in the phase-space diagram in Figure \ref{fig:PSDbox}. 

These radial scales can be compared to those predicted by simulations for the other main perturbing mechanisms. \citet{Byrd90} and \citet{Moore98} consistently underline that gravitational perturbations due to the potential well of the whole cluster and to flyby encounters with other members, whose density rapidly decreases with the clustercentric distance, make harassment particularly efficient in the innermost regions, at $\lesssim$ 0.5 $\times$ $r_{200}$ \citep[e.g.][]{Moore98, Smith10, Bialas15}. Conversely, starvation occurs once a galaxy enters the hot halo surrounding a massive cluster which could be outside the virial radius (up to $R$ $\simeq$ 3-5 $\times$ $r_{200}$ \citep{Bahe13, Ayromlou21}. However the effects of starvation on the star formation activity of galaxies can be delayed compared to the time of virial radius crossing. This could lead to an observed radial distribution of passive objects more similar to that expected for RPS. Moreover stripped or harrassed galaxies can ``backsplash'' the halo after the first crossing and possibly after having lost gas and star formation activity. They can account for up to 40\% of the galaxies observed between 1 and 2.5 virial radii \citep{Wetzel13} further complicating an accurate determination of the radius of influence of the aforementioned processes. We therefore warn that the sole position of a perturbed galaxy is not a unique indicator of the perturbing process, however the position in phase-space combined with accurate star formation indicators and ideally with the reconstruction of the star formation history of the galaxy can help in identifying the processes it underwent.

\subsection{Dependence on structure formation}

Despite the general picture of structure formation and galaxy evolution is becoming increasingly clear, it is still hard to quantify the relative importance of RPS with respect to that of other perturbing mechanisms at different epochs and in different environments. As mentioned in Sec. \ref{subsec:redshift}, we are still not able to trace with no ambiguity how the efficiency of RPS varied with time because of the complex and competitive evolution of the density of the ICM and of the gravitational bounding forces keeping the gas anchored to the stellar disc of galaxies with increasing $z$. 
However, we can intuitively argue that gravitational perturbations were more important than RPS at early epochs. Indeed, there are several indications that the infall of galaxies occurred via smaller groups, characterised by a relatively low velocity dispersion and a high density of objects, where gravitational interactions are more and more efficient in detriment of RPS (pre-processing, \citep{Dressler04, Fujita04}). 
Still controversial, however, is the fraction of galaxies accreted as isolated vs. group members. While the merger-tree used in the 
semi-analytic models of \citet{McGee09} and \citet{DeLucia12} and the hydrodynamic simulations of \citet{Han18} consistently indicate that a 10$^{14.5}$ M$_{\odot}$ mass cluster at $z$=0 has accreted, on average, $\sim$ 40-50\%\ of galaxies of stellar mass $M_{\rm star}$ $\gtrsim$ 10$^8$-10$^9$ $h^{-1}$ M$_{\odot}$ via groups with halo masses $M_{\rm group}$ $>$ 10$^{13}$ M$_{\odot}$, 
the N-body simulations of \citet{Berrier09} rather indicate that 70\%\ of the galaxies are accreted as isolated objects in clusters of similar mass. The same simulations also suggest that less than 12\%\ of the galaxies are accreted within groups with five or more systems, where pre-processing can efficiently perturb their evolution. The merger-tree used by \citet{McGee09} also suggests
that galaxies accreted through groups are, on average, more massive than those accreted as isolated systems. The same models also indicate that the fraction of galaxies falling in as isolated objects was higher in the past than at the present epoch. For reference, we recall that the typical infall rate of galaxies of mass $M_{\rm star}$ $\geq$ 10$^{8}$ M$_{\odot}$ in a cluster like Virgo ($M_{\rm Virgo}$ $\simeq$ 10$^{14}$ M$_{\odot}$) is $\sim$ 200 gal/Gyr \citep{Boselli08, Sorce21}.
Nonetheless we have observational evidence of two RPS galaxies in a cluster of mass $M_{\rm cluster}$ = 2 $\times$ 10$^{14}$ M$_{\odot}$ at $z\sim0.7$ suggesting that, although rare, this process was already at place in these structures. We still do not have sufficient observational data to see whether this process was dominant at these epochs.
Given the self-similar evolution of the ICM density suggested by the existing data to $z$ $\simeq$ 1.8 as discussed in Sec.~\ref{subsec:prophidens}, we can conclude that ram pressure was probably at place in all this redshift range. 

It is interesting to underline that, through its first infall into a cluster, which generally occurs on radial orbits, a gas-rich galaxy suffers first a suppression of gas infall on the disc (starvation), with the later addition of an active removal of the cold ISM via RPS. At its passage close to the cluster core, the galaxy might also suffer a violent gravitational perturbation. This evolutionary scenario is only slightly changed if a galaxy is accreted from within a group, in which case the infalling system is expected to suffer starvation already in the group environment \citep[e.g.][]{Kawata08}, possibly combined with gravitational perturbations. 
A typical example is the Fornax cluster, where galaxies with tails have often perturbed morphologies, suggesting that RPS and gravitational perturbations are both active, as predicted by cosmological hydrodynamic simulations in this halo mass regime \citep[e.g.][]{Marasco16}. As mentioned in Sec. \ref{subsec:mainmech}, the two families of mechanisms have additive and interwined effects, thus contributing both directly and indirectly to the stripping of the gas \citep[e.g.][and references in Sec. \ref{subsec:mainmech}]{Weinmann10, Bahe15, Henriques15, Trayford16, Cortese21}. Gravitational perturbations, indeed, flatten the gravitational potential well of galaxies, thus favoring the RPS process. 

Other interesting objects are the lenticular galaxies inhabiting nearby clusters. While the physical, structural, and kinematical properties of the dwarf spheroidal galaxies can be easily reproduced by an RPS scenario \citep[e.g.][]{Boselli08,Boselli14c}, the origin of massive lenticulars is probably dominated by gravitational interactions. Indeed, the structural properties of massive S0 galaxies and the weak dependence of the morphology segregation effect on galaxy density \citep[e.g.][]{Dressler04} are hardly explained by an RPS transformation of spirals into lenticulars, but rather suggest an active role of gravitational perturbations occurring at early epochs
when these objects were still members of smaller groups (pre-processing, \citep[e.g.][]{Byrd90, Gnedin03a,Dressler04, Fujita04}). This suggests that RPS, if present, was not dominant in the early Universe on the most massive galaxies entering the most massive haloes at that epochs. 
The same considerations also rule out starvation which is a milder process than RPS: in particular, the lack of fresh gas supply on long timescales after the perturbed galaxy becomes satellites of a larger halo, would lead to a decrease of the surface brightness of the inner disc, opposite to what observed in nearby massive lenticulars \citep[e.g.][]{Boselli06a, Boselli22}. Starvation in the main cluster halo, which would require long timescales to be efficient, is also in odd with the observed recent formation of lenticulars, relatively rare at redshift $z$ $\sim$ 0.5 but dominant in local clusters \citep{Dressler97}.
To conclude, these scenarios clearly indicate that the average galaxy population in rich environments is affected by multiple effects and the objects where a single process dominates are generally rare.

\subsection{Effects on the surrounding environment}

In the RPS scenario all the components of the ISM (atomic, molecular, ionised, hot gas, and dust) can be stripped from the disc (see Sec. \ref{subsec:tails}).
The stripped ISM remains multi-phase, as some stripped ISM can cool to form stars and the mixing between the stripped ISM and the hot ICM proceeds. As discussed in Sec.~\ref{subsec:clumping}, \ref{subsec:enrichment} and \ref{subsec:ICL}, RPS pollutes the ICM with the stripped ISM, effectively enriching the ICM with time, adding turbulence in the ICM, inducing density inhomogeneity or clumping in the ICM and increasing the ICL content with star formation in the stripped ISM.
Nevertheless, the late enrichment with RPS can only contribute to a small fraction of the ICM metals ($\sim$ 10\% - 15\%), although the roles of RPS in early enrichment are still unclear, mostly because of the uncertainty in the early enrichment models. The contribution to the ICL by the stars formed within the tails of stripped material of RPS galaxies is also small as discussed in Sec.~\ref{subsec:ICL}.

Starvation, on the other hand, starts to take effect from relatively low-density regions so the galactic halo gas is mixed with the ICM at the outskirts where clumping is stronger than the cluster centre. The halo gas, with low density and the virial temperature of the galaxy, can hardly cool to form stars and may only be quickly heated to replenish the ICM, without leaving long-lasting ICM clumps or adding the ICL content. The halo gas generally has lower metallicity than the disc ISM so its role on the ICM enrichment is small. 

Gravitational perturbations like mergers, tidal interaction and harassment act indifferently on all the different components of the perturbed galaxies (all forms of baryonic and dark matter). For cluster galaxies, the galactic dark matter halo is tidally truncated. Both stars and ISM can be dynamically heated to become unbound in the intracluster space, increasing the ICL and ICM content. It is believed that tidal interactions and mergers contribute to the ICL the most over the hierarchical growth of clusters as discussed in Sec~\ref{subsec:ICL}. 

\section{Conclusions}

The works presented in this review have clearly shown how galaxies suffering an RPS event can be identified with no ambiguity by the presence of extended tails of gas at different phases without a diffuse and evolved stellar counterpart. Galaxies which have been perturbed in the past are generally characterised by radially truncated ISM discs, combined with a normal and extended stellar distribution. Other selection criteria, such as those based on the identification of strong asymmetries in broad-band optical images (jellyfish galaxies), require further multi-wavelength data to ensure they are RPS objects. RPS galaxies where star formation is occurring within the tail would satisfy the jellyfish selection but they are a sub-sample of the full RPS galaxy population.

The results we have collected and presented in this review consistently indicate that RPS is a major mechanism affecting galaxy evolution and it should be always taken into account when the galaxy environment is studied. Systems undergoing or recently perturbed by an RPS event are the dominant galaxy population in nearby massive clusters ($M_{\rm halo}$ $\gtrsim$ 10$^{14}$ M$_{\odot}$), where radio continuum, HI, H$\alpha$ tails or truncated discs are ubiquitous provided that wide field observations are gathered with a sufficient sensitivity. Galaxies with extended tails have been identified down to halo masses of the order of a few 10$^{12}$ M$_{\odot}$, but are generally rare and sometime associated to perturbed stellar morphologies witnessing at the same time an ongoing gravitational perturbations. Galaxies suffering an RPS event have been also identified in clusters up to $z$ $\sim$ 0.7, but the lack of complete and wide field spectroscopic surveys still prevent a firm understanding of how frequent these objects are. The physical properties of galaxies and of the high-density regions, and their expected evolution as predicted by hydrodynamic cosmological simulations, combined with statistical galaxy samples let us think that RPS has a milder impact than other mechanisms (starvation, harassment) on galaxy evolution at higher redshift.

Tuned models and hydrodynamic simulations indicate that the RPS event removes the different component of the ISM outside-in. Whereas in dwarf systems the stripping is generally complete, in massive systems some gas can be retained in the inner region, where the gravitational potential well of the galaxy is the deepest. 
The activity of star formation is reduced by the lack of gas. The perturbed galaxies thus migrate below the main sequence becoming quiescent, red objects without affecting their kinematic properties and the structure of old stellar populations. This quenching process is rapid ($\lesssim$ 1 Gyr), in particular in dwarfs systems where most of the gas is removed at the first infall of the galaxy within the ICM before the pericenter of their orbit. Only under particular configurations such as edge-on stripping, when the perturbed gas is pushed along the disc, there might be a short lived burst of the star formation activity of the perturbed galaxies.

Star formation can be present in the tail of stripped material, but this does not occur in all RPS systems. There are, indeed, several objects with spectacular tails without any associated star forming region. The way the stripped gas can or cannot collapse to form stars is still poorly understood, and probably depends both on large (impact parameters between the galaxy and the surrounding ICM along the orbit) and small (turbulence, magnetic fields, self shielding of the gas within the tail) scale factors. A similar uncertainty (also related to this problem) remains in the way the cold gas stripped from the galaxy disc changes phase once mixed with the surrounding hot ICM. This gas contributes to the pollution of the ICM, but its effect remains marginal compared to the one from supernovae feedback in ellipticals.

%FUTURE OBSERVATIONS

Recent observations are clearly indicating H$\alpha$ narrow-band, HI, and radio continuum imaging surveys as the best way to identify galaxies undergoing an RPS event, provided that deep observations with wide field cameras coupled with sensitive instruments are used. To this regard, a large progress in this field will be made by the new generation radio facilities ASKAP, MeerKAT, and especially SKA, which will provide radio continuum and line observations at unprecedented sensitivity and angular resolutions for hundreds of thousands of galaxies in nearby and high redshift clusters. Significant progress in the characterisation of the properties of the hot ICM and in the study of the gas phase transformation within the tail of stripped material will be gathered by several X-rays space missions such as eRosita and ATHENA, which will also lead to the discovery of thousands new clusters at intermediate redshift.

%OPEN Qs

Despite the fact that our view on the importance of the RPS process in galaxy evolution has significantly improved since this mechanism has been first proposed in the seventies of last century to explain the nature of head-tail radio galaxies, main questions remains unanswered. Some of them are related to the physics which governs the phase transition of the cold gas stripped from the disc. Indeed new observations and models are required to explain why the tails are not always observed simultaneously in the different cold atomic, molecular, ionised, and hot gas components and understand whether this evidence is physical or just resulting from an observational bias. The same improvements in our understanding of the physics of the RPS process might also be necessary to understand under which conditions the stripped gas collapses or not to form stars. The two physical processes of gas phase transformation and star formation, might be intimately related to the presence, strength and orientation of magnetic fields. Finally, it is still not totally clear which is the role of RPS in galaxy evolution at high redshift, where deep multifrequency data at high angular and spectral resolution are still lacking for significant samples of galaxies. Upcoming multiplexed spectroscopic facilities (e.g. MOONS or ERIS at VLT, PFS at Subaru, JWST) will dramatically increase the size and quality of the observational samples at intermediate to high-$z$, helping us to solve the remaining pieces of the RPS puzzle.

\appendix

\section{Table of galaxies suffering RPS in nearby clusters}

Table \ref{tab:RPSgalaxies} shows the properties of galaxies suffering RPS in the nearby clusters Virgo, Coma, Norma and A1367. The description of the columns is as follows: \\
Column 1: galaxy name.\\
Column 2: morphological type: in order of priority: NED, GoldMine \citep{Gavazzi03}.\\
Column 3: nuclear classification, in order of priority from MUSE IFS or SDSS fiber spectra and published in \citet{Gavazzi11, Gavazzi13b, Gavazzi18}.\\
Column 4: stellar masses are derived using the calibration of \citet{Zibetti09} for a \citet{Chabrier03} IMF using the $i$ and $g$-band magnitudes taken in order of priority from \citet{Cortese12}, 
Gavazzi et al., in prep., or for a few objects from dedicated observations. The assumed distance of each object is that given in Table 1, with the exception of the Virgo cluster for which 
distances are taken as those of the different subgroups to which each galaxy belongs (see \citealt{Gavazzi99}).\\
Column 5: undergoing perturbation: RP = ram pressure, M = merging, H = harassment. When multiple perturbing mechanisms are at place, the first one reported has been identified as the dominant.\\
Column 6: projected distance from the X-rays centre of the cluster, in units of $r_{200}$. For the Virgo cluster, this is the distance from M87 even for those objects probably belonging to other cluster substructures.\\
Column 7: heliocentric velocities from targeted observations (see column 12), GoldMine \citet{Gavazzi03} or NED.\\
Column 8: electron density of the ICM at the projected distance of the cluster from the X-rays centre of the cluster, derived from Fig. \ref{ne_all}, in units of $10^{-4}~{\rm cm}^{-3}$\\
Column 9: HI-deficiency parameter, taken in order of preference from \citet{Boselli14d}, GoldMine \citep{Gavazzi03}, or from the dedicated references given in column 10.\\
Column 10: references to dedicated works on each single object, as follows. References for Virgo: AB11: \citet{Abramson11}, AB14: \citet{Abramson14}, AB16: \citet{Abramson16}, AB12: \citet{Arrigoni-Battaia12}
B05: \cite{Boselli05}, B06a: \citet{Boselli06a}, B16: \citet{Boselli16}, B16a: \citet{Boselli16a}, B18a: \citet{Boselli18a}, B18b: \citet{Boselli18b}, B21a: \citet{Boselli21}, B21b: Boselli et al. in prep., CH07: \citet{Chung07}, CH09: \citet{Chung09}, CO06: \citet{Cortes06}, CR20: \citet{Cramer20}, C05: \citet{Crowl05}, CK08: \citet{Crowl08}, DB08: \citet{Duc08}, F11: \citet{Fumagalli11}, F18: \citet{Fossati18}, HA07: \citet{Haynes07}, HE10: \citet{Hester10}, K04: \citet{Kenney04}, K14: \citet{Kenney14}, J13: \citet{Jachym13}, L20: \citet{Longobardi20a}, P10: \citet{Pappalardo10}, SO17: \citet{Sorgho17}, V: VESTIGE survey private comm., VE99: \citet{Veilleux99}, V99: \citet{Vollmer99}, V03: \citet{Vollmer03}, V03a: \citet{Vollmer03a}, V04a: \citet{Vollmer04a}, V05: \citet{Vollmer05}, V05a: \citet{Vollmer05a}, V06: \citet{Vollmer06}, V08: \citet{Vollmer08}, V08a: \citet{Vollmer08a}, V09: \citet{Vollmer09}, V18: \citet{Vollmer18}, V21: \citet{Vollmer21}, YO02: \citet{Yoshida02}, YO04: \citet{Yoshida04}
References for Norma: F16: \citet{Fossati16}, FU14: \citet{Fumagalli14}, LA22: \citet{Laudari22}, J14: \citet{Jachym14}, J19: \citet{Jachym19}, SU06: \citet{Sun06}, SU07: \citet{Sun07}, SU10: \citet{Sun10}, ZH13: \citet{Zhang13}
References for Coma:
BD86: \citet{Bothun86}, CH20: \citet{Chen20}, CR19: \citet{Cramer19}, CR21: \citet{Cramer21}. F12: \citet{Fossati12}, G89: \citet{Gavazzi89},G18: \citet{Gavazzi18a}, K15: \citet{Kenney15}, V01a: \citet{Vollmer01a}, Y10: \citet{Yagi10} 
References for A1367:
B94: \citet{Boselli94a}, C17: \citet{Consolandi17}, CO06: \citet{Cortese06}, F19: \citet{Fossati19}, P22: \citet{Pedrini22}, G78: \citet{Gavazzi78}, G84: \citet{Gavazzi84}, G87: \citet{Gavazzi87}, G89: \citet{Gavazzi89}, G01a: \citet{Gavazzi01}, G01b: \citet{Gavazzi01a}, G03: \citet{Gavazzi03a}, G17: \citet{Gavazzi17}, N82: \citet{Nulsen82}, Q00: \citet{Quilis00}, RO21: \citet{Roberts21}, S10: \citet{Scott10}, S12: \citet{Scott12}, S13: \citet{Scott13}, S15: \citet{Scott15}, S18: \citet{Scott18}, S22: \citet{Scott22}, Y17: \citet{Yagi17}

%\begin{landscape}{angle=90}
\begin{table}
\caption{Galaxies suffering RPS in nearby clusters }
{\tiny
\begin{tabular}{lcccrccrcll}
\hline

Name      	& Type		    & Nuc. & log$M_{\rm star}$ & Process	& $R/r_{200}$& c$z$	        & $n_{\rm{e}}$		&    HI-def	& Ref.          \\  
		    &		        &          & M$_{\odot}$&		    &		     & km s$^{-1}$	& $\frac{\mathrm{cm}^{-3}}{10^4}$	& 		    &		        \\    
\hline
\hline
Virgo       &               &          &            &           &           &               &           &           &               \\
\hline
NGC4254		& SA(s)c	    &  COMP     & 10.39			& H,RP		& 1.07		& 2405		& 0.6			&  0.06		& V05,HA07,DB08,B18a,CH09		\\ %V307, H102
NGC4294		& SB(s)cd	    &    HII    & 9.23			& RP		& 0.75		& 355		& 1.0			&  0.04		& CH07,CH09		\\ %V465,H110
NGC4299     & SAB(s)dm      &    HII    & 8.82          & RP        & 0.73      & 232       & 1.1           &  -0.11    & CH07,CH09   \\%V491
NGC4302		& Sc            &    LIN    & 10.44			& RP		& 0.94		& 1150		& 0.7			&  0.50		& CH07,CH09,B16a		\\ %V497, H113
NGC4330		& Scd?		    &    HII    & 9.52			& RP		& 0.63		& 1563		& 1.3			&  0.92		& CH07,CH09,AB11,F18,L20,V21		\\ %V630, H124 
NGC4388		& SA(s)b        &    AGN    & 10.14			& RP		& 0.38		& 2515		& 2.7			&  0.98		& VE99,YO02,V03a,YO04,CK08,CH09,P10,V18		\\ %V836, H144
NGC4396		& SAd	        &    HII    & 9.25			& RP		& 1.03		& -122		& 0.7			&  0.20		& CH07,CH09,B16a		\\ %V865, H148
NGC4402		& Sb	        &    HII    & 10.04			& RP		& 0.41		& 230		& 2.4			&  0.83		& C05,CK08,CH09,AB14,AB16,B16a,CR20		\\ %V873,H149
NGC4424		& SB(s)a?	    &    HII    & 10.17			& RP,M		& 0.92		& 437		& 0.8			&  0.98		& CO06,CH07,CK08,CH09,SO17,B18b		\\ %V979, H159
NGC4438     & SA(s)0a pec   &    LIN    & 10.66         & H,RP      & 0.29      & 104       & 3.9           &  1.20     & B05,V05a,V09      \\ %V1043, H163
NGC4469     & SB(s)a?       &    COMP   & 10.64         & RP        & 1.08      & 587       & 0.6           &  1.88     & B16a,B21b  \\  %V1190, H176    
NGC4470     & Sa?           &    HII    & 9.20          & RP        & 1.36      & 2341      & 0.4           &  0.32     & B16a,V     \\ %V1205, H177
NGC4491     & SB(s)a        &    HII    & 9.55          & RP        & 0.27      & 497       & 4.4           &  $>$1.27  & B16a,V     \\ %V1326,H184
NGC4501		& SA(rs)b	    &    LIN    & 10.98			& RP		& 0.61		& 2282		& 1.4			&  0.58		& B16a,V08a		\\ %V1401,H190
NGC4506     & Sa pec        &    HII    & 9.39          & RP        & 0.32      & 737       & 3.4           &  $>$1.31  & B16a,V     \\ %V1419,H192
NGC4522		&SB(s)cd        &    HII    & 9.38			& RP		& 0.97		& 2329		& 0.7			&  0.68		& K04,V06,V08,CK08,CH09,AB16,B16a,L20		\\%V1516,H197
NGC4548     & SB(rs)b       &    LIN    & 10.74         & RP        & 0.71      & 479       & 1.1           &  0.94     & V99   \\ %V1615,H208
NGC4569		& SAB(rs)ab	    &    LIN    & 10.66			& RP		& 0.50		& -221		& 1.8			&  1.05		& V04a,B06a,CK08,B16,B16a \\%V1690,H217
NGC4654		& SAB(rs)cd	    &    HII    & 10.14			& RP,H		& 0.99		& 1038		& 0.7			&  -0.05	& V03,CH07,CH09,B16a,L20		\\%V1987,H247
IC3105      & Im:           &    HII    & 7.89          & RP        & 0.98      & -167      & 0.7           &  0.41     & V     \\ %V241
IC3412      & Irr           &    HII    & 8.36          & RP        & 0.72      & 728       & 1.1           &  1.06     & V     \\ %V1179
IC3418      & IBm           &    PAS    & 9.06          & RP        & 0.30      & 38        & 3.7           &  $>$2.36  & H10, F11,J13,K14,V       \\
IC3476		& IB(s)m	    &    HII    & 9.03  		& RP		& 0.51		& -170		& 1.8			&  0.66		& B21a		\\%V1450,H193
UGC7636     & Im            &    -      & 8.20          & RP,H      & 1.32      & 473       & 0.5           &  0.85     & AB12  \\ %V1249
%NEW ADDED TO BE CHECKED
%4438 ?????? MESSA
%IC3412??????? 1179
%4491  VCC1326
% STELLAR MASSES AND HIDEF ARE FROM HRS OTHERWISE GUVICS EXCEPT V1249 recalculated using Zibetti from Arrigoni-Battaia 12; QA SED HRS when QF>0.5
\hline
\hline
Norma       &               &           &               &           &           &               &           &           &                \\
\hline
ESO137-001	& SBc?		    &    HII    & 9.81			& RP		& 0.18		& 4461		& 8.5			& 	-	    & SU06,SU07,SU10,FU14,J14,F16,J19	\\
ESO137-002	& S0?    &     -     & 10.54  		& RP		& 0.13 		& 5691   	& 12.0 			& 	-	    & SU10,ZH13,LA22	\\
\hline
\hline
Coma       &               &           &               &           &           &               &            &            &         \\
\hline
NGC4848		& Sab		    &   COMP     & 10.32		& RP		& 0.49		& 7193		& 2.5			&	0.25    &BD86,G89,V01a,F12,G18,CH20,RO21	\\ %160055
NGC4853     & S0            &   COMP     & 10.37        & RP        & 0.46      & 7676      & 2.8           &   -       &Y10,G18,RO21        \\ %160068
NGC4858		& Sb	        &	HII	     & 9.40      	& RP		& 0.25    	& 9430	    & 8.5		    &	0.96    &BD86,Y10,G18,CH20,RO21	\\ %160213
NGC4911		& Sa	        &	LIN	     & 10.89     	& RP		& 0.26	   	& 7995	    & 7.9			&   0.31    &Y10,G18	\\ %160260
NGC4921		& Sb	        &	LIN      & 11.34     	& RP		& 0.33	   	& 5481	    & 5.3		    &   0.73    &K15,G18,CR21	\\ %160095
IC837       & S0a           &   HII      & 10.00        & RP        & 1.40      & 7230      & 0.3           &   0.40    &RO21       \\%160041
IC3913      & Sc            &   HII      & 9.82         & RP        & 0.99      & 7525      & 0.6           &   0.18    &RO21   \\  %160026
IC3949		& Sa		    &	COMP     & 10.42     	& RP		& 0.27    	& 7553	    & 7.7		    &	1.01	&Y10,G18,CH20,RO21	\\ %160212
IC4040		& Sdm	        &	HII	     & 9.65      	& RP		& 0.16    	& 7637	    &16.2		    &	0.45	&BD86,Y10,G18,CH20,RO21	\\ %160252
CGCG160-020 & BCD           &   HII      & 9.22         & RP        & 0.90      & 4968      & 0.7           &   0.26    &RO21         \\
CGCG160-033 & S0a           &   AGN      & 10.02        & RP        & 1.18      & 6249      & 0.4           &   -       &RO21   \\ %GMP5226
CGCG160-058 & Sbc           &   HII      & 10.07        & RP        & 0.77      & 7616      & 0.9           &   0.17    &RO21   \\ %GMP4437
CGCG160-064 & Pec           &   HII      & 9.16         & RP        & 0.70      & 7368      & 1.2           &   0.92    &RO21   \\ %Mrk56
CGCG160-067 & Pec           &   HII      & 9.51         & RP        & 0.77      & 7664      & 1.0           &  -0.18    &RO21   \\ %Mrk57
CGCG160-073	& Sa	        &	HII      & 9.74      	& RP		& 0.35    	& 5435	    & 4.6		    &	0.69	&BD86,Y10,G18,CH20,RO21		\\%Mrk58
CGCG160-076 & Sc            &   HII      & 9.28         & RP        & 0.59      & 5358      & 1.7           &  -0.32    &RO21      \\ %GMP3253
CGCG160-086 & Sb            &   HII      & 9.31         & RP        & 0.32      & 7476      & 5.6           &   0.59    &BD86,CH20,RO21   \\ %GMP2599,KUG1258+279A
CGCG160-098 & Sc            &   HII      & 9.96         & RP        & 0.70      & 8762      & 1.2           &   0.31    &BD86,RO21   \\ %GMP2073
CGCG160-106 & S0a           &   HII      & 9.91         & RP        & 0.55      & 6902      & 1.9           &   0.34    &BD86,RO21   \\ %N4926A GMP1616
CGCG160-108 & Sb            &   HII      & 9.49         & RP        & 0.54      & 8202      & 2.0           &   0.89    &RO21   \\ %GMP1576
CGCG160-127 & Sc            &   HII      & 9.27         & RP        & 1.15      & 5500      & 0.4           &  -0.05    &RO21   \\ %GMP455 quite marginal
CGCG160-243	& BCD		    &	HII	     & 9.26      	& RP		& 0.10   	& 5316	    &25.9		    &	-	    &Y10,J17,G18,CR19,CH20,RO21	\\ %D100, GMP2910
GMP2601     & BCD           &   HII      & 8.96         & RP        & 0.43      & 5602      & 3.1           &   -       &RO21   \\ %
GMP2923	    & BCD		    &	PSB	     & 8.75      	& RP		& 0.18    	& 8726	    &14.3		    &	-	    &Y10,G18		\\
GMP3016 	& S..		    &	-	     & 7.49      	& RP		& 0.10    	& 7770	    &25.4		    &	-	    &Y10,G18		\\
GMP3071 	& BCD	        &	HII	     & 8.90      	& RP		& 0.20    	& 8975	    &12.2		    &	-	    &Y10,G18		\\
GMP3271 	& Pec	        &	HII	     & 9.20      	& RP		& 0.36    	& 5007	    & 4.5		    &	-	    &Y10,G18,RO21	\\
GMP3618     & Sa            &   HII      & 9.46         & RP        & 0.78      & 8412      & 0.9           &   -       &RO21   \\ %
GMP4060 	& Sm	        &	PSB	     & 9.12      	& RP		& 0.34    	& 8753	    & 4.9		    &	-	    &Y10,G18	\\
GMP4106     & Pec           &   HII      & 8.69         & RP        & 1.11      & 7467      & 0.4           &  -0.05    &RO21   \\ %
GMP4232     & S..	        &	-	     & 7.39      	& RP		& 0.49	   	& 7285	    & 2.4		    &	-	    &Y10,G18		\\
GMP4351     & BCD           &   HII      & 9.18         & RP        & 0.69      & 7447      & 1.2           &  -0.24    &RO21   \\
GMP4555     & Pec           &   HII      & 9.75         & RP        & 0.46      & 8142      & 2.8           &   -       &G18,CH20,RO21   \\%J125757+280342
GMP4570     & Pec           &   HII      & 8.53         & RP        & 0.46      & 4579      & 2.8           &   0.27    &G18,CH20       \\%J125756+275930
GMP4629     & BCD           &   HII      & 8.47         & RP        & 0.51      & 6935      & 2.2           &   -       &G18,CH20       \\%J125750+281013
%J125757.71+280342.4& Pec    &   HII      & 9.75         & RP        & 0.36      & 8142      &               &   0.53    &G18       \\ % IS GMP4555 or KUG1255+283!!!
%GMP3509       &               &           &               & RP        &           & 7000      &               &   -       &RO21   \\ %VERY MARGINAL
%J130545+285216&               &           &               & RP        &           & 7664      &               &   -       &RO21   \\ %VERY MARGINAL
%KUG1257+278   &               &           &               & RP        &           & 5012      &               &   -       &CH20   \\ %VERY MARGINAL
%
\hline
\hline
A1367       &               &           &               &           &           &           &           &       &        \\
\hline
UGC6697		& Pec		    &   HII     & 10.13			& RP		& 0.38		& 6727		&	5.1		&	0.00 &N82,G84,G87,G89,B94,G95,G01b,S10,S18,C17,Y17	\\
UGC6697N    & Pec           &   HII     & 8.78          & RP, H     & 0.37      & 7542      &   5.0     &   -    &C17    \\
CGCG097-026	& Pec		    &   HII     & 9.86			& H, RP		& 2.24		& 6195		&	0.1		&  -0.66 &S12,S22	\\ 
CGCG097-073	& Pec		    &   HII     & 9.27			& RP, H		& 0.57		& 7298		&	2.2		&  -0.09 &G87,G89,B94,Q00,G01a,Y17,P22,RO21		\\
CGCG097-079	& Pec		    &   HII     & 9.08			& RP, H		& 0.52		& 7026		&	2.6		&   0.17 &G87,G89,B94,G01a,S13,S15,Y17,P22,RO21		\\
CGCG097-092 & Sb            &   HII     & 9.62          & RP        & 0.56      & 6536      &   2.3     &   0.18 &Y17    \\
CGCG097-093 & Sc            &   HII     & 9.22          & RP, H     & 0.20      & 7298      &   13.0    &   0.47 &Y17    \\
CGCG097-114 & Pec           &   HII     & 9.22          & H, RP     & 0.05      & 8425      &   48.1    &  -0.32 &G03,C06,F19,RO21     \\
CGCG097-120 & Sa            &   LIN     & 10.53         & RP        & 0.08      & 5620      &   36.1    &   0.90 &G03,F19     \\
CGCG097-122 & Pec           &   HII     & 10.11         & RP, H     & 0.32      & 6721      &   6.2     &   0.35 &Y17    \\
CGCG097-125 & S0a           &   HII     & 10.32         & M, RP     & 0.08      & 8311      &   36.0    &  -0.05 &G03,C06,F19     \\
FGC1287		& Sdm		    &	HII	    & 9.81       	& RP, H		& 1.58		& 6825		&	0.2	    &   0.00 &S12	\\ %calculated using HI Scott et al 2012 only in the galaxy, 3.7x10^9
J114432.1+200623 & BCD   &   HII     & 9.87          & RP        & 0.43      & 7214      &   3.9     &   -    &G17    \\ %2MASX
\noalign{\smallskip}
\hline
\end{tabular}

\label{tab:RPSgalaxies}}
\end{table}
%\end{landscape}

\begin{acknowledgements}
We warmly thank the two anonymous referees for their accurate reading of the text and for their constructive comments and suggestions which helped improving the completeness and clarity of the manuscript.
We warmly thank S. Boissier, G. Consolandi, G. Gavazzi, A. Longobardi, A. Pedrini, P. Serra, S. Tonnesen, M. Yagi for constructive discussions and for giving us access to different set of multifrequency data and for their help in the preparation of some figures presented along the text.
We also thank C. Ge and S. Laudari on the help on several figures.
AB acknowledges financial support from "Programme National de Cosmologie and Galaxies" (PNCG) funded by CNRS/INSU-IN2P3-INP, CEA and CNES, France.
MF acknowledges funding from the European Research Council (ERC) (grant agreement No 757535).
MS acknowledges support provided by the SAO grants GO6-17111X and GO0-21118X, the NSF grant 1714764 and the USRA grant 09-0221.
Part of the results presented in this work have been gathered or collected from archival datasets by the VESTIGE survey team. We thus warmly thank all members of this collaboration in allowing us to use and reproduce some of their work and results obtained during this project.

\end{acknowledgements}


\bibliographystyle{aa}

\end{document}